\documentclass[11pt]{article}  

\usepackage{amsmath, amssymb, amsthm}
\usepackage{geometry}
\usepackage{graphicx}
\usepackage{algorithm}
\usepackage{algpseudocode}
\usepackage{array}
\usepackage{subcaption}
\usepackage{tikz}
\usepackage{fancyhdr}
\usepackage{enumitem}
\usepackage{authblk}
\usepackage{empheq}
\usepackage[numbers]{natbib}  

\theoremstyle{plain}
\newtheorem{theorem}{Theorem}[section]
\newtheorem{conjecture}[theorem]{Conjecture}

\theoremstyle{definition}
\newtheorem{definition}[theorem]{Definition}

\theoremstyle{remark}

\title{The $Z$-curve as an $n$-Dimensional Hypersphere: Properties and Analysis}

\author{
    \begin{minipage}[t]{0.45\textwidth}
        \centering
        Diego Vázquez González \\
        \texttt{d11215810@mail.ntust.edu.tw}
    \end{minipage}
    \begin{minipage}[t]{0.45\textwidth}
        \centering
        Hsing-Kuo Pao \\
        \texttt{pao@mail.ntust.edu.tw}
    \end{minipage}
}

\affil{
    Department of Computer Science and Information Engineering\\
    National Taiwan University of Science and Technology \\
    Taiwan
}

\begin{document}

\date{}
\maketitle

\begin{abstract}
In this research, we introduce an algorithm that produces what appears to be a new mathematical object as a consequence of projecting the \( n \)-dimensional \( Z \)-curve onto an \( n \)-dimensional sphere. The first part presents the algorithm that enables this transformation, and the second part focuses on studying its properties.


\end{abstract}


\section{Introduction}

Space-filling curves (SFCs), such as the Peano curve~\cite{peano1990courbe}, Hilbert curve~\cite{hilbert1935}, and the $Z$-curve~\cite{morton1966computer}, are discrete fractals formed by a continuous line that visits every cell in an $n$-dimensional unit cube exactly once. Each cell is assigned a unique number, and when connected sequentially by this line, a self-repeating pattern emerges that covers the entire space without intersecting itself. Initially, the hypercube contains $2^D$ cells. Each cell is subdivided into $2^D$ new cells, resulting in a total of $2^{D \times K}$ smaller cells, where $D$ is the number of dimensions and $K$ represents the number of subdivision levels (referred to in this research as the scaling factor). Each new cell is assigned a unique number from \( 0 \) to \( 2^{D \times K} - 1 \), and we denote this range of numbers by \( \mathbf{X}^D_K \). Each type of space-filling curve—whether Peano, Hilbert, or $Z$-curve—has its own method for assigning numbers to the initial cells and the cells resulting from subsequent subdivisions. These numbering methods determine the specific connection patterns within the hypercube when the cells are connected.\\

The $SFC$ that this research focuses on is the $Z$-curve, also known as the Morton curve or Lebesgue curve. In this curve, we assign specific directions to the bits, which determine the value of each cell, and there is always one bit per dimension. In 2D, the initial four cells correspond to the smallest number of space-directional divisions—up-right, up-left, down-right, and down-left, so we only need 2 bits to define those positions, one for the $X$ axis and the other for the $Y$ axis. The bits 0 and 1 specify the directions for up/down and left/right in both dimensions, and the curve moves according to how these bits assign the directions through the different dimensions. When the parameter \( K \) increases, continuing with the 2D case, each portion of space or cell is subdivided into four smaller cells, with each new cell receiving directions based on the original directions assigned to the bits. For instance, a cell initially oriented up-right subdivides into four new cells: up-right, up-left, down-right, and down-left. To calculate the number for each new cell, the bits corresponding to the new directions are added to the right of the binary representation of the previous cell. For example, if the original cell value was \( 01 \), after subdivision, the new cells would be assigned values \( 0111 \), \( 0101 \), \( 0110 \), and \( 0100 \), respectively. The resulting string of bits is then converted to a decimal number. By connecting these decimal numbers in sequence with a continuous thread, we can observe the emergence of the Z-pattern that characterizes the $Z$-curve.

\begin{table}[H]
    \centering
    \caption{Values of the $Z$-curve for \( K = 3 \)}
    \begin{tabular}{|c|c|c|c|c|c|c|c|}
        \hline
        21 & 23 & 29 & 31 & 53 & 55 & 61 & 63 \\ \hline
        20 & 22 & 28 & 30 & 52 & 54 & 60 & 62 \\ \hline
        17 & 19 & 25 & 27 & 49 & 51 & 57 & 59 \\ \hline
        16 & 18 & 24 & 26 & 48 & 50 & 56 & 58 \\ \hline
        5  & 7  & 13 & 15 & 37 & 39 & 45 & 47 \\ \hline
        4  & 6  & 12 & 14 & 36 & 38 & 44 & 46 \\ \hline
        1  & 3  & 9  & 11 & 33 & 35 & 41 & 43 \\ \hline
        0  & 2  & 8  & 10 & 32 & 34 & 40 & 42 \\ \hline
    \end{tabular}
    \label{tab:values}
\end{table}

As we continue fragmenting the space, we increase the precision in defining regions, starting with the larger area, like up-right, and then subdividing within that, for example, focusing on bottom-left within the up-right region. This process works similarly to a coordinate system. In fact, the $Z$-curve integrates its own coordinate system, known as the Morton coordinates~\cite{morton1966computer}. Morton coordinates encode multi-dimensional points into a single scalar value by interleaving the binary representations of each coordinate. This binary representation relies on two parameters: the dimension \( D \) and the expansion factor \( K \), which together result in \( D \times K \) bits. If the number has fewer than \( D \times K \) bits, this chain of bits will add leading zeros to reach the desired length. After obtaining the final binary representation, we split the chain of bits into blocks of size \( D \), with each bit corresponding to a specific dimension. The dimensions follow the order \([X, Y, Z, \dots]\), and we enclose each block in square brackets for clarity:

\[
[B_{1,1}, B_{2,1}, \ldots, B_{D,1}][B_{1,2}, B_{2,2}, \ldots, B_{D,2}][\ldots][B_{1,K}, B_{2,K}, \ldots, B_{D,K}]
\]

\begin{figure}[H]
    \centering
    \includegraphics[width=0.7\textwidth]{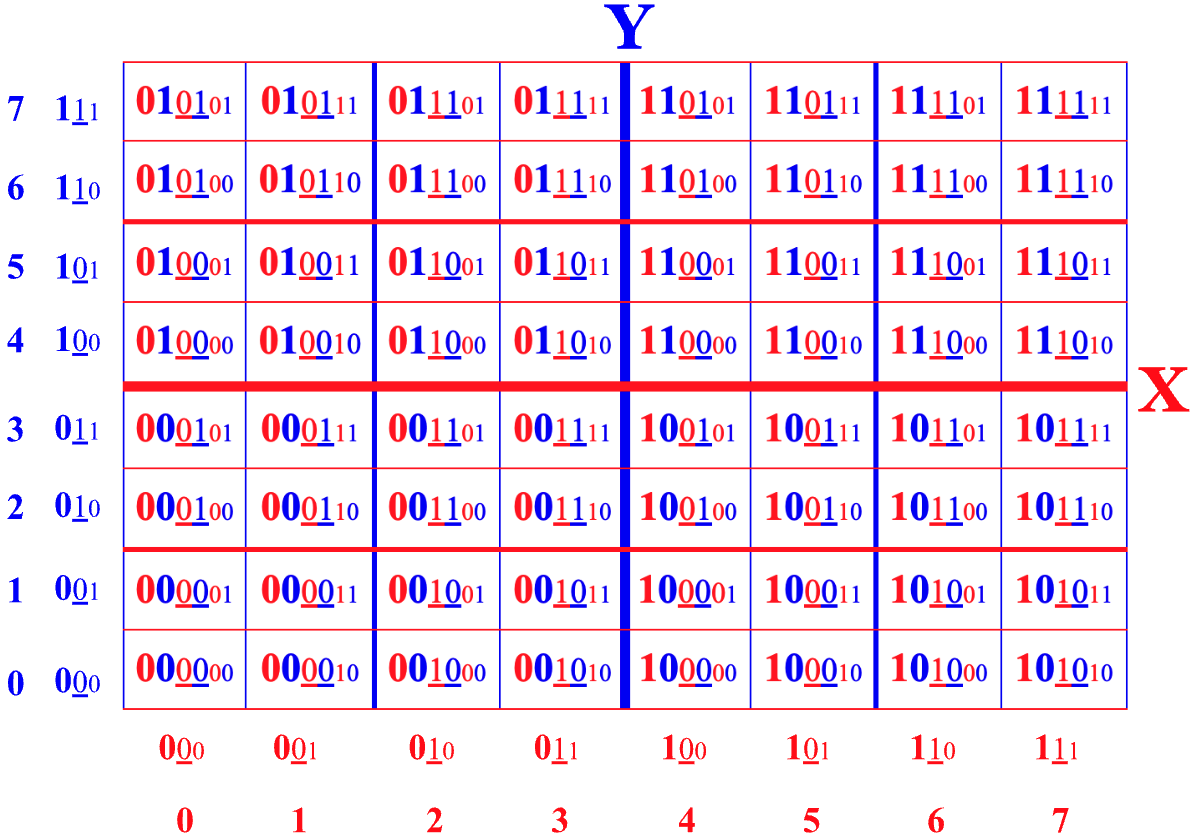}
\caption{The image shows how the values in \( \textbf{X}_3^2 \), when following the Z-order pattern, result from the intersection of the binary representations of the numbers associated with each row and column, creating the Morton coordinate system. To enhance readability and interpretation, different typographical styles are used for the bits in each dimensional component based on their position.}

\end{figure}

For example, to represent the number 47 with \( D = 2 \) and \( K = 3 \), the corresponding Morton coordinate is:

\[
[10][11][11] \,
\]

\noindent Later in this work, we will organize this binary representation using a matrix, where each column corresponds to the bits of the same dimension, and the number of rows corresponding to \( K \) :
\[
\begin{bmatrix}
B_{1,1} & B_{2,1} & \cdots & B_{D,1} \\
B_{1,2} & B_{2,2} & \cdots & B_{D,2} \\
\vdots & \vdots & \ddots & \vdots \\
B_{1,K} & B_{2,K} & \cdots & B_{D,K} \\
\end{bmatrix} \, 
\]
\noindent Following the example of the number 47 with \( D = 2 \) and \( K = 3 \), the matrix form of the Morton coordinates is:
\[
\begin{bmatrix}
1 & 0 \\
1 & 1 \\
1 & 1 \\
\end{bmatrix} \, 
\]

\section{Methodology}

\begin{table}[H]
\centering
\begin{tabular}{|c|l|}
\hline
\textbf{Symbol} & \textbf{Meaning} \\ \hline 
$D$ & Dimension of the space \\ \hline
$K$ & Scaling factor or resolution level \\ \hline
$BD$ & BitDistance \\ \hline
$\mathbf{X}_K^D$ & Set of numbers from $0$ to $2^{D \times K} - 1$ \\ \hline
$\mathbb{T}$ & Subset of numbers in \( \mathbf{X}_K^D \) \\ \hline
$\mathcal{O}$ & Subset of numbers in $\mathbb{T}$ \\ \hline
$S(\mathcal{X})$ & Sum of all elements in the input object \(\mathcal{X} \) \\ \hline
$S$ and $B$ & Scaling factors, where $S < B $ \\ \hline
$\phi^D_{(S,B)}$ & Mapping function from \( X^D_S \) to \( X^D_B \) \\ \hline
$\Psi_K^D$ & Constant related to the sum of the elements in \( \mathbb{T} \) \\ \hline
$\Upsilon^D_K$ & Total number of orthotopes in \( \mathbf{X}_K^D \) \\ \hline
$SG$ & SubsetGenerator \\ \hline
\end{tabular}
\caption{Notation Guide: Symbols and Their Meanings}
\end{table}

This section has two parts. First, we will explain the concept of BitDistance and its relationship with the $Z$-curve. Next, we will use the Morton coordinates and BitDistance to project the numbers from an $N$-dimensional $Z$-curve onto an $N$-dimensional sphere.

\subsection{BitDistance}

\begin{definition}[BitDistance]
\label{def:bitdistance}
The \textit{BitDistance} (\( BD \)) represents the distance a dimensional component of a number moves, determined by the directions and weights assigned to its bits.
\end{definition}

In Morton coordinates, each binary number contains \( D \) strings of bits, each of length \( K \) bits. The BitDistance is calculated in each dimension, resulting in \( D \) BitDistances. The position of each bit in the sequence of each dimension determines the magnitude of the movement, with the leftmost bits contributing more heavily to the distance. These contributions are weighted according to the inverse powers of \( 2 \). Each bit’s weight can be positive or negative, depending on the assigned direction, with the values 1 and 0 being arbitrarily assigned positive or negative weights. The directions assigned to the bits in up to \( 3D \) are as follows:

\begin{itemize}
    \item Bit = 1: Positive weights, $\{$ Right, Up, Forward $\}$.
    \item Bit = 0: Negative weights, $\{$ Left, Down, Backward $\}$.
\end{itemize}

Given a chain of bits \(b_0, b_1, b_2, \ldots, b_{K-1}\) representing the Morton coordinates in the dimension \(D\), we calculate the BitDistance for that dimension using the following expression:

\begin{equation}
BD_D(\{b_i\}) = \sum_{i=0}^{K-1} \left( (-1)^{1 - b_i} \cdot \frac{1}{2^{i+1}} \right)
\end{equation}

Here, \(b_i\) denotes the \(i\)-th bit in the binary representation for dimension \(D\), and \(K\) is the total number of bits. We apply this function to each dimension in the Morton coordinates to obtain the corresponding Cartesian coordinates.\\

Obtaining the $BD$ for each dimensional component of each number, we construct a vector that positions the integers from \(0\) to \(2^{D \times K} - 1\) within a \(D\)-dimensional cube \( \left[ 0, 1 \right]^D \), where the numbers follow the $Z$ pattern.\\

\begin{figure}[h!]
    \centering
    \includegraphics[width=0.5\textwidth]{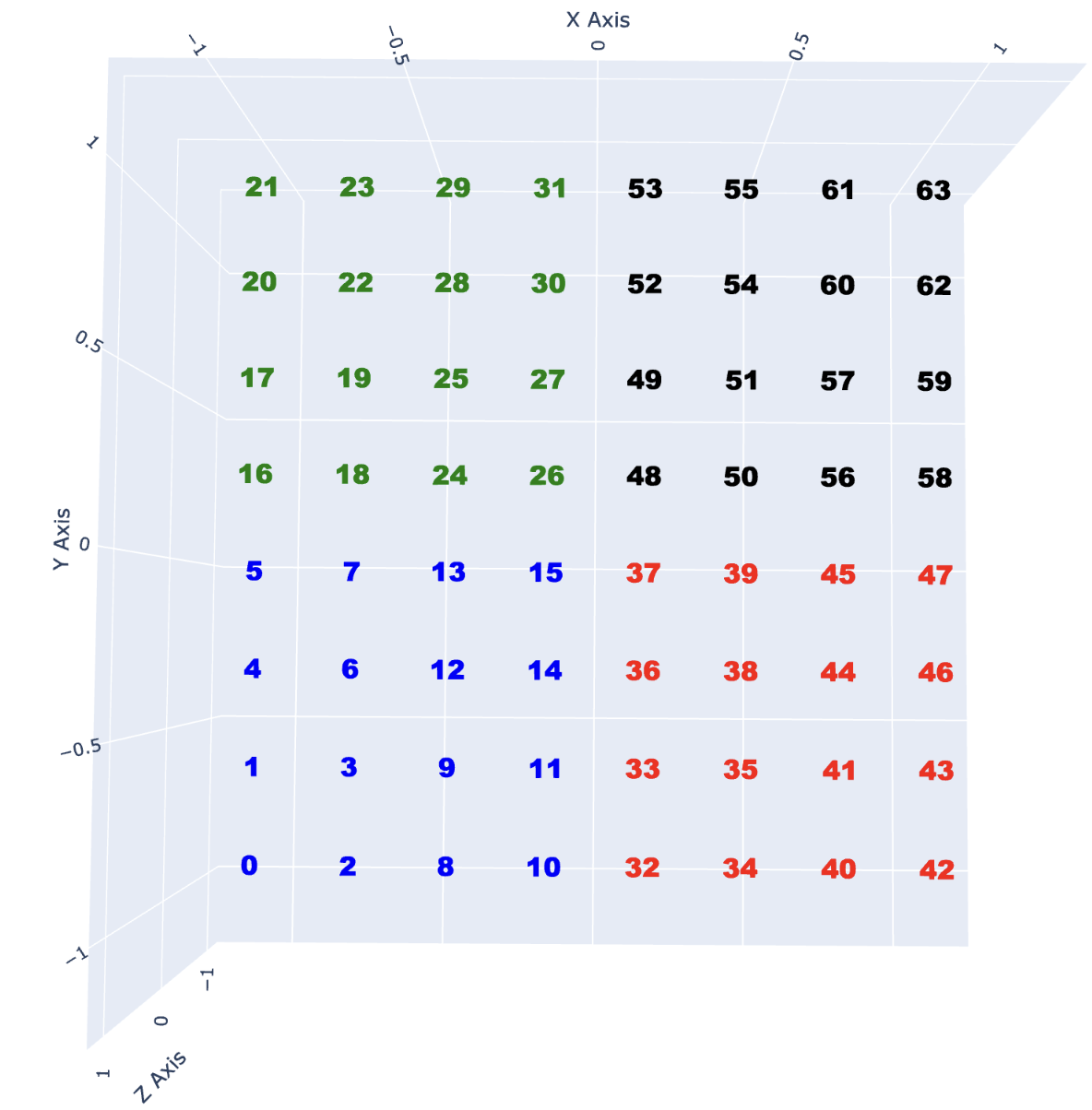}
    \caption{This image plots the set of numbers in $\mathbf{X}^2_3$, showing where each number is placed on the Cartesian plane after calculating the $BD$ in each dimension. We can observe that after being plotted, they are placed in the space following the $Z$-pattern. The numbers that start with the same $D$ bits share the same color, corresponding to whether they begin with 00, 01, 10, or 11.}

\end{figure}

\subsection{Projecting onto an $N$-Dimensional Sphere}

The $Z$-curve forms a fractal, exhibiting a self-repeating structure. A point or number on the fractal takes different relative positions depending on the zoom level. For example, a point may appear in the upper-right at one scale but shift to the bottom-left when zooming in. Zooming in, we focus only on numbers with the same leftmost \( D \) bits as the target number. Each zoom step ignores or removes 1 bit per dimension, halving the space size in each dimension. Consequently, the new space includes only the numbers that match the same \( D \times \text{Zoom} \) bits as the target number. Therefore, when calculating the position of a number after zooming in using the $BD$, we ignore the bits already excluded during the zoom process.\\

The absolute position of a point, represented by the $BD$ calculated taking into consideration all the bits, and the relative position, determined by the $BD$ after ignoring at least one bit per dimension, are independent. We will specify the number of bits ignored in the calculation to distinguish between the BitDistance ($BD$) calculated using all the bits and the BitDistance that ignores $D \times n$ bits. Each time we reference a BitDistance that ignores certain bits, we denote it as $BD^n$, where \( n \) indicates the number of bits excluded per dimension, with \( n \) ranging from 0 to $K-1$. For instance, when ignoring the first $D$ bits in the calculation, we label it as $BD^1$.\\

To link these positions and encapsulate all possible locations of a point $P$ into a single value, we sum over all potential positions it can have. However, this direct summation assumes equal contribution from each position. To address this, we will :

\begin{enumerate}

\item For each number in $\mathbf{X}^D_K$, we progressively ignore one leftmost bit per dimension. Each time we ignore one bit per dimension, we calculate $BD^n_D$. Starting with an initial weight of \( \frac{1}{2} \) assigned to $BD$ calculated using all the bits, each subsequent BitDistance calculation reduces the weight by half as we ignore more bits. This process is expressed by the formula:

\[
\sum_{n=0}^{K-1} \left(\frac{1}{2}\right)^{(n+1)} \times BD^n_D
\]

\item Normalize the BitDistance values across all dimensional components. We do it in the $BD$ we calculated using all the bits and for all the other $BD$ that miss at least one bit per dimension. To normalize each $BD^n_D$, we apply the Pythagorean theorem. First, calculate the hypotenuse by summing the squares of the BitDistances values for each dimension and then taking the square root of this sum:

\[
H = \sqrt{ \sum_{i=1}^{D} (BD^n_i)^2 }
\]

We obtain a total of $K$ hypotenuses, one for each BitDistance at each zoom level, $BD^n$. Then, we use each of these hypotenuses to normalize the corresponding $BD^n$. By applying this process to all the numbers in $\mathbf{X}^D_K$, we distribute them across the surface of the \(n\)-dimensional sphere.

\begin{figure}[H]
    \centering
    \begin{minipage}[b]{0.43\textwidth}
        \centering
        \includegraphics[width=\textwidth]{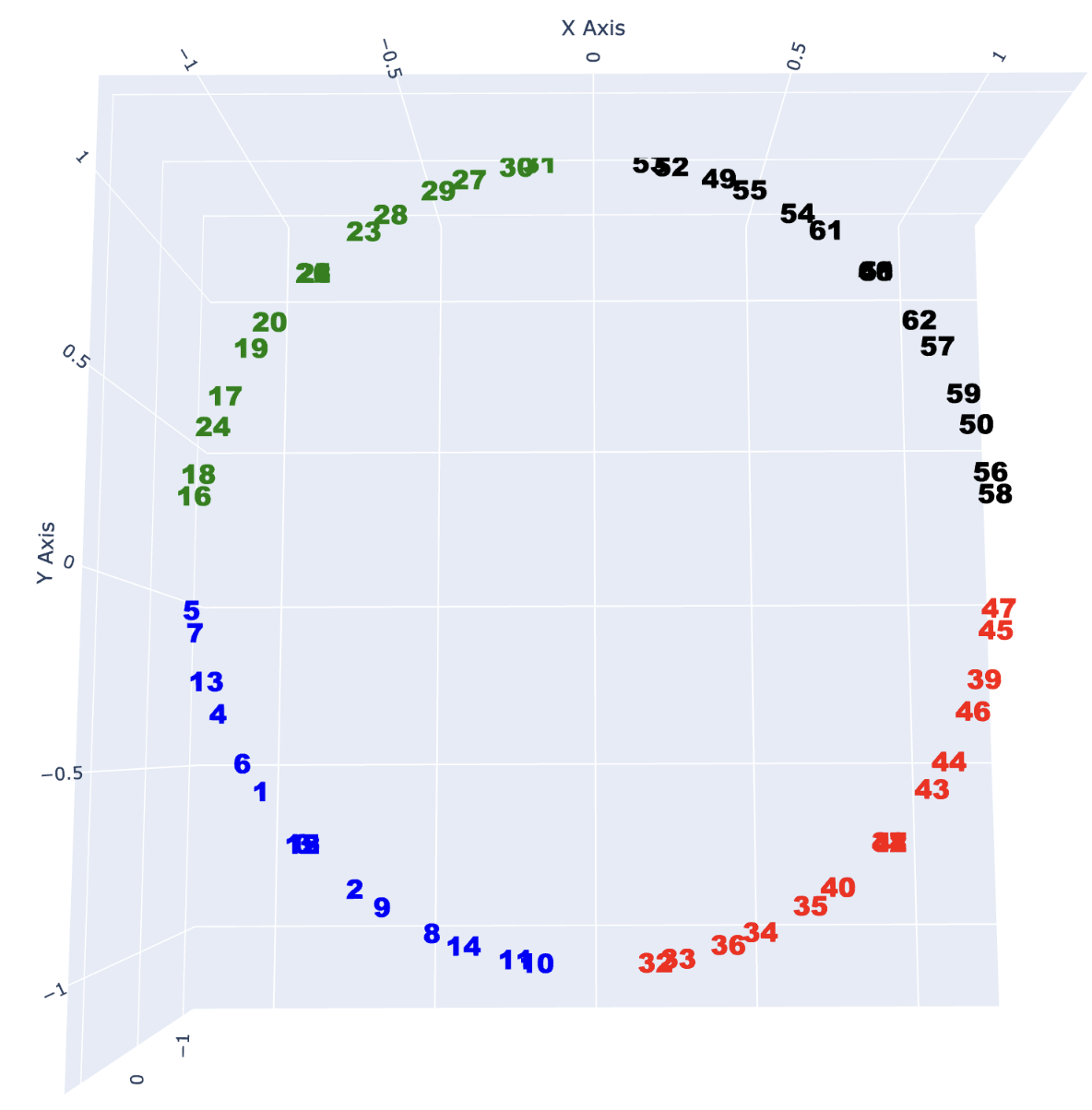}
    \end{minipage}
    \hspace{0.05\textwidth}
    \begin{minipage}[b]{0.43\textwidth}
        \centering
        \includegraphics[width=\textwidth]{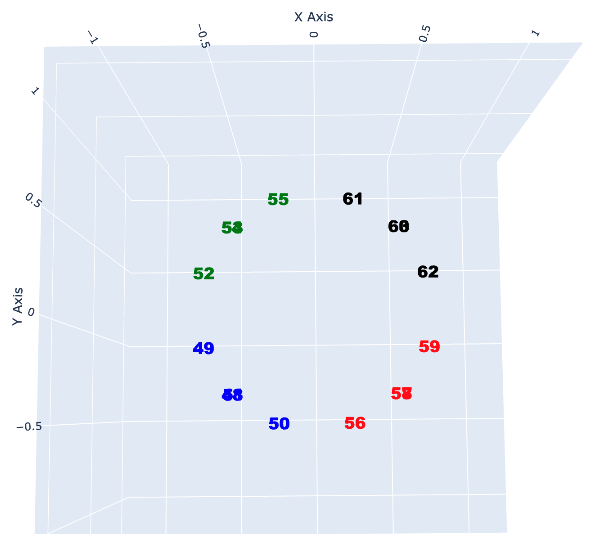}
    \end{minipage}
\caption{These images show the numbers in $\mathbf{X}^2_3$ after calculating and normalizing their BitDistance. The color of each number corresponds to whether it starts with 00, 01, 10, or 11. The image on the right displays all numbers in $\mathbf{X}^2_3$, while the image on the left shows only the numbers that start with 11.}

\end{figure}

  In the previous images, multiple numbers share the same position. These numbers correspond to those on the diagonal of the $2D$ $Z$-curve; all of them converge in the same location. This occurs until the final zoom level is reached, when there will be $2^D$ numbers, each of them with a unique position.

    \begin{figure}[H]
    \centering
    \includegraphics[width=0.45\textwidth]{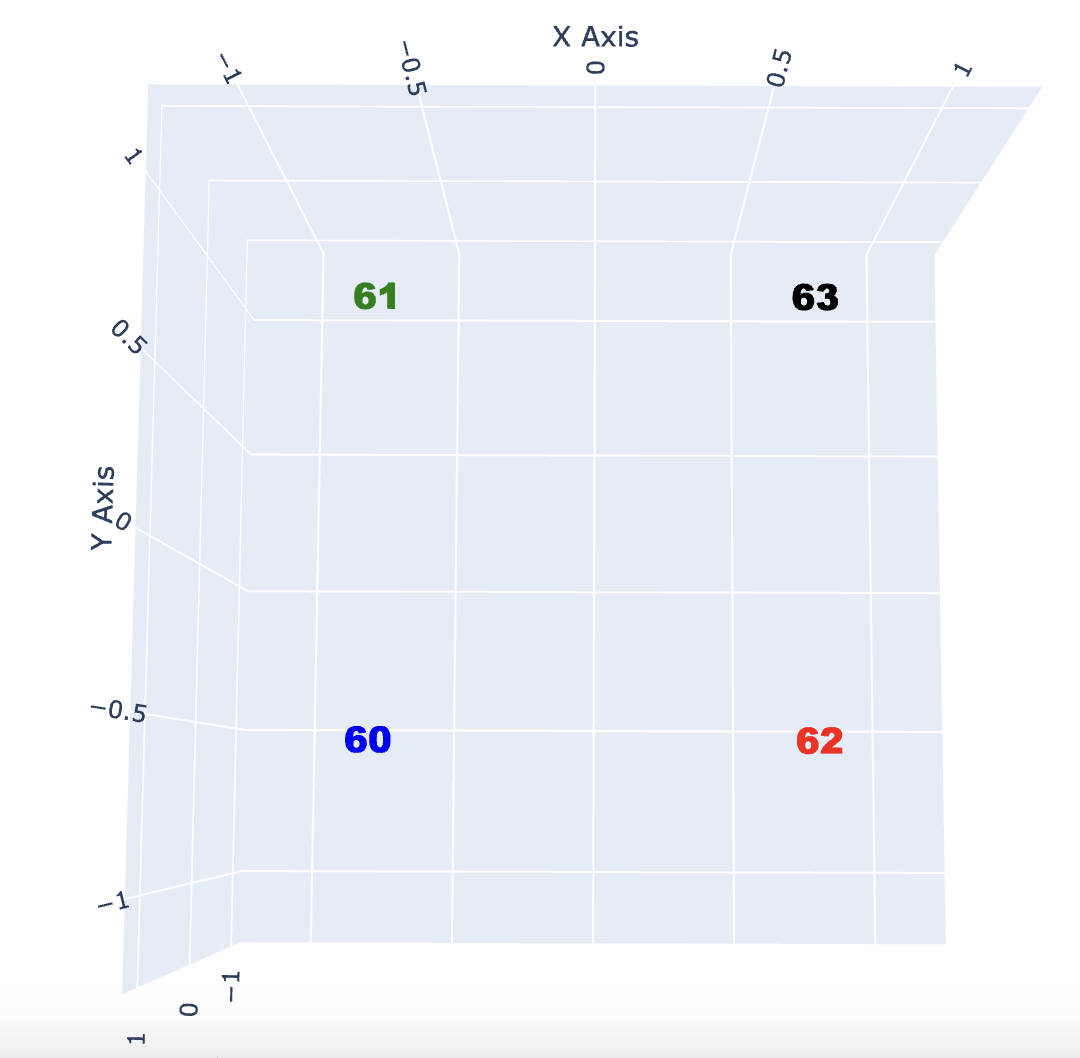}
    \caption{This image shows the numbers $\{$60, 61, 62, and 63 $\}\in\mathbf{X}^2_3$ after zooming on one of them two times. The color of the number depends on whether the numbers start with 00, 01, 10, or 11 after ignoring two bits in their dimensional components X and Y.}
    \end{figure}

    \item Multiply the normalized BitDistances across all dimensional components by the corresponding weights, as defined in point 1. In this step, the \( K \)-th term of the sequence of inverses of powers of 2 scales the normalized BitDistances. 

    \item Add the $K$ weighted BitDistances for each number across the same dimensional component to calculate the final coordinates.

\end{enumerate}

The overall process can be visualized as a concatenation of $K$ \( D \)-dimensional spheres or circles for simplicity. Each radius extends from the origin or from the end of the previous circle's radius to the position of the number $N$ we zoom in on the surface. The first circle, with a radius of \( \frac{1}{2} \), contains all the numbers in $\mathbf{X}^D_K$ and the radius points to the position of the given number $N$ on its surface. The second circle, with a radius of \( \frac{1}{4} \), contains \( \frac{1}{2^D} \) of the total numbers from the previous circle and extends from the end of the first circle’s radius to the new position of $N$ on this smaller circle. This process is repeated a total of $K$ times.

\begin{figure}[H]
    \centering
    \includegraphics[width=0.7\textwidth]{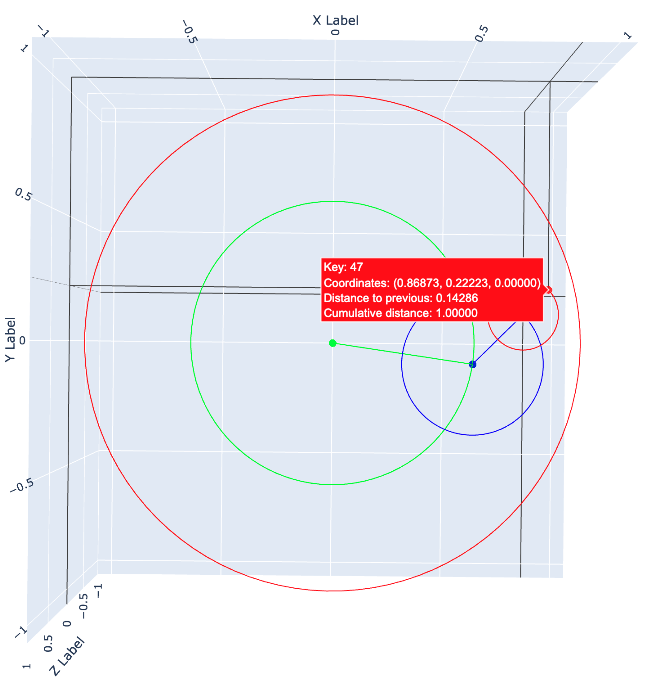}
    \caption{On the surface of the green circle are all the numbers contained in $\mathbf{X}^2_3$, and the green radius connects the origin to the position of the number 47. The surface of the blue circle contains numbers that share the first two bits with 47, where the BitDistance is calculated ignoring the first two bits. The blue radius connects the position of 47 in the green circle to its position in the blue circle. The same idea applies to the final smaller circle.}
    \label{fig:route_point}

\end{figure}

Finally, when summing the radius of the $K$ spheres, the total sum only converges to 1 if \( K \) is very large. To address this, we introduce the correction factor \( S_K \), which adjusts the weight applied to the normalized BitDistance to ensure that the sum of the inverses of powers of 2 converges to 1. To adjust the weight, ensuring that the sum converges to one, we use:

\[
1 = \sum_{n=1}^{K} \frac{1/2^n}{S_K}, \quad \text{where} \quad S_K = \sum_{m=1}^{K} \frac{1}{2^m}
\]

Here, \( S_K \) represents the sum of all terms \( \frac{1}{2^m} \) up to \( K \), ensuring that each term \( \frac{1}{2^n} \) is proportionally scaled so that the total sum equals 1, even when \( K \) is finite.\\

\begin{algorithm}
\caption{Algorithm to Project a Number into a Sphere }
\label{alg:scaled_coordinates}
\begin{algorithmic}[1]

    \State \textbf{Global Variables:}
    \State $D \gets$ 3 
    \State $K \gets$ n  
    \State CorrectionF $\gets$ sum of inverse powers of 2 from 1 to K
    \vspace{0.5em}  

    \Function{Get3DCoordinates}{$n$}

    \State  \texttt{binary = bin(n)[2:].zfill(D * K)} \Comment{Convert \(n\) to binary with a length of \(3 \times K\) bits}

    \vspace{1em}  

    \State Initialize $(X_{total}, Y_{total}, Z_{total}) \gets (0, 0, 0)$ 
    
    \State Initialize counter $\gets 1$

    \While{len(binary) $>$ 0}

    \State $X_{Bits} \gets$ binary[0::3]   \Comment{Extract the first bit from each 3-bit block}
    \State $Y_{Bits} \gets$ binary[1::3]   \Comment{Extract the second bit from each 3-bit block}
    \State $Z_{Bits} \gets$ binary[2::3]   \Comment{Extract the third bit from each 3-bit block}

        \vspace{0.5em}  

        \State $length \gets 1 / (2^{counter})$
        \State $CorrectedL \gets length / CorrectionF$
        
        \vspace{0.5em}  
        \Comment{Calculate BitDistance for each dimension}
         \State $X_{BD} \gets$ BitDistance($X_{Bits}$)
         \State $Y_{BD} \gets$ BitDistance($Y_{Bits}$)
         \State $Z_{BD} \gets$ BitDistance($Z_{Bits}$)

        \vspace{0.5em}  
        \Comment{Calculate hypotenuse and normalize the $BD$s}

        \State $H \gets (X_{BD}^2 + Y_{BD}^2 + Z_{BD}^2)^{0.5}$
        \vspace{0.5em}  

        \State $(X_{norm},Y_{norm},Z_{norm})  \gets (X_{BD},Y_{BD},Z_{BD}) / H$
        \vspace{0.5em}  

        \Comment{Compute weighted $BD$s and sum to the right dimension}
        \State $(X_{weighted},Y_{weighted},Z_{weighted}) \gets  (X_{norm},Y_{norm},Z_{norm})\cdot CorrectedL $
            
        \vspace{0.5em}  

        \State  $(X_{total},Y_{total},Z_{total}) \gets (X_{total},Y_{total},Z_{total}) +
        (X_{weighted},Y_{weighted},Z_{weighted})$
        \vspace{1em}  
        
        \Comment{Increment the counter and remove the leftmost block of bits}
        \State counter $\gets$ counter +1            
        \State \texttt{binary = binary[3:]} 

    \EndWhile

    \State \Return $(X_{total}, Y_{total}, Z_{total})$
\EndFunction
\end{algorithmic}
\end{algorithm}

In the following table will be applied the described algorithm in the number 47$\in \mathbf{X}^2_3$: 

\begin{table}[H]
    \centering
    \caption{This table illustrates the steps involved in calculating the Morton codes and BitDistances for the X and Y components, along with the corresponding hypotenuse (H) and normalized BitDistances. The calculations are performed for the number 47$\in \mathbf{X}^2_3$.}
    \begin{tabular}{|r|r|r|r|r|c|r|r|}
        \hline
        Morton XY & Morton X & Morton Y & $BD_X$ & $BD_Y$ & H & Norm $BD_X$ & Norm $BD_Y$ \\ \hline
        [10][11][11] & 111 & 011 & 0.875 & -0.125 & 0.883883 & 0.989949 & -0.141421 \\ \hline
        [11][11] & 11 & 11 & 0.75 & 0.75 & 1.060660 & 0.707107 & 0.707107 \\ \hline
        [11] & 1 & 1 & 0.5 & 0.5 & 0.707107 & 0.707107 & 0.707107 \\ \hline
    \end{tabular}

    \label{tab:example_table}
\end{table}

The next step is to multiply the BitDistances of each dimension by their corresponding weights and apply the correction factor \( S_K \), which is 0.875 when \( K = 3 \). Finally, the weighted BitDistances across all dimensions are summed up to calculate the final position.

\begin{table}[H]
    \centering
    \caption{Normalized and weighted BitDistances, along with the accumulative sums for X and Y components.}
    \begin{tabular}{|r|r|r|r|r|r|r|}
        \hline
        Weigth & Norm $BD_X$ & Norm $BD_Y$ & W. $BD_X$ & W. $BD_Y$ & Acc. $X$ & Acc. $Y$ \\ \hline
        0.57142 & 0.989949 & -0.141421 & 0.565677 & -0.080811 & 0.565677 & -0.080811 \\ \hline
        0.28571 & 0.707107 & 0.707107 & 0.202028 & 0.202028 & 0.767704 & 0.121217 \\ \hline
        0.14285 & 0.707107 & 0.707107 & 0.101010 & 0.101010 & 0.868714 & 0.222227 \\ \hline
    \end{tabular}
    \label{tab:bd_table}
\end{table}

\noindent When the algorithm processes all the numbers in $\mathbf{X}^D_K$, it assigns to each of them a coordinate within a \( D \)-dimensional sphere of radius 1, creating the circular projection of the $Z$-curve. The following images show the results for dimensions 2 and 3 with \( K \) set to 1, 2, and 3.\\

The images below show the circular projections of $\mathbf{X}^2_1, \mathbf{X}^2_2 $ and $\mathbf{X}^2_3$ :
    
    \begin{figure}[H]
        \centering
        \begin{subfigure}[b]{0.45\textwidth}
            \includegraphics[width=\textwidth]{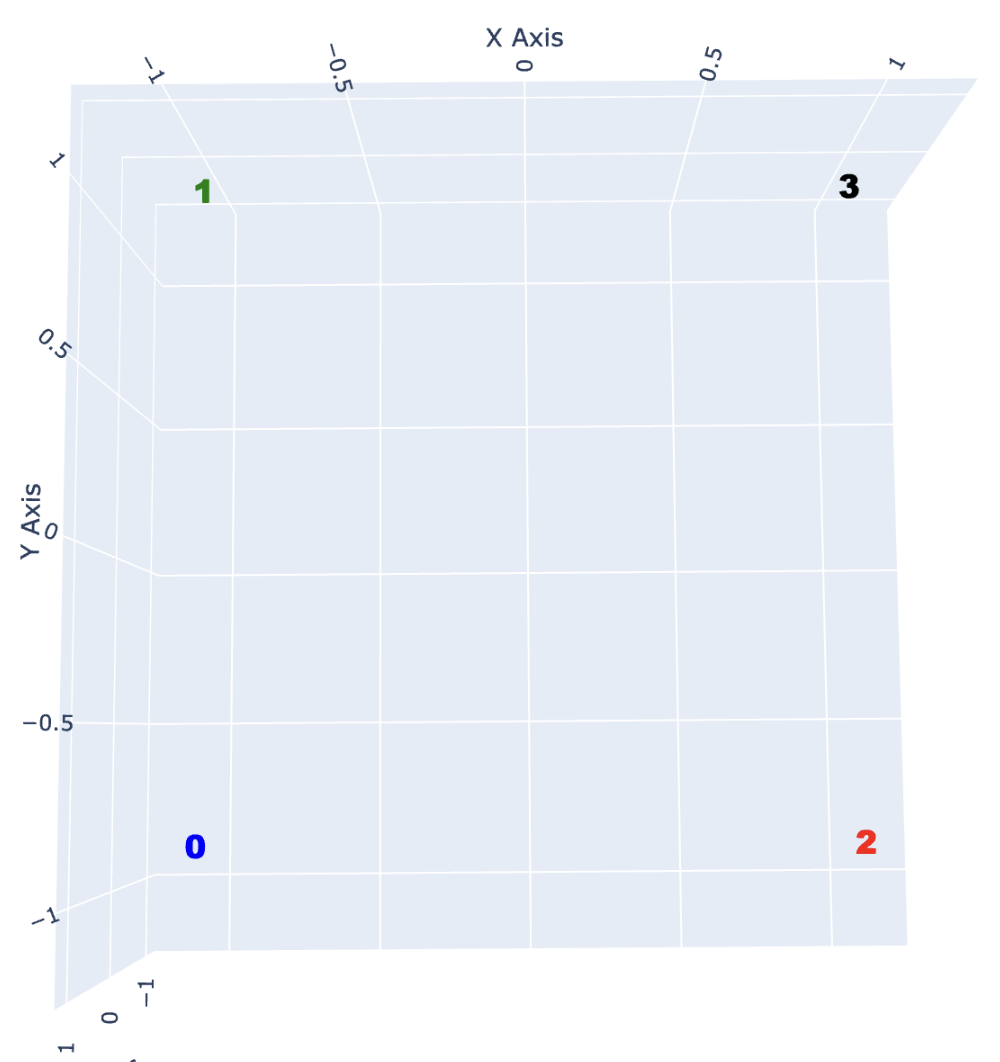} 
        \end{subfigure}
        \begin{subfigure}[b]{0.45\textwidth}
            \includegraphics[width=\textwidth]{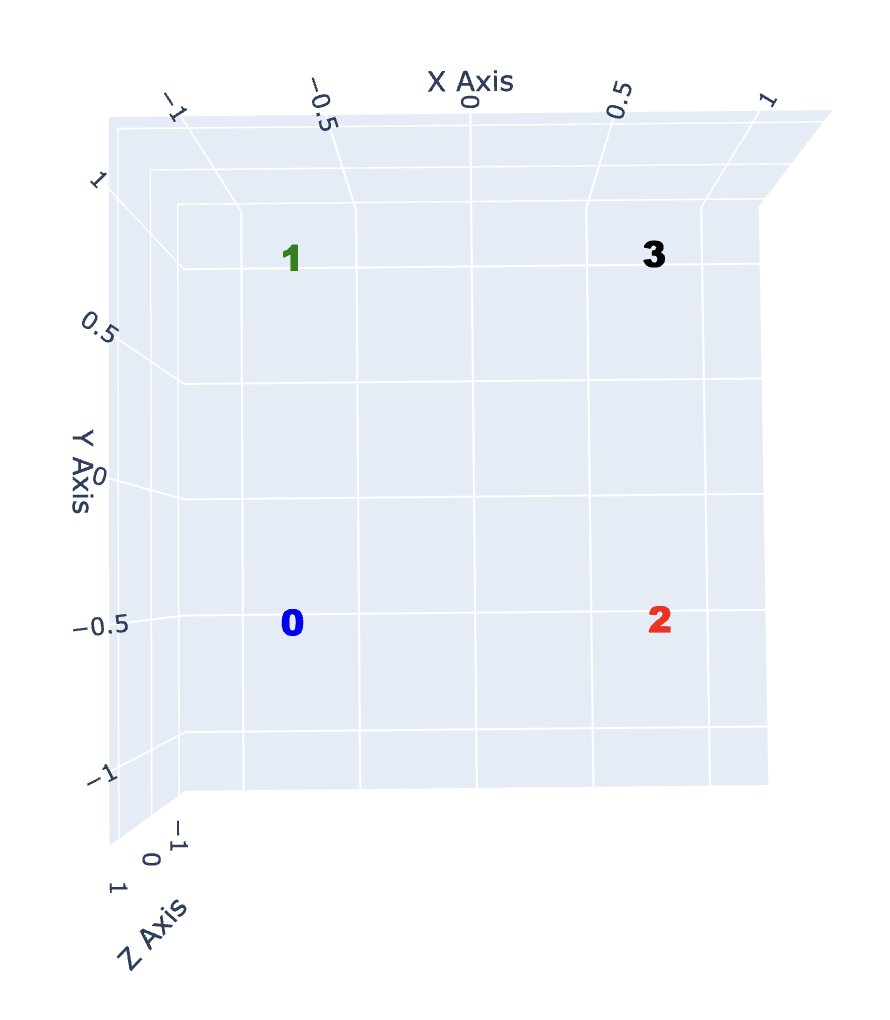} 
        \end{subfigure}
        \caption{In the left image are the values in $\mathbf{X}^2_1$ after calculating their BitDistance. In the right image, the values in $\mathbf{X}^2_1$ had been projected into a circle. The color of the number depends on whether the number starts with 00, 01, 10, or 11. }
    \end{figure}

    \begin{figure}[H]
        \centering
        \begin{subfigure}[b]{0.45\textwidth}
            \includegraphics[width=\textwidth]{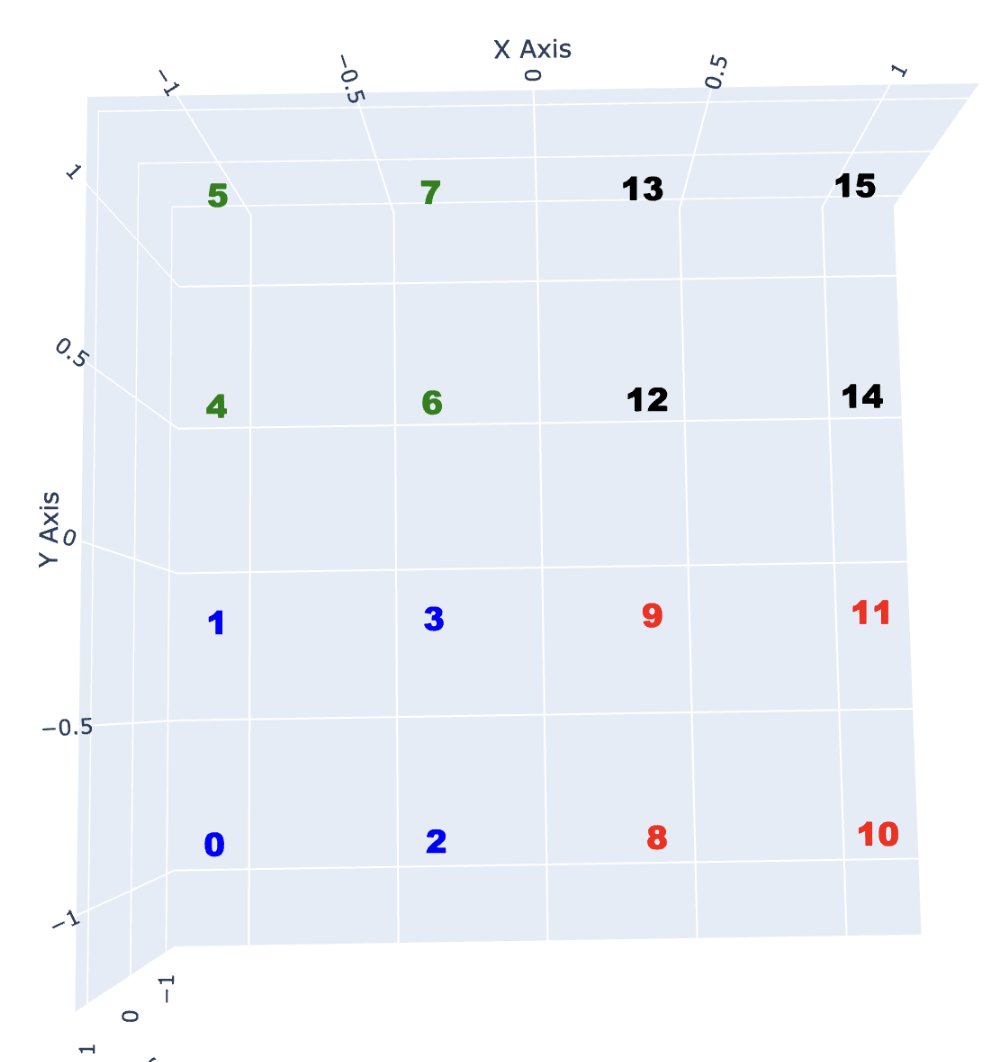} 
        \end{subfigure}
        \begin{subfigure}[b]{0.45\textwidth}
            \includegraphics[width=\textwidth]{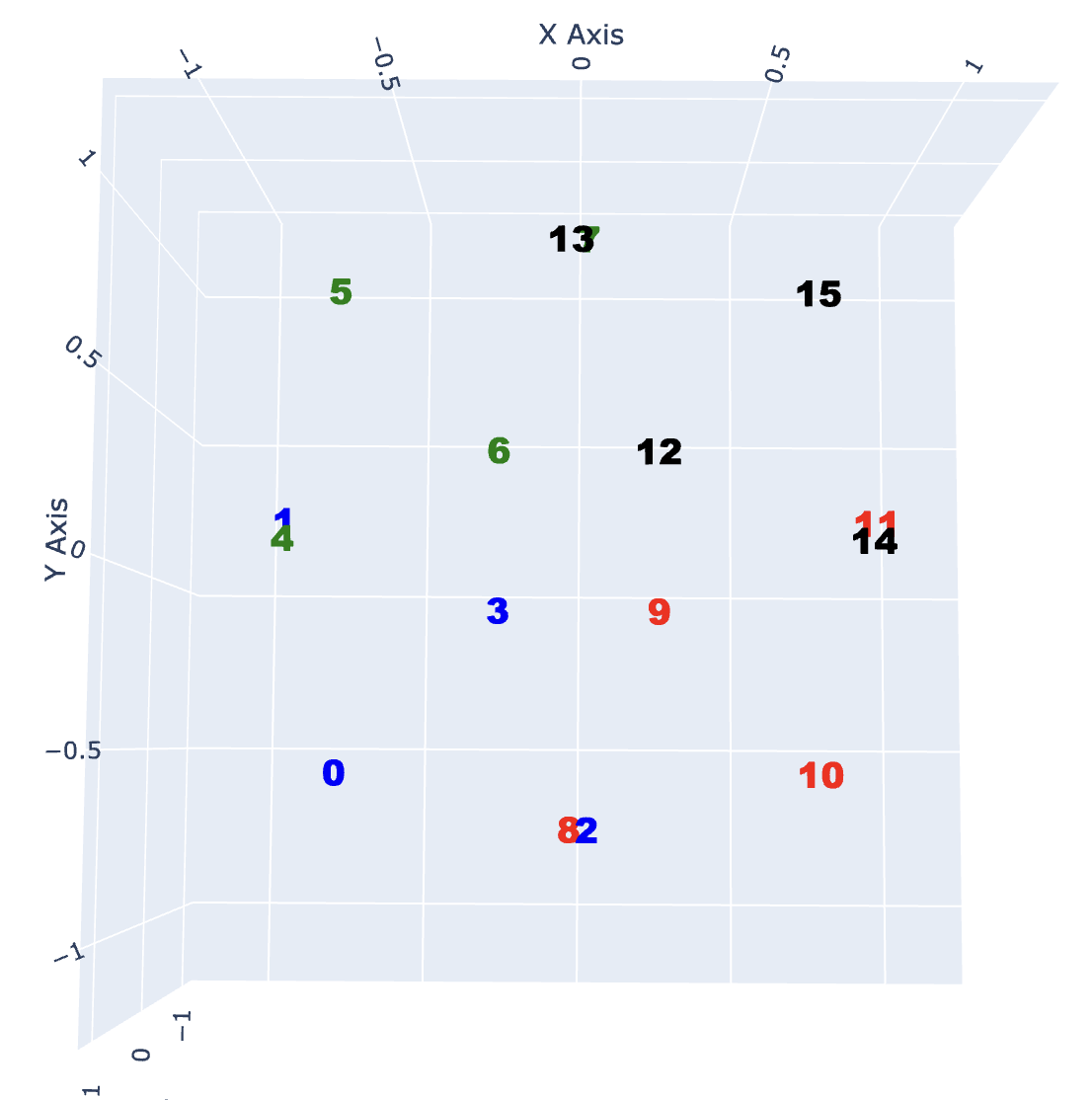} 
        \end{subfigure}
        \caption{In the left image are the values in $\mathbf{X}^2_2$ after calculating their BitDistance. In the right image, the values in $\mathbf{X}^2_2$ had been projected into a circle. The color of the number depends on whether the number starts with 00, 01, 10, or 11.}
    \end{figure}

    \begin{figure}[H]
        \centering
        \begin{subfigure}[b]{0.45\textwidth}
            \includegraphics[width=\textwidth]{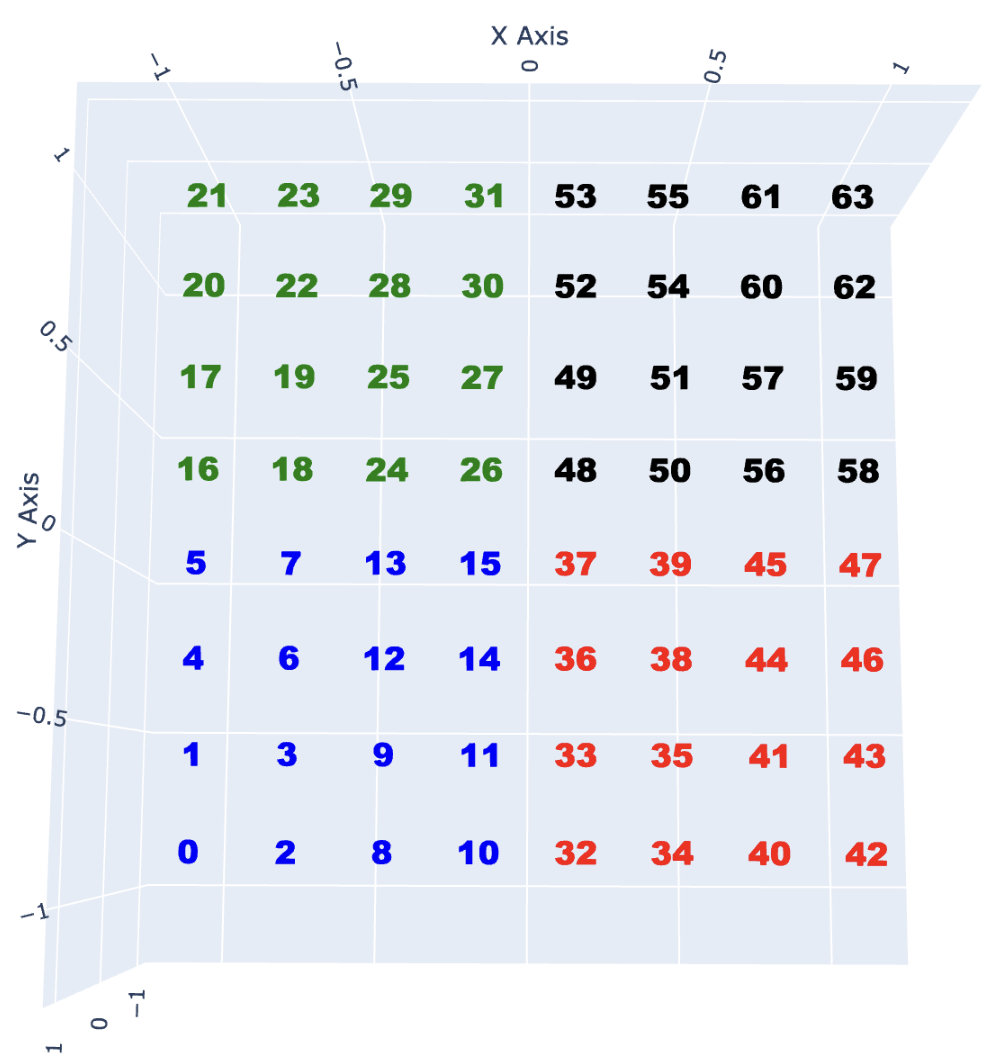} 
        \end{subfigure}
        \begin{subfigure}[b]{0.45\textwidth}
            \includegraphics[width=\textwidth]{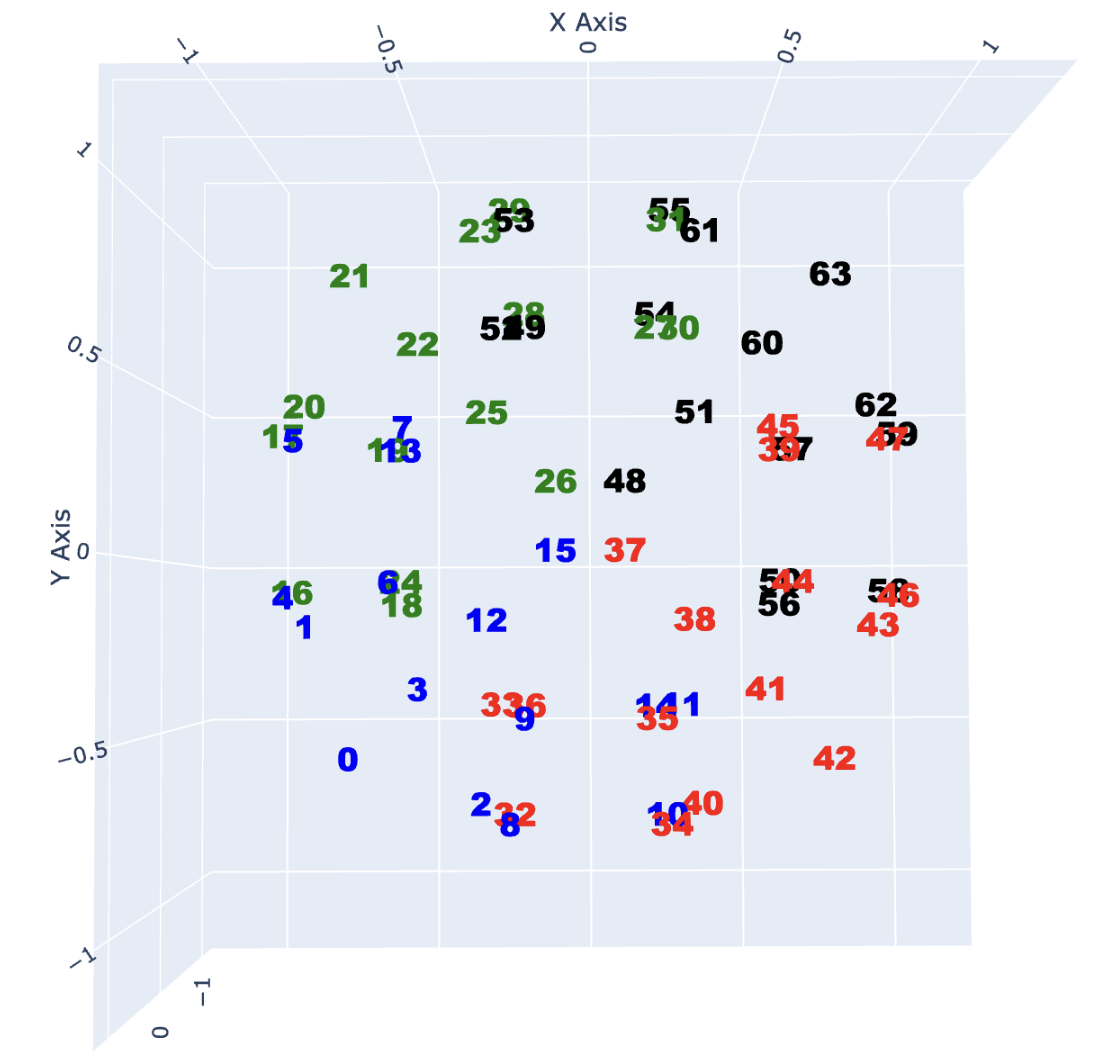} 
        \end{subfigure}
                \caption{In the left image are the values in $\mathbf{X}^2_3$ after calculating their BitDistance. In the right image, the values in $\mathbf{X}^2_3$ had been projected into a circle. The color of the number depends on whether the number starts with 00, 01, 10, or 11.}
    \end{figure}

The images below show the circular projections $\mathbf{X}^3_1$, $\mathbf{X}^3_2$, and $\mathbf{X}^3_3$. In dimension 3, except when $K=1$, we include two images to help visualize the sets more clearly.

    \begin{figure}[H]
        \centering
        \includegraphics[width=0.45\textwidth]{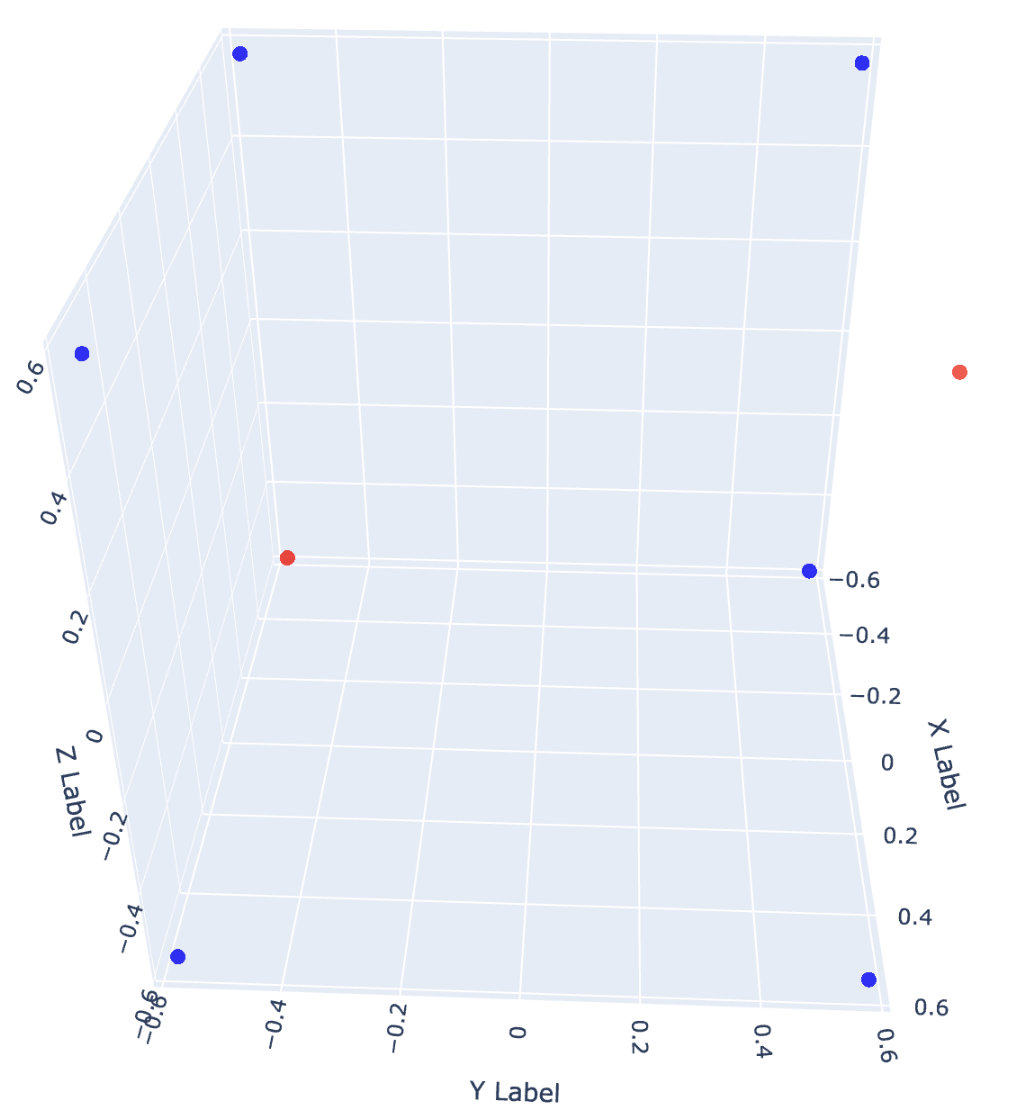}
        \caption{This image shows the values in $\mathbf{X}^3_1$ after being projected into a sphere.} 
    \end{figure}

    \begin{figure}[H]
        \centering
        \begin{subfigure}[b]{0.45\textwidth}
            \includegraphics[width=\textwidth]{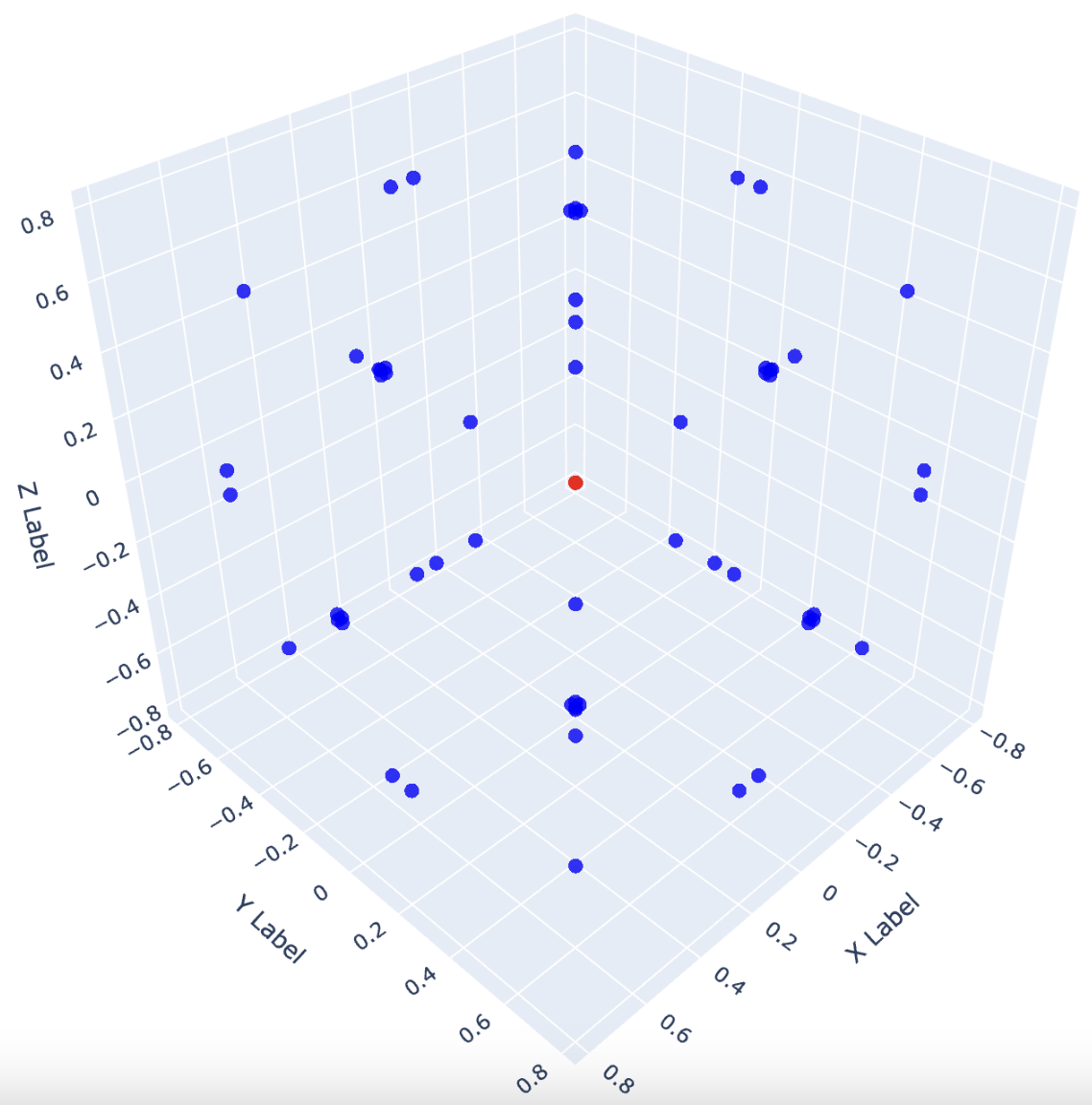} 
        \end{subfigure}
        \begin{subfigure}[b]{0.45\textwidth}
            \includegraphics[width=\textwidth]{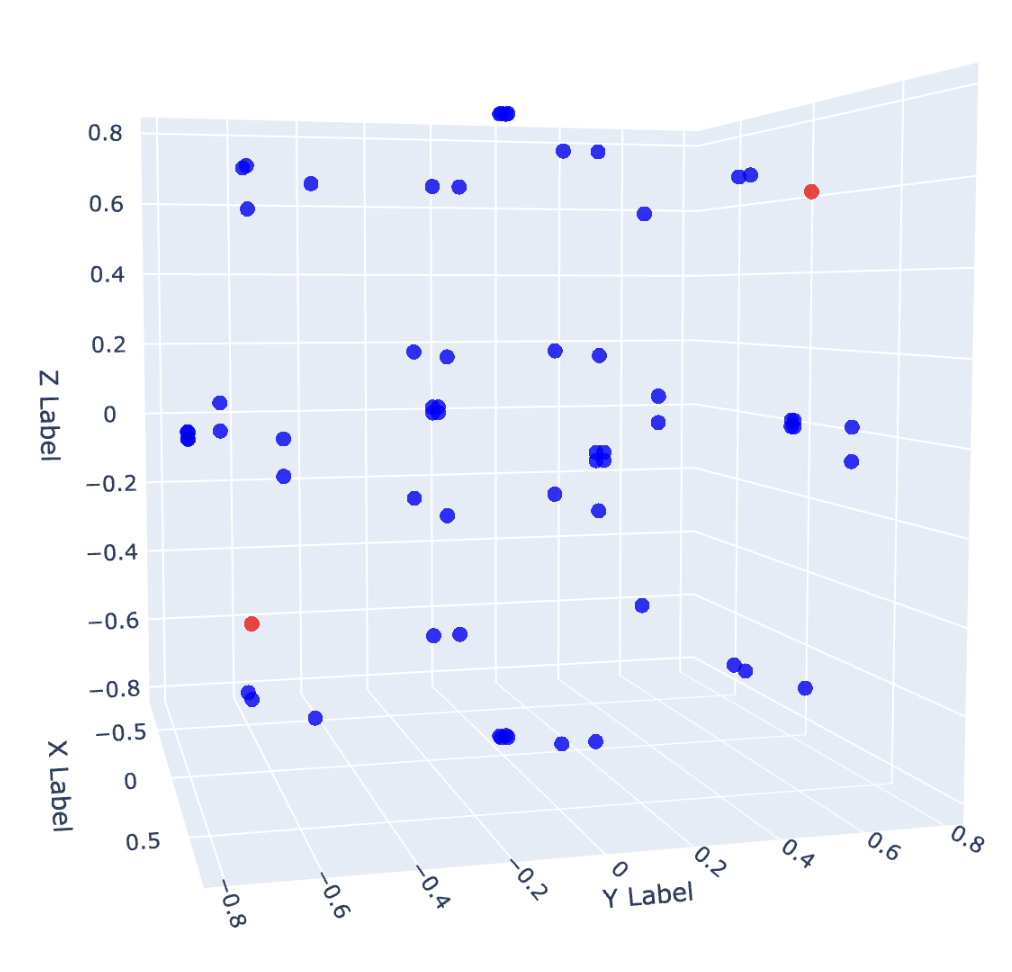} 
        \end{subfigure}
        \caption{The two images show the values in $\mathbf{X}^3_2$ after being projected into a sphere in different perspectives.} 
    \end{figure}

    \begin{figure}[H]
        \centering
        \begin{subfigure}[b]{0.45\textwidth}
            \includegraphics[width=\textwidth]{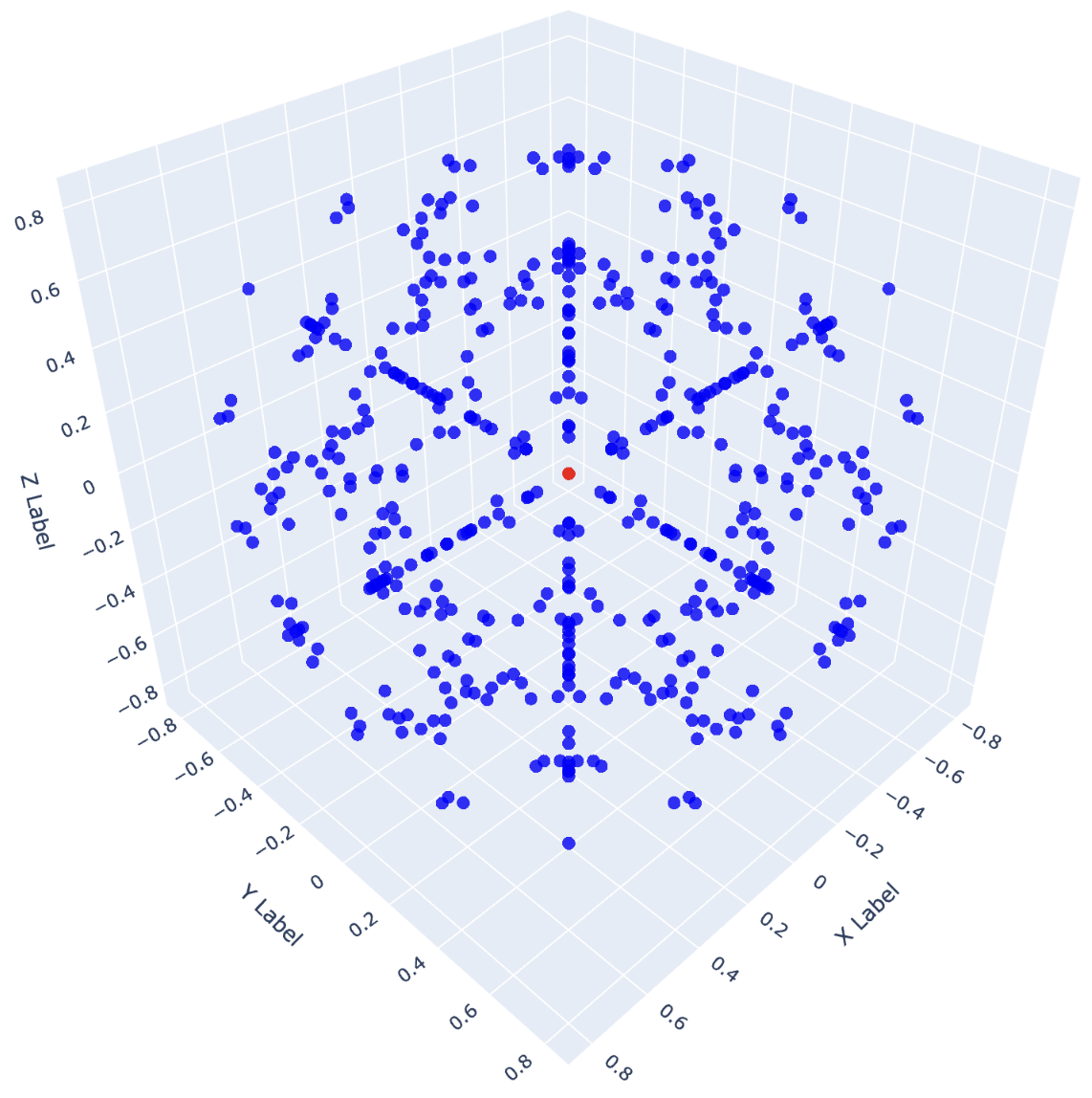} 
        \end{subfigure}
        \begin{subfigure}[b]{0.45\textwidth}
            \includegraphics[width=\textwidth]{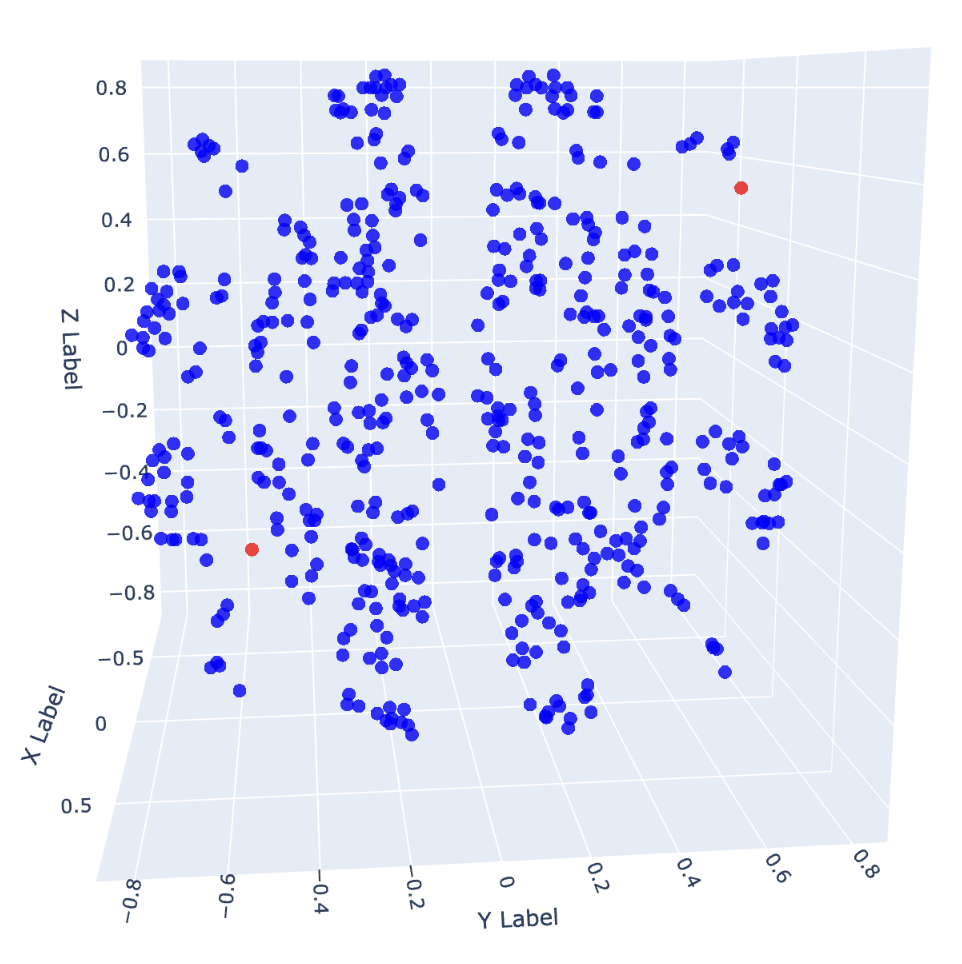} 
        \end{subfigure}
          \caption{The two images show the values in $\mathbf{X}^3_3$ after being projected into a sphere in different perspectives.} 
    \end{figure}  
    
\section{Analysis and Properties}

In this section, we will study the properties of the circular projection of $\mathbf{X}^D_K$, focusing mainly on dimensions 2 and 3. We will do it through the object called subset.

\begin{definition}[Subset]
\textit{A subset \( \mathbb{T} \) is defined as the set of elements in \( \mathbf{X}^D_K \) that share the same radius after being projected onto a \( D \)-dimensional sphere.}

\end{definition}

In this section, we will study these objects and the subsets by dividing the study into three parts, each focusing on a different aspect.




\subsection{Numerical Properties of the Elements in $\mathbb{T}$}


\subsubsection{The Number of Elements in $\mathbb{T}$}

We observe that \textbf{the number of elements a subset $\mathbb{T}$ contains, denoted by $|\mathbb{T}|$, is always \( 2^D \) or a multiple of it}. When \( K = 1 \), there exists only one subset containing all the \( 2^{D \times 1} \) elements. As \( K \) increases, additional subsets appear, but this emergence follows a non-linear pattern. It means that a subset \( \mathbb{T} \) in a given \( \mathbf{X}^D_K \) (when \( K > 1 \)) may contain \( 2^D \times n \) elements without necessarily exist  a subset of size \( 2^D \times (n-1) \). The observed subset sizes for dimensions \( D = 2 \) and \( D = 3 \), based on computations for values of \( K \) up to 11 and 7, respectively, are:

\begin{itemize}
    \item For \( D = 2 \), the sizes of subsets \( \mathbb{T} \subset \mathbf{X}^2_K \) are \( |\mathbb{T}| \in \{4, 8\} \).
    \item For \( D = 3 \), the sizes of subsets \( \mathbb{T} \subset \mathbf{X}^3_K \) are \( |\mathbb{T}| \in \{8, 16, 24, 32, 48\} \).
\end{itemize}

Tables \ref{tab:num_sub_set_2} and \ref{tab:num_sub_set_3} show the number of subsets of various sizes for different values of $K$ in dimensions 2 and 3.

\begin{table}[H]
\centering
\caption{The number of subsets with 4 and 8 elements for different values of $K$ when $D=2$.}
\begin{tabular}{|c|c|c|}
\hline
\textbf{$K$} & \textbf{$|\mathbb{T}| = 4$} & \textbf{$|\mathbb{T}| = 8$} \\
\hline
1 & 1 & 0 \\
\hline
2 & 2 & 1 \\
\hline
3 & 4 & 6 \\
\hline
4 & 8 & 28 \\
\hline
... & ... & ... \\
\hline
\end{tabular}
\label{tab:num_sub_set_2}
\end{table}

\begin{table}[H]
\centering
\caption{The number of subsets with 8, 16, 24, 32, 40, and 48 elements for different values of $K$ when $D=3$.}
\begin{tabular}{|c|c|c|c|c|c|c|}
\hline
\textbf{$K$} & \textbf{$|\mathbb{T}| = 8$} & \textbf{$|\mathbb{T}| = 16$} & \textbf{$|\mathbb{T}| = 24$} & \textbf{$|\mathbb{T}| = 32$} & \textbf{$|\mathbb{T}| = 40$} & \textbf{$|\mathbb{T}| = 48$} \\
\hline
1 & 1 & 0 & 0 & 0 & 0 & 0 \\
\hline
2 & 3 & 1 & 1 & 0 & 0 & 0 \\
\hline
3 & 7 & 3 & 9 & 0 & 0 & 4 \\
\hline
4 & 16 & 19 & 48 & 11 & 0 & 45 \\
\hline
... & ... & ... & ... & ... & ... & ... \\
\hline
\end{tabular}
\label{tab:num_sub_set_3}
\end{table}

\subsubsection{Summation of the Elements in $\mathbb{T}$}

We observe a consistent pattern in the summation of all elements in any subset \( \mathbb{T} \subset \mathbf{X}^D_K \). Specifically, the total sum \( S(\mathbb{T}) \) is always equal to a constant \( \Psi^D_K \) multiplied by a factor determined by the size of the subset. If the subset contains \( |\mathbb{T}| = 2^D \times n \) elements, then the sum is \( \Psi^D_K \) multiplied by \( n \). The constant \( \Psi^D_K \) is defined by:

\begin{equation}
    \Psi^D_K = 2^{D(K+1)-1} - 2^{D-1}
\end{equation}

\noindent This observation leads us to the following conjecture:

\begin{conjecture}
\label{conj:subset_sum}
For any subset \( \mathbb{T} \subset \mathbf{X}^D_K \) with \( |\mathbb{T}| = 2^D \times n \) elements, the sum of its elements is given by:

\begin{equation}
    S(\mathbb{T}) = \Psi^D_K \times n
    \label{eq:total_sum}
\end{equation}
\end{conjecture}

\noindent We have evaluated this conjecture empirically for various values of \( D \) and \( K \), with results for dimensions 2 and 3 presented in Tables \ref{tab:sum_D2} and \ref{tab:sum_D3}.

\begin{table}[H]
\centering
\caption{Summation results for subsets of size 4 and 8 across different values of $K$ when $D=2$.}
\begin{tabular}{|c|c|c|}
\hline
\textbf{$K$} & \textbf{$|\mathbb{T}| = 4$} & \textbf{$|\mathbb{T}| = 8$} \\
\hline
1 & 6 & 0 \\
\hline
2 & 30 & 60 \\
\hline
3 & 126 & 252 \\
\hline
4 & 510 & 1020 \\
\hline
... & ... & ... \\
\hline
\end{tabular}
\label{tab:sum_D2}
\end{table}

\begin{table}[H]
\centering
\caption{Summation results for subsets of size 8, 16, 24, 32, and 48 across different values of $K$ when $D=3$.}
\begin{tabular}{|c|c|c|c|c|c|}
\hline
\textbf{$K$} & \textbf{$|\mathbb{T}| = 8$} & \textbf{$|\mathbb{T}| = 16$} & \textbf{$|\mathbb{T}| = 24$} & \textbf{$|\mathbb{T}| = 32$} & \textbf{$|\mathbb{T}| = 48$} \\
\hline
1 & 28 &  &  &  &  \\
\hline
2 & 252 & 504 & 756 &  &  \\
\hline
3 & 2044 & 4088 & 6132 &  & 12264 \\
\hline
4 & 16380 & 32760 & 49140 & 65520 & 98280 \\
\hline
... & ... & ... & ... & ... & ... \\
\hline
\end{tabular}
\label{tab:sum_D3}
\end{table}


\subsubsection{Equivalent Position and the Minimum Number of Knowable Subsets}
\label{subsubsec:eq_pos}

Estimating the number of subsets of different sizes within a set \( \mathbf{X}^D_K \) using only the parameters \( D \) and \( K \) is challenging. Specifically, we cannot predict whether there are \( n \) subsets of size \( 2^D \times A \) or \( m \) subsets of size \( 2^D \times B \) after applying the circular projection to the set \( \mathbf{X}^D_K \). However, we observe that certain \textbf{regions} in the $Z$-curve consistently form subsets with the same number of elements, independently of the value of $K$ or the numerical value within each cell. For example, the values located at the corners of the $Z$-curve, where is always include the zero. Independently of the value of \( K \), these regions consistently constitute a subset with exactly \( 2^D \) elements. Initially, this suggests that when a parent cell divides into \( 2^D \) new child cells, one of them will always preserve its equivalent position, maintaining a subset size consistent with its parent.

\begin{definition}[Equivalent Position]
The \textit{equivalent position} is the unique cell among the \( 2^D \) subdivisions of a parent cell that retains the same relative position as the parent cell in its original space.
\end{definition}

To determine the equivalent position of a number in \( \mathbf{X}^D_S \) and map it to its corresponding position in \( \mathbf{X}^D_B \), we define a mapping function \( \phi \). In this context, \( S \),"small" \( K \), represents the scaling factor of the source grid \( \mathbf{X}^D_S \). The target grid \( \mathbf{X}^D_B \) has as scaling factor \( B \), "big" \( K \), defined as \( B = S + n \), where \( n \) is the increment in the scaling factor. The function \( \phi \) enables the translation of positions between different scaling factors within the same \( D \)-dimensional space by mapping elements from the smaller grid to their equivalent positions in the larger grid, and it is given by :

\begin{equation}
\phi^D_{(S,B)}(x) = 2^{D(B - S + 1)} \cdot \left\lfloor \frac{x}{2^D} \right\rfloor + \Omega \cdot (x \bmod 2^D),
\label{eq:position_eq}
\end{equation}

where \( \Omega \) is a number whose binary representation is given by:

\[
\Omega = \underbrace{(00\cdots 0)1}_{ D-1 \text{ zeros followed by a 1}} \text{ repeated } (B - S + 1) \text{ times}.
\]

\begin{table}[H]
\centering
\caption{Values of $\Omega$ for $(B - S) \in \{1, 2, 3, 4\}$ across dimensions \( D \in \{1, 2, 3\} \).}
\begin{tabular}{|c|c|c|c|c|c|c|c|}
\hline
$B-S$ & \multicolumn{2}{c|}{Dimension = 1} & \multicolumn{2}{c|}{Dimension = 2} & \multicolumn{2}{c|}{Dimension = 3} \\ \hline
 & Binary & Decimal & Binary & Decimal & Binary & Decimal \\ \hline
1 & 11 & 3 & 0101 & 5 & 001001 & 9 \\ \hline
2 & 111 & 7 & 010101 & 21 & 001001001 & 73 \\ \hline
3 & 1111 & 15 & 01010101 & 85 & 001001001001 & 585 \\ \hline
4 & 11111 & 31 & 0101010101 & 341 & 001001001001001 & 4681 \\ \hline
5 & ... & ... & ... & ... & ... & ... \\ \hline
\end{tabular}
\label{your_label}
\end{table}

The underlying idea of this function is to divide the target grid \( \mathbf{X}^D_B \) into equal size sub-grids, where each sub-grid size is determined by the difference between \( B \) and \( S \), specifically \( 2^{D(B - S + 1)} \). The multiplication that accompanies it indicates in which of the sub-grids is located the equivalent position of the number $x$, and \( \Omega \) represents the distance, \textbf{following the Z-pattern} with orientation defined based on the direction given to the bits $0$ and $1$ in each dimension, between two consecutive equivalent positions within the same sub-grid. Lastly, in each sub-grid we can find up to \( 2^D \) possible locations that are valid solutions. The coefficient accompanying \( \Omega \) indicates which of these \( 2^D \) possible locations within the sub-grid will correspond to the function's input.\\

To illustrate this idea, we will show the result of computing $\phi^2_{(2,4)}(\textbf{X}^2_2)$. The function for these parameters will look like:

\[
\phi^2_{(2,4)}(x) = 2^{2(4-2+1)} \cdot  \left\lfloor \frac{x}{2^2} \right\rfloor  + 010101 \cdot (x \mod 2^2)
\]
\[
\phi^2_{(2,4)}(x) = 64 \cdot  \left\lfloor \frac{x}{4} \right\rfloor  + 21 \cdot (x \mod 4)
\]

\begin{table}[H]
    \caption{Results after compute $\phi^2_{(2,4)}(\textbf{X}^2_2)$. }

    \centering
    \begin{tabular}{|c|c|c|c|c|c|c|c|c|}
    \hline
    $\mathbf{X}^2_2$ Values & 0 & 1 & 2 & 3 & 4 & 5 & 6 & 7 \\ \hline
    Equivalent in $\mathbf{X}^2_4$ & 0 & 21 & 42 & 63 & 64 & 85 & 106 & 127 \\ \hline
    \hline
    $\mathbf{X}^2_2$ Values & 8 & 9 & 10 & 11 & 12 & 13 & 14 & 15 \\ \hline
    Equivalent in $\mathbf{X}^2_4$ & 128 & 149 & 170 & 191 & 192 & 213 & 234 & 255 \\ \hline
    \end{tabular}
\end{table}

\begin{figure}[H]
\centering
\includegraphics[width=10 cm]{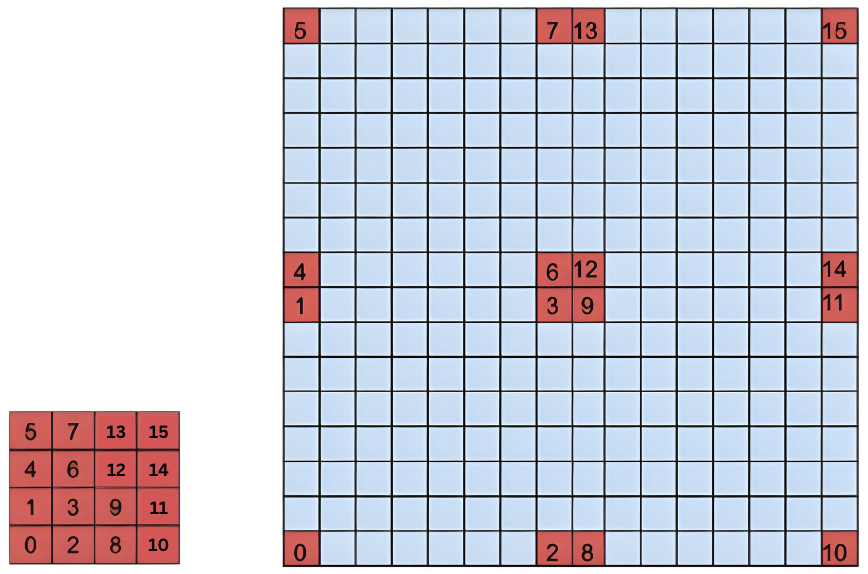}
\caption{
The left image shows the set of numbers in $\mathbf{X}^2_2$ arranged in the $Z$-order pattern. In the right image, only the positions equivalent to the numbers in $\mathbf{X}^2_2$ are marked in red, highlighting where these numbers project in $\mathbf{X}^2_4$.}

\label{framework}  
\end{figure}

According to Equation \ref{conj:subset_sum}, the sum of all elements within the same subset is always a multiple of \( \Psi^D_K \). Therefore, when calculating \( \mathbb{T}' = \phi^D_{(S,B)}(\mathbb{T}) \), the sum of the elements in $\mathbb{T}'$ should also be a multiple of the new \( \Psi^D_B \) value corresponding to the updated \( K \).\\


In dimension 2, for every subset \( \mathbb{T} \subset \mathbf{X}^D_K \), when we compute \( \mathbb{T}' = \phi^2_{(K,K+1)}(\mathbb{T}) \), the resulting \( \mathbb{T}' \) is always a valid subset in \( \mathbf{X}^2_{K+1} \). This has been verified up to \( K = 8 \). This suggests that the number of subsets of each size in \( \mathbf{X}^2_K \) is at least as large as in \( \mathbf{X}^2_{K-1} \). This implies that all subsets in \( \mathbf{X}^2_S \) can generate other valid subsets in any other \( \mathbf{X}^2_{B} \) applying \( \phi^2_{(S,B)} \).\\


In dimension 3, as shown in Table \ref{tab:num_sub_set_3}, the number of subsets of each size is always greater in \( \mathbf{X}^3_{K+1} \) than in \( \mathbf{X}^3_{K} \). This initially suggests that, as in dimension 2, when we compute $\phi^D_{(S,B)}(\mathbb{T})$, the resulting $\mathbb{T}'$ would also be a valid subset in $\mathbf{X}^D_B$. However, in dimension 3, this pattern breaks down. When we compute \( \phi^D_{(S,B)}(\mathbb{T}) \) the resulting subset \( \mathbb{T}' \) does not always exist in \( \mathbf{X}^D_B \). Moreover, we also observed that even if the resulting subset \( \mathbb{T}' = \phi^3_{(S,B)}(\mathbb{T}) \) does not exist in \( \mathbf{X}^3_{B} \), it still holds true that \( S(\mathbb{T}') = \Psi^D_B \) or a multiple of \( \Psi^D_B \). This leads us to the following conjecture:

\begin{conjecture}
\label{conj:eq_pos_alw}
For any valid subset \( \mathbb{T} \subset \mathbf{X}^D_S \): 
\[
S(\phi^D_{(S,B)}(\mathbb{T})) = \Psi^D_B \times \frac{|\mathbb{T}|}{2^D}
\]

\end{conjecture}

The following tables show how many subsets \( \mathbb{T} \), after applying \( \phi \), result in valid subsets for different subset sizes.

\begin{table}[H]
\centering
\caption{This table shows the number of 8-element subsets in \( \mathbf{X}^3_S \) that produce a valid 8-element subset in \( \mathbf{X}^3_B \) after applying the \( \phi \) function.}

\begin{tabular}{|c|c|c|c|c|c|c|c|c|}
\hline
K & B=1 & B=2 & B=3 & B=4 & B=5 & B=6 & B=7 & B=8 \\
\hline
S=1 &  & 1/1 & 1/1 & 1/1 & 1/1 & 1/1 & 1/1 & 1/1 \\
\hline
S=2 & &  & 2/3 & 2/3 & 2/3 & 3/3 & 2/3 & 2/3 \\
\hline
S=3 & & & & 4/7  & 5/7 & 4/7 & 5/7 & 4/7 \\
\hline
S=4 & & & &  &9/16 & 10/16 & 10/16 & 9/16 \\
\hline
S=5 & & & & &  & 22/49 & 23/49 & 21/49 \\
\hline
\end{tabular}
\end{table}

\begin{table}[H]
\centering
\caption{This table shows the number of 16-element subsets in \( \mathbf{X}^3_S \) that produce a valid 16-element subset in \( \mathbf{X}^3_B \) after applying the \( \phi \) function.}
\begin{tabular}{|c|c|c|c|c|c|c|c|c|}
\hline
K & B=1 & B=2 & B=3 & B=4 & B=5 & B=6 & B=7 & B=8 \\
\hline
S=1 &  &  &  &  &  &  &  &  \\
\hline
S=2 & & & 0/1 & 0/1 & 0/1 & 1/1 & 0/1 & 0/1 \\
\hline
S=3 & & &  &0/3  & 1/3 & 0/3 & 1/3 & 0/3 \\
\hline
S=4 & & & &  & 1/19 & 2/19 & 5/19& 3/19 \\
\hline
S=5 & & & & &  & 11/127 & 19/127 & 11/127 \\
\hline
\end{tabular}
\end{table}

\begin{table}[H]
\centering
\caption{This table shows the number of 24-element subsets in \( \mathbf{X}^3_S \) that produce a valid 24-element subset in \( \mathbf{X}^3_B \) after applying the \( \phi \) function.}
\begin{tabular}{|c|c|c|c|c|c|c|c|c|}
\hline
K & B=1 & B=2 & B=3 & B=4 & B=5 & B=6 & B=7 & B=8 \\
\hline
S=1 & & & & & & & & \\
\hline
S=2 & &  & 1/1 & 1/1 & 1/1 & 1/1 & 1/1 & 1/1 \\
\hline
S=3 & & &  & 7/9 & 8/9 & 7/9 & 9/9 & 8/9 \\
\hline
S=4 & & & & & 43/48 & 36/48 & 44/48 & 40/48  \\
\hline
S=5 & & & & & & 168/207 & 185/207 & 182/207 \\
\hline
\end{tabular}
\end{table}

\begin{table}[H]
\centering
\caption{This table shows the number of 32-element subsets in \( \mathbf{X}^3_S \) that produce a valid 32-element subset in \( \mathbf{X}^3_B \) after applying the \( \phi \) function.}
\begin{tabular}{|c|c|c|c|c|c|c|c|c|}
\hline
K & B=1 & B=2 & B=3 & B=4 & B=5 & B=6 & B=7 & B=8 \\
\hline
S=1 & & & & & & & & \\
\hline
S=2 & &  &  & &  &  &  &  \\
\hline
S=3 & & &  &  &  &  &  &  \\
\hline
S=4 & & & & & 0/11 & 4/11 & 3/11 & 2/11  \\
\hline
S=5 & & & & & & 12/94 & 23/94 & 12/94 \\
\hline
\end{tabular}
\end{table}

\begin{table}[H]
\centering
\caption{This table shows the number of 48-element subsets in \( \mathbf{X}^3_S \) that produce a valid 48-element subset in \( \mathbf{X}^3_B \) after applying the \( \phi \) function.}
\begin{tabular}{|c|c|c|c|c|c|c|c|c|}
\hline
K & B=1 & B=2 & B=3 & B=4 & B=5 & B=6 & B=7 & B=8 \\
\hline
S=1 & & & & & & & & \\
\hline
S=2 & &  &  & &  &  &  &  \\
\hline
S=3 & & &  & 4/4 &4/4  & 1/4 & 4/4 & 3/4 \\
\hline
S=4 & & & & & 39/45 & 33/45 & 35/45 & 36/45  \\
\hline
S=5 & & & & & & 370/466 & 374/466 & 374/466 \\
\hline
\end{tabular}
\end{table}





\subsubsection{Summation of the Elements in $\mathbf{X}^D_K$}

We have previously established that the sum of the elements in any subset $\mathbb{T} \subset \mathbf{X}^D_K$ with size $|\mathbb{T}| = 2^D \times n$ is a multiple of the constant $\Psi^D_K$, defined by:

\begin{equation}
    \Psi^D_K = 2^{D(K+1)-1} - 2^{D-1}
\end{equation}

\noindent This naturally leads us to consider the total sum of all elements in $\mathbf{X}^D_K$, denoted by $S(\mathbf{X}^D_K)$. We assert that $S(\mathbf{X}^D_K)$ is also a multiple of $\Psi^D_K$. To compute $S(\mathbf{X}^D_K)$, we utilize the formula for the sum of the first $N$ natural numbers:

\begin{equation}
    S(\mathbf{X}^D_K) = \sum_{n=0}^{2^{D \times K}-1} n = \frac{(2^{D \times K} - 1)(2^{D \times K})}{2}
    \label{eq:sum_X}
\end{equation}

\noindent Since the sum of all subsets within $\mathbf{X}^D_K$ is a multiple of $\Psi^D_K$, there exists an integer $\Upsilon$ such that:

\begin{equation}
    S(\mathbf{X}^D_K) = \Psi^D_K \times \Upsilon
    \label{eq:S_X}
\end{equation}

\noindent After substituting the corresponding expressions of $S(\mathbf{X}^D_K)$ and $\Psi^D_K$ into Equation \ref{eq:S_X}, we have:

\begin{equation}
    \frac{(2^{D \times K} - 1)(2^{D \times K})}{2} = \left( 2^{D(K+1)-1} - 2^{D-1} \right) \times \Upsilon
    \label{eq:sum_equation}
\end{equation}

\begin{theorem}
\label{thm:upsilon_value}
For any dimension $D$ and scaling factor $K$, the integer $\Upsilon$ satisfying Equation \ref{eq:sum_equation} is given by:

\begin{equation}
    \Upsilon = 2^{D(K-1)}
    \label{eq:upsilon}
\end{equation}
\end{theorem}

This $\Upsilon$ tells us the total number of times we can count $\Psi^D_K$ in $S(\mathbf{X}^D_K)$. To count one $\Psi^D_K$, we have to sum all the elements of a given subset size $|\mathbb{T}|= 2^D \times 1 $. To count 2 $\Psi^D_K$, we have to sum all the elements of a given subset size $|\mathbb{T}|= 2^D \times 2 $. We can observe that the quotient between S($|\mathbb{T}|$) and $\Psi^D_K$ is the same as the quotient between $|\mathbb{T}|$ and $2^D$. This statement is exposed after we divide bought sides of the Equation \ref{eq:total_sum} by $\Psi^D_K$ : 

\begin{equation}
    \frac{S(\mathbb{T})}{\Psi^D_K} = \frac{|\mathbb{T}|}{2^D}
    \label{eq:equality}
\end{equation}

\noindent The total number of times we can count \( \Psi^D_K \) or \( 2^D \) across all subsets of any \( \mathbf{X}^D_K \) is \( \Upsilon \). To verify this statement, we refer to Tables \ref{tab:num_sub_set_2} and \ref{tab:num_sub_set_3}, which show how often subsets of different sizes appear in dimensions two and three for various values of \( K \). We will add a column to these tables to count the total occurrences of \( 2^D \) for each value of \( K \). For each subset of different size, we multiply the number of times a subset appears by the right-hand side of Equation \ref{eq:equality}, then sum the results obtained from all subset sizes. This total should match the value of \( \Upsilon = 2^{D(K - 1)} \).

\begin{table}[H]
\centering
\caption{The column 'Counted $2^D$' shows how many times $2^D$ can be counted for different values of $K$. The results match the values of $\Upsilon$ after substituting $D$ and $K$ into Equation \ref{eq:upsilon}.}
\begin{tabular}{|c|c|c|l|}
\hline
\textbf{$K$} & \textbf{$|\mathbb{T}| = 4$} & \textbf{$|\mathbb{T}| = 8$} & \textbf{Counted $2^D$} \\
\hline
1 & 1 & 0 & $1 \times 1 + 0 \times 2 = 1$ \\
\hline
2 & 2 & 1 & $2 \times 1 + 1 \times 2 = 4$ \\
\hline
3 & 4 & 6 & $4 \times 1 + 6 \times 2 = 16$ \\
\hline
4 & 8 & 28 & $8 \times 1 + 28 \times 2 = 64$ \\
\hline
... & ... & ... & ... \\
\hline
\end{tabular}

\end{table}

\begin{table}[H]
\centering
\caption{The column 'Counted $2^D$' shows how many times $2^D$ can be counted for different values of $K$. The results match the values of $\Upsilon$ after substituting $D$ and $K$ into Equation \ref{eq:upsilon}.}

\begin{tabular}{|c|c|c|c|c|c|c|c|l|}
\hline
& \textbf{$|\mathbb{T}|$} & 8 & 16 & 24 & 32 & 40 & 48 & \textbf{Counted  $2^D$} \\
\hline
\textbf{$K$} &  &   &  &  &  &  &  & \\
\hline
1 &  & 1 &  &  &  &   &  & $1 \times 1 = 1$ \\
\hline
2 &  & 3 & 1 & 1 &  &   &  & $3 \times 1 + 1 \times 2 + 1 \times 3 = 8 $ \\
\hline
3 &  & 7 & 3 & 9 &  &   & 4 & $7 \times 1 + 3 \times 2 + 9 \times 3 + 0 \times 4 + 0 \times 5 + 4 \times 6 = 64 $ \\
\hline
4 &  & 16 & 19 & 48 & 11 &   & 45 & $16 \times 1 + 19 \times 2 + 48 \times 3 + 11 \times 4 + 0 \times 5 + 45 \times 6 = 512 $ \\
\hline
... &  & ... & ... & ... & ... &   & ... & ... \\
\hline
\end{tabular}

\end{table}

In the previous two tables, we can observe that the number of times we count $2^D$  in all the $ \mathbb{T} \subset \mathbf{X}^D_K $ follows the progression of $\Upsilon$, $2^{D\times (K-1)}$.\\

\subsection{Spatial Properties of $\mathbb{T}$}

This subsection explores how the elements within a subset relate to one another in the space. In 2D and 3D, when plotting the elements of a subset, we consistently observe that they form the same pattern of shapes. However, the proportions of these shapes are arbitrary, which makes it challenging to work with them without establishing an explicit reference or framework.\\

To address this, we observe that each point shares its dimensional components with $n$ other points. We use this property to group points in coplanes, $\Pi$, that captures all the numbers with a common dimensional component. All the coplanes must contain the following information :

\begin{enumerate}
    \item The two dimensions that define the coplane, \( A \) and \( B \), are represented as \( \Pi^{AB} \).

    \item The number of elements \( n \) in the coplane \( \Pi^{AB} \) is denoted as \( \Pi^{AB}_n \).
\end{enumerate}


In dimension two, only exists one coplane, $\Pi^{XY}$ for all the subsets in $\mathbf{X}^2_K$. The number of elements of $\Pi^{XY}$ can be 4, $\Pi^{XY}_4$, or 8 , $\Pi^{XY}_8$. The following images show two coplanes, one with four elements and another with eight elements, along with the path each point in the subset followed before reaching its final position, as seen on Image \ref{fig:route_point}.

\begin{figure}[H]
    \centering
    \begin{minipage}{0.425\textwidth}
        \centering
        \includegraphics[width=\textwidth]{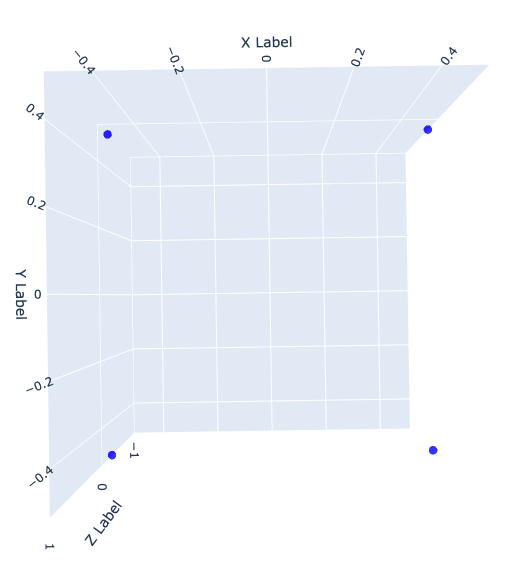}
    \end{minipage}
    \hfill
    \begin{minipage}{0.425\textwidth}
        \centering
        \includegraphics[width=\textwidth]{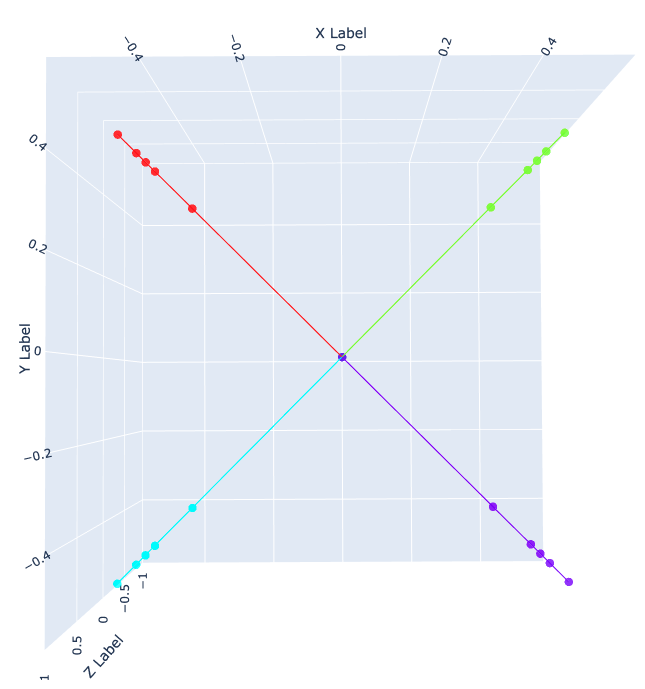}
    \end{minipage}
    \caption{
        Plot and routes of the points in \( \mathbb{T} = \{ 358, 972, 51, 665 \} \) $\subset$ \( \mathbf{X}^2_5 \) with radius 0.6774193548387097.
    }

\end{figure}

\begin{figure}[H]
    \centering
    \begin{minipage}{0.495\textwidth}
        \centering
        \includegraphics[width=\textwidth]{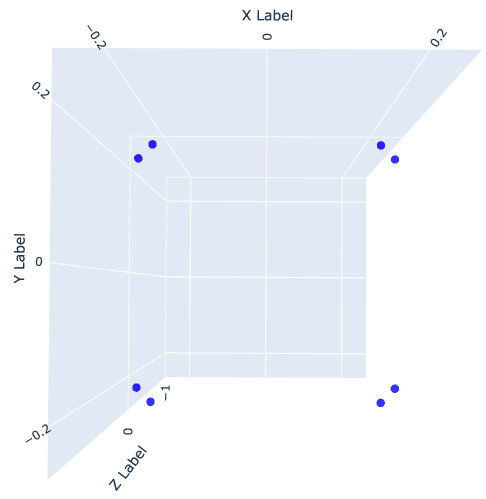}
    \end{minipage}
    \hfill
    \begin{minipage}{0.495\textwidth}
        \centering
        \includegraphics[width=\textwidth]{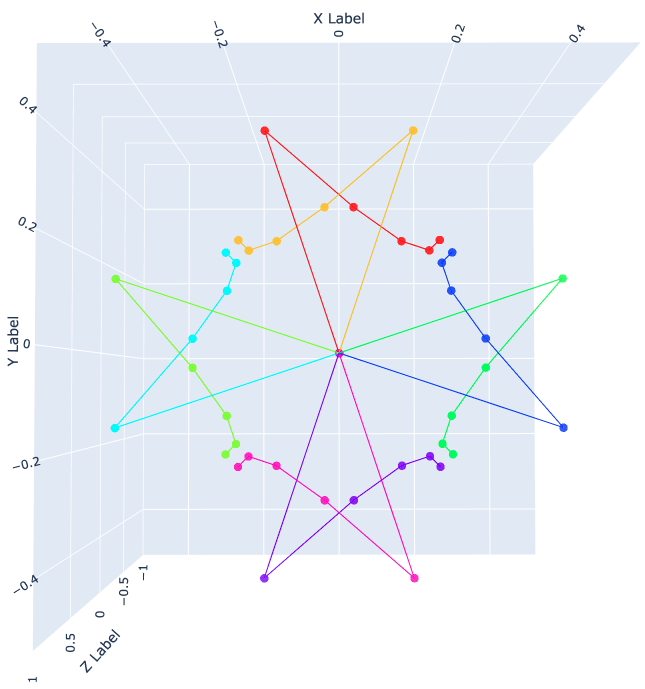}
    \end{minipage}
    \caption{
        Plot and routes of the points in \( \mathbb{T} = \{ 427, 769, 424, 770, 253, 599, 254, 596 \} \) $\subset$ \( \mathbf{X}^2_5 \) with radius 0.33208795317357703.
    }
\end{figure}

In dimension 3, we use two-dimensional objects, coplanes, to describe the structure of three-dimensional subsets. This approach has revealed the following key observations: 

\begin{enumerate}
    \item There are six types of coplanes in the subsets, denoted as $\Pi^{AB}_n$, where $A, B \in \{X, Y, Z\}$ with $A \ne B$, and $n \in \{4, 8\}$. These coplanes include $\Pi^{XY}_4$, $\Pi^{XY}_8$, $\Pi^{XZ}_4$, $\Pi^{XZ}_8$, $\Pi^{YZ}_4$, and $\Pi^{YZ}_8$.
    \item Every point $P \in \mathbb{T}$ lies at the intersection of 3 different coplanes. If we apply any operation to a point, the specific coplane over which the operation is performed must be specified.

    \item Each coplane that intersects any point has a corresponding mirror coplane, $\overline{\Pi^{AB}}$. In the mirror coplane, the coordinates in dimensions $A$ and $B$ remain the same as in the original coplane. The elements in the mirror coplane share the same values in dimensions $A$ and $B$, but their values in dimension $C$ are inverted, as would be applied the NOT gate on them.

\end{enumerate}

Now, we will show how the subsets of different sizes appear when we plot them in dimension 3. Subsets with the same number of elements consistently form the same structures, with only the distances between points changing. However, showing all possible structures for each subset size is not feasible. Instead, we will plot for each subset size one subset and display it from two different perspectives. Below each plotted subset, two additional images from the same perspectives will illustrate each point's paths before reaching its final position, as shown in Figure \ref{fig:route_point}. To simplify interpretation, the corresponding spheres to the different segments are omitted, and the route segments are colored based on the first three bits of the numbers in their Morton coordinates.

\begin{itemize}
    \item Visualization of a subset size 8: 

    \begin{figure}[H]
    \centering
        \includegraphics[width=0.495\textwidth]{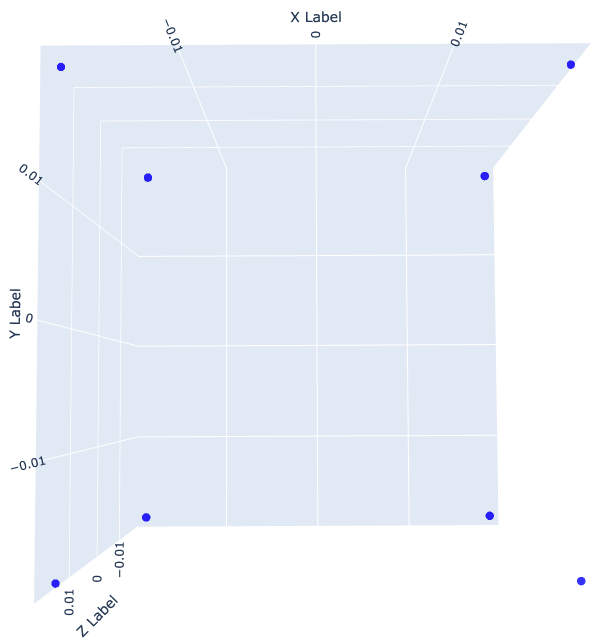}
        \includegraphics[width=0.495\textwidth]{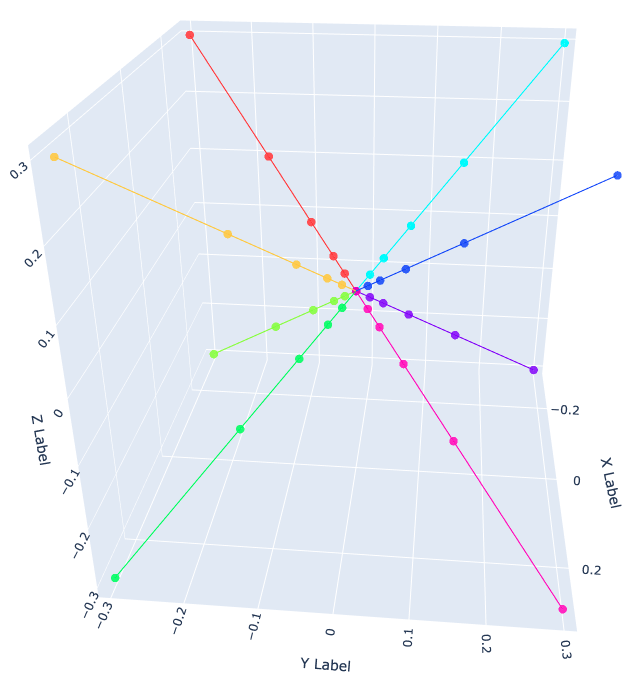}
       \caption{
        Plot and routes of the points in \( \mathbb{T} = \{ 47606, 21650, 4095, 18139, 14628, 28672, 11117, 25161 \} \) $\subset$ \( \mathbf{X}^3_5 \) with radius 0.03225806451612912.
    }
    \end{figure}

    \item Visualization of a subset size 16: 

\begin{figure}[H]
    \centering
    \begin{minipage}{0.495\textwidth}
        \centering
        \includegraphics[width=\textwidth]{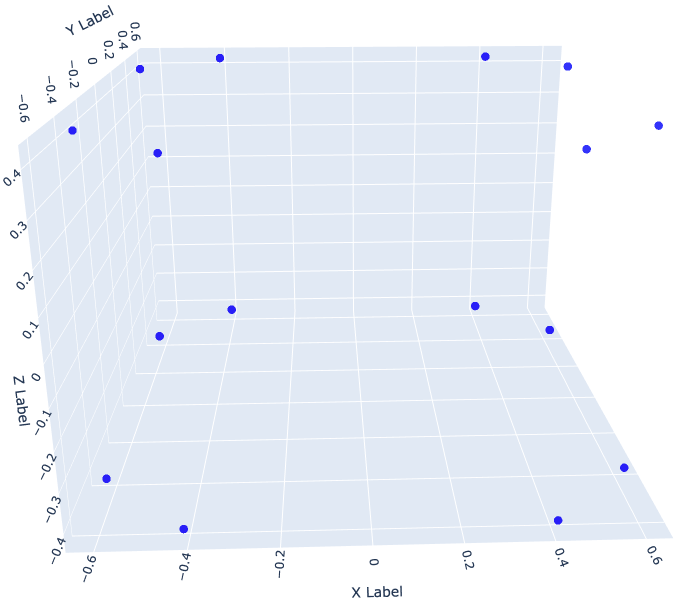}
    \end{minipage}
    \hfill
    \begin{minipage}{0.495\textwidth}
        \centering
        \includegraphics[width=\textwidth]{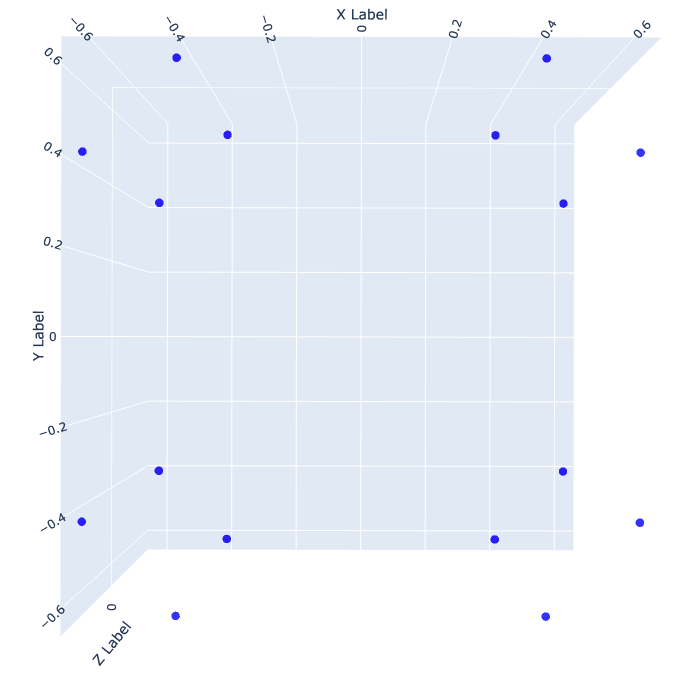}
    \end{minipage}
        
    \begin{minipage}{0.495\textwidth}
        \centering
        \includegraphics[width=\textwidth]{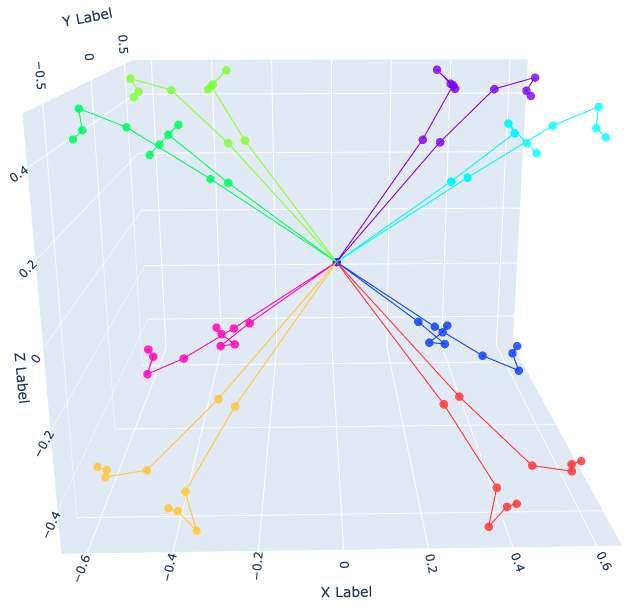}
    \end{minipage}
    \hfill
    \begin{minipage}{0.495\textwidth}
        \centering
        \includegraphics[width=\textwidth]{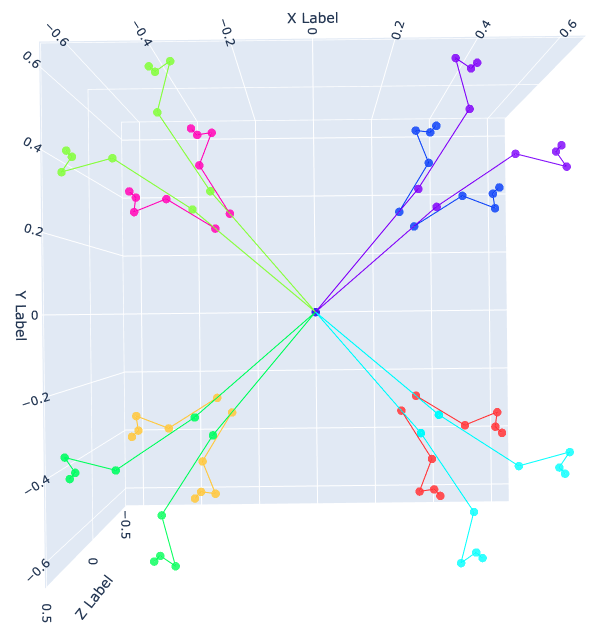}
    \end{minipage}

    \caption{
        Plot and routes of the points in $\mathbb{T}$ = $\{$ 4944, 23156, 281, 18493, 4832, 23492, 169, 18829, 13938, 32598, 9275, 27935, 14274, 32486, 9611, 27823 $\}$ $\subset \mathbf{X}^3_5$ with radius 0.8551178077162921.
    }
\end{figure}

    \item Visualization of a subset size 24: 

\begin{figure}[H]
    \centering
    \begin{minipage}{0.495\textwidth}
        \centering
        \includegraphics[width=\textwidth]{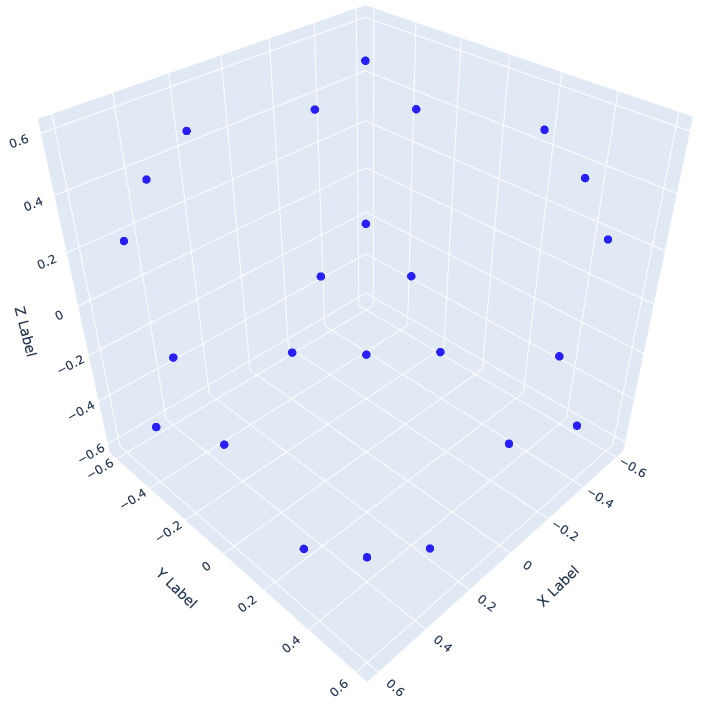}
    \end{minipage}
    \hfill
    \begin{minipage}{0.495\textwidth}
        \centering
        \includegraphics[width=\textwidth]{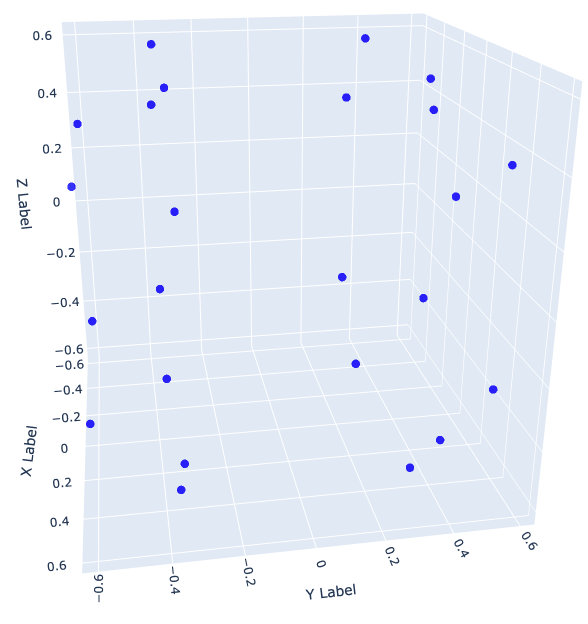}
    \end{minipage}
        
    \begin{minipage}{0.495\textwidth}
        \centering
        \includegraphics[width=\textwidth]{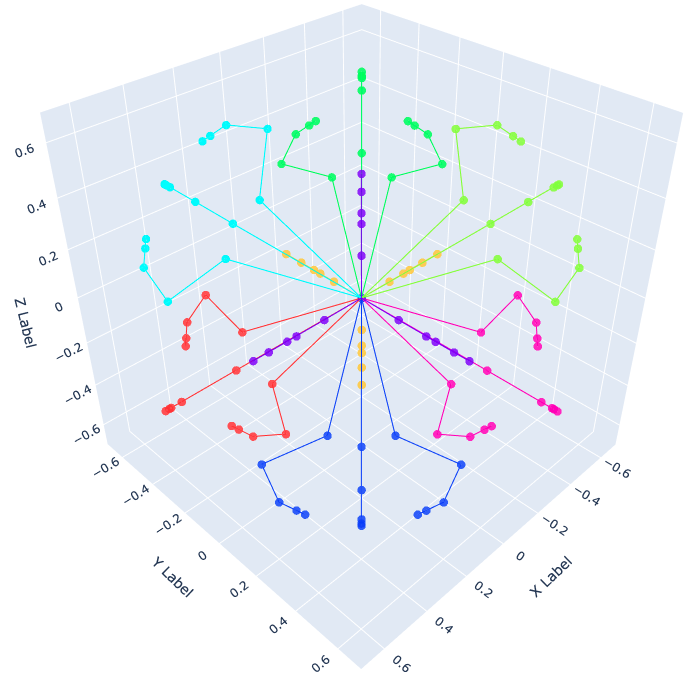}
    \end{minipage}
    \hfill
    \begin{minipage}{0.495\textwidth}
        \centering
        \includegraphics[width=\textwidth]{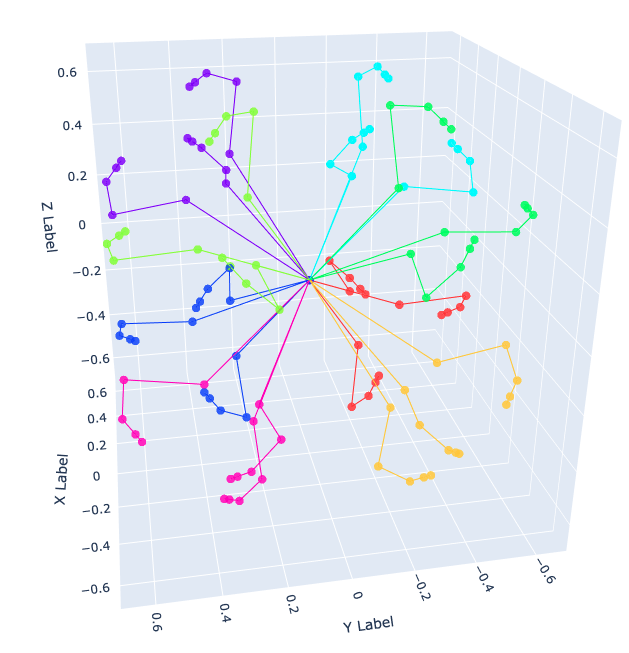}
    \end{minipage}

    \caption{
        Plot and routes of the points in $\mathbb{T}$ = $\{$ 6235, 20863, 2578, 17206, 7744, 22372, 5229, 23881, 1572, 20224, 3081, 17709, 15058, 29686, 12543, 31195, 8886, 27538, 10395, 25023, 15561, 30189, 11904, 26532 $\}$ $\subset \mathbf{X}^3_5$ with radius 0.7861258176340167.
    }
\end{figure}

    \item Visualization of a subset size 32: 

\begin{figure}[H]
    \centering
    \begin{minipage}{0.495\textwidth}
        \centering
        \includegraphics[width=\textwidth]{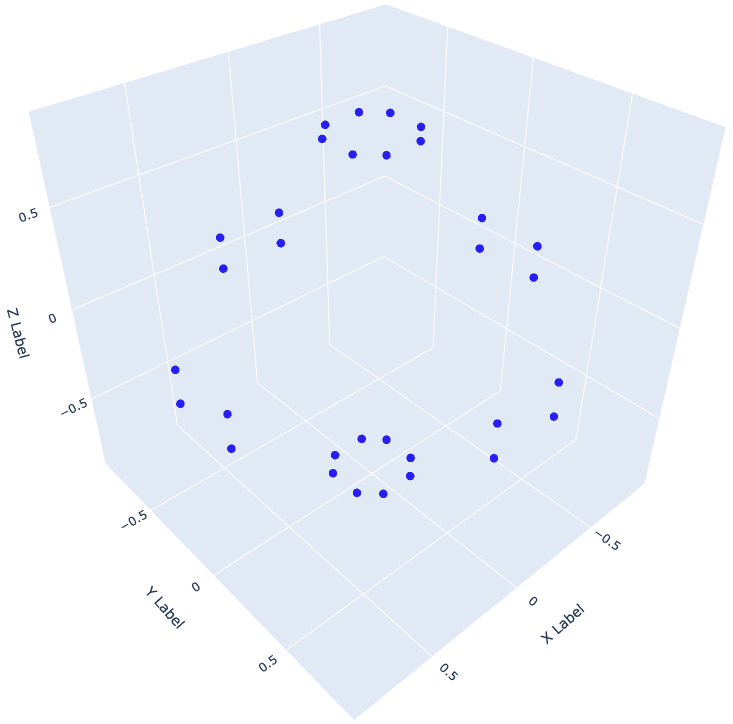}
    \end{minipage}
    \hfill
    \begin{minipage}{0.495\textwidth}
        \centering
        \includegraphics[width=\textwidth]{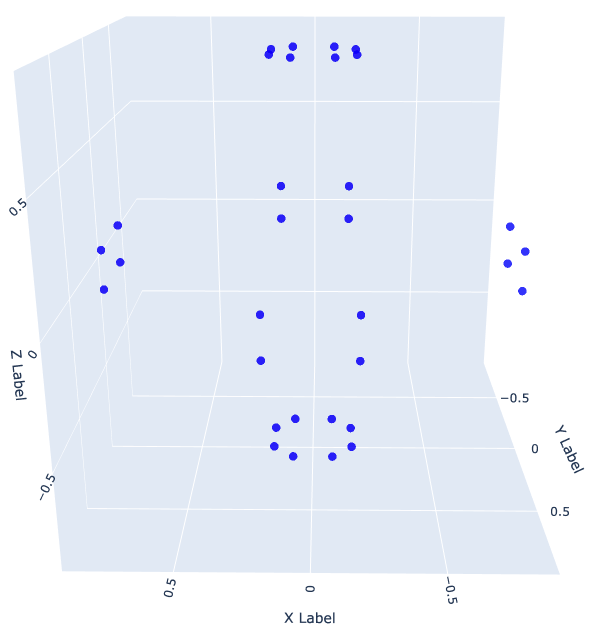}
    \end{minipage}
        
    \begin{minipage}{0.495\textwidth}
        \centering
        \includegraphics[width=\textwidth]{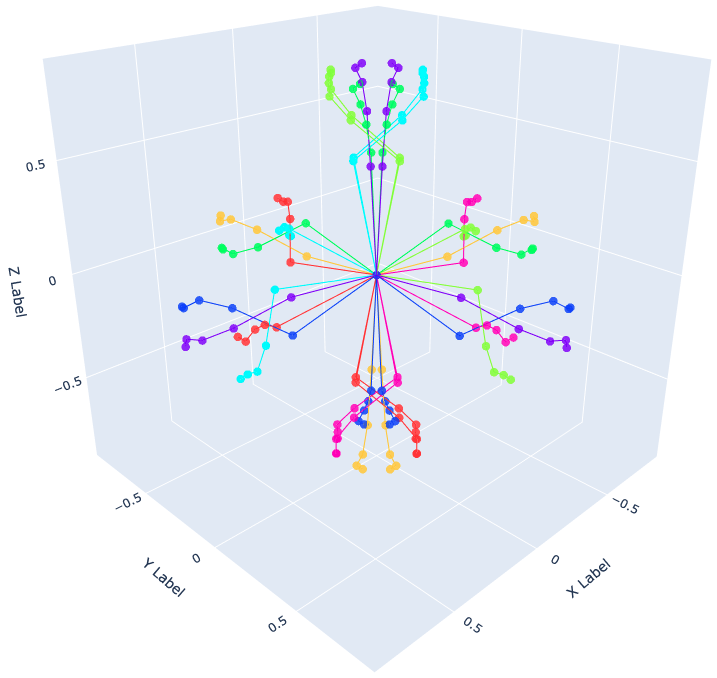}
    \end{minipage}
    \hfill
    \begin{minipage}{0.495\textwidth}
        \centering
        \includegraphics[width=\textwidth]{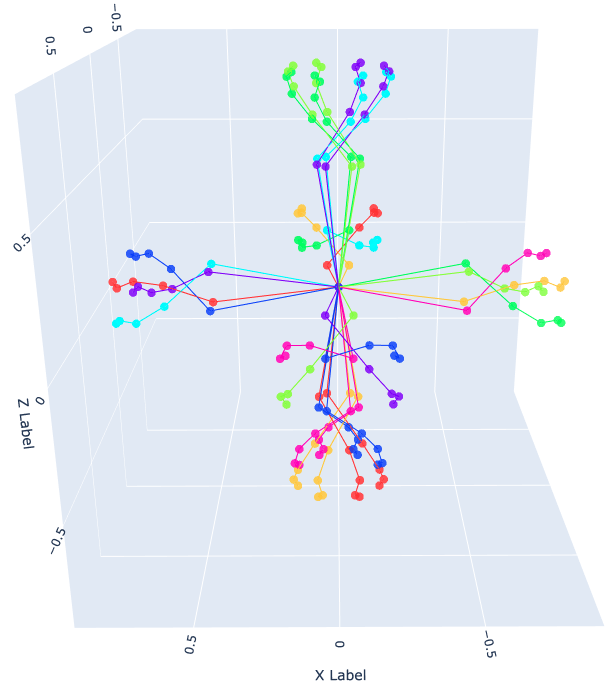}
    \end{minipage}

    \caption{
        Plot and routes of the points in $\mathbb{T}$ = $\{$ 6440, 20492, 2913, 16965, 8171, 22223, 3490, 17542, 8157, 22265, 5272, 23996, 1745, 20469, 3476, 17584, 15183, 29291, 12298, 31022, 8771, 27495, 10502, 24610, 15225, 29277, 10544, 24596, 15802, 29854, 12275, 26327 $\}$ $\subset \mathbf{X}^3_5$ with radius 0.8748984638909747.
    }
\end{figure}
\newpage
    
    \item Visualization of a subset size 48: 

\begin{figure}[H]
    \centering
    \begin{minipage}{0.495\textwidth}
        \centering
        \includegraphics[width=\textwidth]{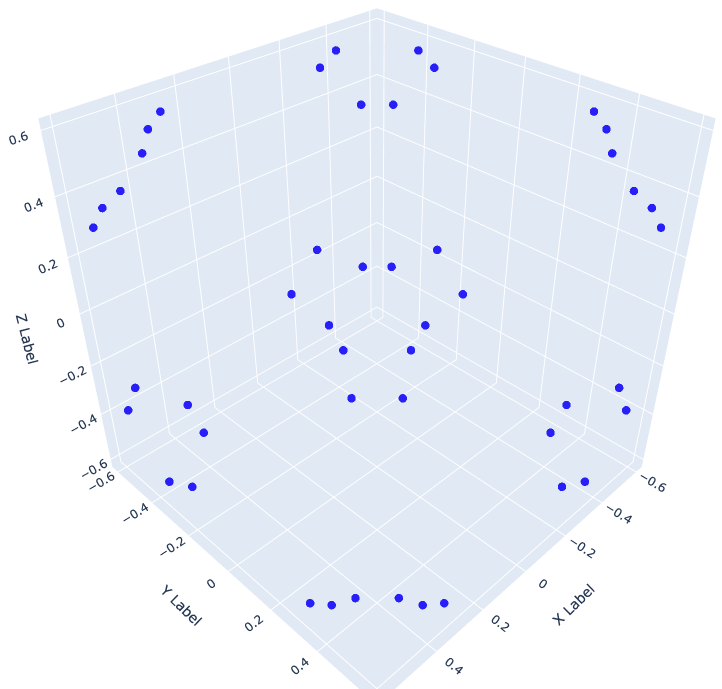}
    \end{minipage}
    \hfill
    \begin{minipage}{0.495\textwidth}
        \centering
        \includegraphics[width=\textwidth]{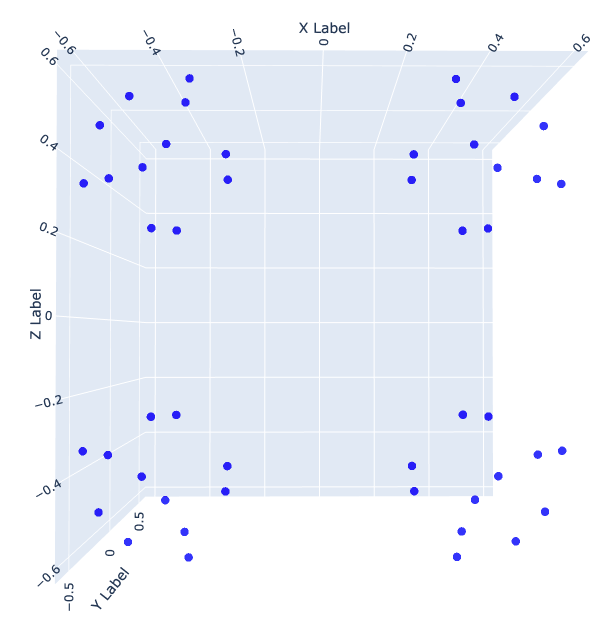}
    \end{minipage}
        
    \begin{minipage}{0.495\textwidth}
        \centering
        \includegraphics[width=\textwidth]{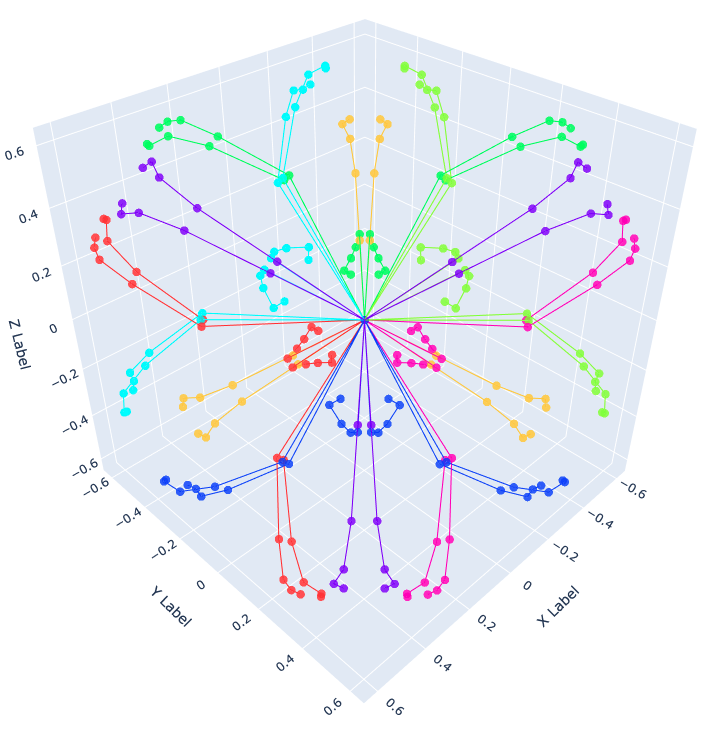}
    \end{minipage}
    \hfill
    \begin{minipage}{0.495\textwidth}
        \centering
        \includegraphics[width=\textwidth]{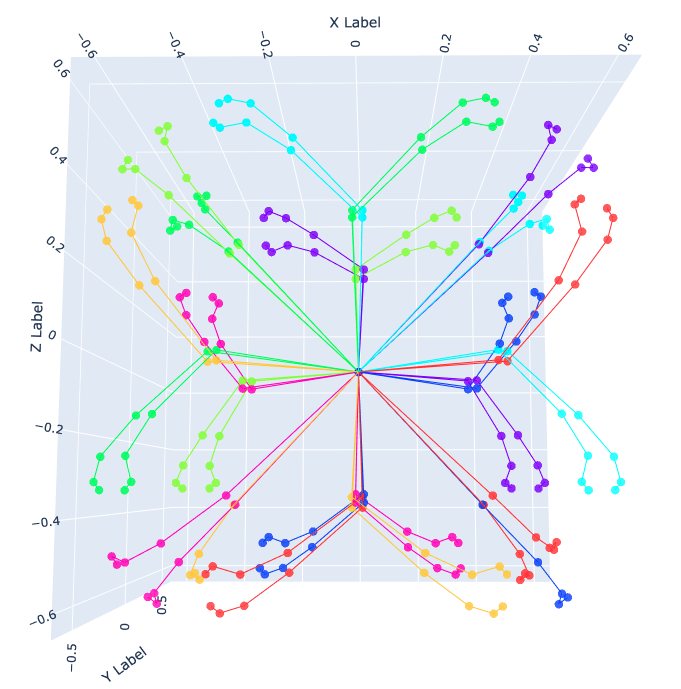}
    \end{minipage}

    \caption{
            Plot and routes of the points in $\mathbb{T}$ = $\{$ 7015, 21059, 4130, 22790, 619, 19279, 2350, 16394, 7036, 21080, 4116, 22832, 605, 19321, 2357, 16401, 5882, 24542, 5847, 24563, 1182, 19898, 1203, 19863, 12904, 31564, 12869, 31585, 8204, 26920, 8225, 26885, 16366, 30410, 13446, 32162, 9935, 28651, 11687, 25731, 16373, 30417, 13488, 32148, 9977, 28637, 11708, 25752 $\}$ $\subset \mathbf{X}^3_5$ with radius 0.8582254615975.
    }
    \label{fig:subset_48_plot}

\end{figure}  
    
\end{itemize}
\newpage
\subsubsection{Operations in $\Pi^{AB}_n$}

Now, we will examine how each point relates to the other points with which it shares a coplane. The number of elements in each coplane a point belongs to can vary. For instance, one coplane containing point \( P \) might have eight elements, while another coplane that intersects $P$, only has 4 points. This variation affects how any point \( P \) interacts with other points in each coplane, meaning its relationships can differ across coplanes.\\

To better understand these relationships, we first have to represent each point \( P \) in the coplane using its Morton coordinates in matrix form, represented as $[P]$. This notation is particularly useful because any change applied to a column affects all the bits corresponding to the same dimension. Using this approach, we can easily generate all the other points in the same coplane by applying transformations in the columns of bits, and since every point in $\mathbb{T}$ belongs to three coplanes, we can generate all the other points in $\mathbb{T}$.\\




\textit{To do it, we will make use of the operations in the groups} \( D_4 \) \textit{and} \( D_2 \)\cite{rotman2012introduction} \textit{using their matrix representations as they are described in representation theory}\cite{fulton2013representation}. To illustrate how the matrix representation of the group actions in \( D_4 \) and \( D_2 \) operates on the Morton coordinates, we will use a matrix \( M \) of size \( D \times D \), which represents any operation applied to the Morton coordinates. Each row and column in the matrix represents a dimension. The matrix has as many rows and columns as there are dimensions in the Morton coordinates representation of the point it operates on. For example, this is a matrix that operates in dimension 3:

\[
M = \begin{array}{c|ccc}
  & X & Y & Z \\
\hline
X & 0 & 1 & 0 \\
Y & -1 & 0 & 0 \\
Z & 0 & 0 & 1 \\
\end{array}
\]\\

Now we will show how the group actions, affect Morton coordinates by going through the following three basic interaction cases:

\begin{enumerate}

\item \textbf{No action, $e$ }: This operation consist in the matrix $M$ with 1 for any entry where $i=j$, the identity matrix.

    \emph{Example}:
    \[
    [P] = \begin{pmatrix}
    0 & 1 & 1 \\
    1 & 0 & 0\\
    1 & 0 & 1\\
    1 & 0 & 0\\
    \end{pmatrix}
    \quad \text{and} \quad
    M^e = \begin{pmatrix}
    1 & 0 & 0\\
    0 & 1 & 0\\
    0 & 0 & 1 \\
    \end{pmatrix}
    \]\\

    Resulting in:\\
\[       [P] = \begin{pmatrix}
    0 & 1 & 1 \\
    1 & 0 & 0\\
    1 & 0 & 1\\
    1 & 0 & 0\\
    \end{pmatrix}\]

\item \textbf{Negation, $M^{\neg A}$}: In the matrix \( M \), if \( M_{i, i} = -1 \), it indicates the application of the logic gate NOT to the bits in column \( i \) of the point [$P$].

    \emph{Example}:
    \[
    [P] = \begin{pmatrix}
    0 & 1 & 1 \\
    1 & 0 & 0\\
    1 & 0 & 1\\
    1 & 0 & 0\\
    \end{pmatrix}
    \quad \text{and} \quad
    M^{\neg Y} = \begin{pmatrix}
    1 & 0 & 0\\
    0 & -1 & 0\\
    0 & 0 & 1 \\
    \end{pmatrix}
    \]\\

    Resulting in:\\
\[   [P] = \begin{pmatrix}
    0 & 0 & 1 \\
    1 & 1 & 0\\
    1 & 1 & 1\\
    1 & 1 & 0\\
    \end{pmatrix}\]

\item \textbf{Swap, $M^{S(A,B)}$}: This operation swaps two columns of [$P$]. In $M^{S(A, B)}$, the letters $A$ and $B$ represent the dimensions that switch places.

    \begin{itemize}
        \item If \( M_{i,i} = 1 \), no change happens in dimension \( i \).
        \item If \( M_{i,j} = 1 \) (where \( i \neq j \)), the bits in dimension \( i \) switch places with those in dimension \( j \), and vice versa. If \( M_{i,j} = 1 \), then \( M_{j,i} = 1 \).
    \end{itemize}

This operation works by swapping the columns of bits between dimensions \( i \) and \( j \).

\emph{Example}:
\[
[P] = \begin{pmatrix}
0 & 1 & 1 \\
1 & 0 & 0 \\
1 & 0 & 1 \\
1 & 0 & 0 \\
\end{pmatrix}
\quad \text{and} \quad
M^{S(X,Y)} = \begin{pmatrix}
0 & 1 & 0 \\
1 & 0 & 0 \\
0 & 0 & 1 \\
\end{pmatrix}
\]

    Resulting in:\\

\[
[P] = \begin{pmatrix}
1 & 0 & 1 \\
0 & 1 & 0 \\
0 & 1 & 1 \\
0 & 1 & 0 \\
\end{pmatrix}
\]

\end{enumerate}


Now, we will see the operations that each coplane accepts based on its number of elements and explain how they generate new points within the same coplane. Additionally, we will examine how these operations impact the position of each point in space. Following this, we will present the matrix representations of the group actions in \(D_4\), which also encompass the group actions of \(D_2\), and explain each operation.

\begin{enumerate}

    \item \textbf{Identity ($M^{e}$)}: The identity operation leaves all points unchanged. This is represented by the identity matrix, which places 1 in every diagonal entry and 0 elsewhere. This operation does not alter the position of any points.
    
    \[
    M^{e} = \begin{pmatrix} 1 & 0 \\ 0 & 1 \end{pmatrix}, \quad
    M^{e} = \begin{pmatrix} 1 & 0 & 0 \\ 0 & 1 & 0 \\ 0 & 0 & 1 \end{pmatrix}
    \]
\item \textbf{Inversion ($M^{I}$)}: The inversion operation reflects the position of $[P]$ over both dimensions in the coplane. This transformation matrix includes entries of $-1$ for the two coplane dimensions and $1$ for any other dimension. This operation can always be applied to both \( \Pi^{AB}_4 \) and \( \Pi^{AB}_8 \). That every point within these coplanes has an inverse in the two-dimensional components means that given a point $[P]\in\Pi^{AB}_n$, the point that is the same as $[P]$ with the $A$, $B$, and $AB$ dimensional components negated will always lie on the same coplane. 
The matrices $M$ that represent this idea in dimensions 2 and 3 are:

    \[
M^{\text{I}} = \begin{pmatrix} -1 & 0 \\ 0 & -1  \end{pmatrix}, \quad
M^{\text{I}} = \begin{pmatrix} -1 & 0 & 0 \\ 0 & -1 & 0 \\ 0 & 0 & -1 \end{pmatrix}
\]

    \item \textbf{Reflection ($M^{\sigma(A)}$)}: This operation maps a point to another point whose position is the original point reflected across a specific dimension. The transformation matrix is the identity matrix, with a $-1$ in the entry corresponding to the dimension that is reflected. This operation works without restrictions because the same points as $[P]\in \Pi^{AB}_n$ with dimensional components $A$ or $B$ negated lie on the same coplane. We can use this operation freely in $\Pi^{AB}_4$ and $\Pi^{AB}_8$. The matrices $M$ that represent this idea are :

\[
M^{\sigma(X)} = \begin{pmatrix} -1 & 0 & 0 \\ 0 & 1 & 0 \\ 0 & 0 & 1 \end{pmatrix}, \quad
M^{\sigma(Y)} = \begin{pmatrix} 1 & 0 & 0 \\ 0 & -1 & 0 \\ 0 & 0 & 1 \end{pmatrix}, \quad
M^{\sigma(Z)} = \begin{pmatrix} 1 & 0 & 0 \\ 0 & 1 & 0 \\ 0 & 0 & -1 \end{pmatrix}
\]

Reflections can include swap operations; however, this reflection is defined with respect to a function rather than across an axis. Specifically, the effect of this operation corresponds to a reflection through the function \( f(A) = A \). It can only be applied in coplanes with eight elements or coplanes with four elements where the two-dimensional components are exactly the same or one is the other negated. The matrices $M$ that represent this idea are :\\

\begin{minipage}{0.3\textwidth}
\[
M^{S(X,Y)} = \begin{pmatrix}
0 & 1 & 0 \\
1 & 0 & 0 \\
0 & 0 & 1 \\
\end{pmatrix}
\]
\end{minipage}%
\begin{minipage}{0.3\textwidth}
\[
M^{S(Y,Z)} = \begin{pmatrix}
1 & 0 & 0 \\
0 & 0 & 1 \\
0 & 1 & 0 \\
\end{pmatrix}
\]
\end{minipage}%
\begin{minipage}{0.3\textwidth}
\[
M^{S(X,Z)} = \begin{pmatrix}
0 & 0 & 1 \\
0 & 1 & 0 \\
1 & 0 & 0 \\
\end{pmatrix}
\]
\end{minipage}

\vspace{0.5cm} 

 In a coplane with eight elements, the function does not pass through any of the elements. There are four elements at bought sizes of the function, and each element always has mirroring elements on the other side of the function. If we apply this operation to a point in a coplane where the dimensional components are either identical or one is the negation of the other, two points of the square lie exactly on the function, causing them to reflect onto themselves. The other two points are reflected across the function.\\


    \item \textbf{Rotation ($M^R$)}: The rotation operation transforms $[P]$ by swapping two dimensions and negating one of them. We can freely perform this operation in $\Pi^{AB}_8$ without restrictions. However, in  $\Pi^{AB}_4$, we can not freely use the operation rotation; to use it, the two-dimensional components we are operating must be the same or one the negation of the other.
    The matrices that represent this transformation correspond to the powers of the imaginary unit $i$ in representation theory, the cyclic group $C_4$~\cite{rotman2012introduction}.\\
    
    A single application of this rotation moves the point counterclockwise by $90^\circ$; applying it twice rotates by $180^\circ$, three times by $270^\circ$, and after four applications, the point returns to its original position. The matrices $M$ that represent this idea are : 

    \[
    M^{R} = \begin{pmatrix} 0 & -1 \\ 1 & 0 \end{pmatrix}; \quad
    (M^{R})^2 = \begin{pmatrix} -1 & 0 \\ 0 & -1 \end{pmatrix}; \quad
    (M^{R})^3 = \begin{pmatrix} 0 & 1 \\ -1 & 0 \end{pmatrix}; \quad
    (M^{R})^4 = \begin{pmatrix} 1 & 0 \\ 0 & 1 \end{pmatrix}
    \]

    The previous matrices represent the rotation on a coplane where each point only has two-dimensional components. When there are three-dimensional components, the dimension $C$ of the coplane $\Pi^{AB}$ remains unchanged, represented by a 1 in the position [C, C] of the matrix. In the dimensions $A$ and $B$, the matrix performs the corresponding operations swap and negation.

\end{enumerate}

The set of operations accepted for the coplane with eight elements corresponds to the dihedral group \( D_4 \). The coplane with four elements can accept all the group of actions \( D_4 \) when the two-dimensional components receiving the actions are either identical or one is the negation of the other. Otherwise, it cannot accept rotations or swaps and behaves as the group \( D_2 \).\\

The operations in the groups \(D_2\) and \(D_4\) generate elements within the same coplane and preserve the physical meaning of each operation. This means that if we apply symmetry along the Y-axis to some elements of the coplane, the resulting elements will remain in the coplane and truly symmetric along that axis. We can visualize this operation through the following three examples : 

\begin{figure}[H]
\centering
\includegraphics[width=8 cm]{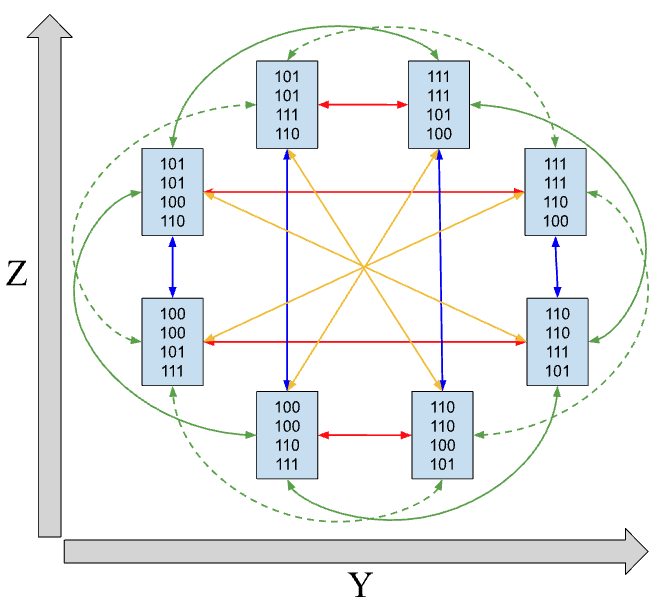}
\caption{This image shows the actions of the group $D_4$ on the elements in the first coplane $\Pi^{ZY}_8$ from the subset with radius 0.8592024201004166 and size $|\mathbb{T}|=48 \in \mathbf{X}^3_4$. The green lines represent rotations, the red and blue lines represent reflections, and the yellow lines represent inversion.}
\end{figure}

\begin{figure}[H]
\centering
\includegraphics[width=8 cm]{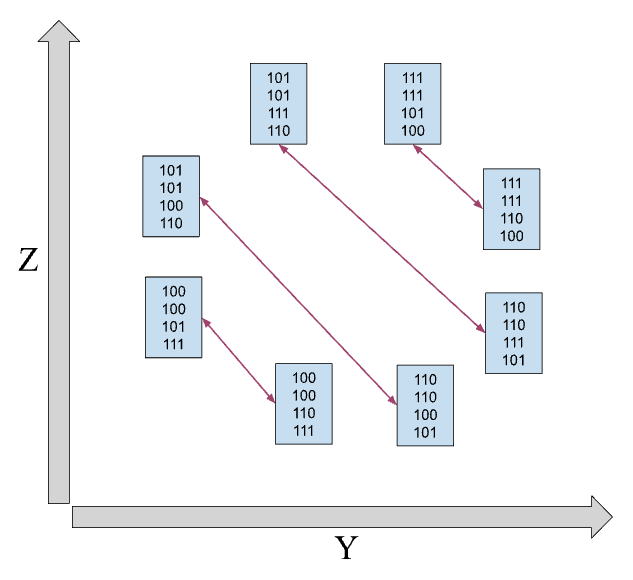}
\caption{This image shows the effect of the operation swap or \( sr \in D_4 \) on the elements in the first coplane $\Pi^{ZY}_8$ from the subset with radius 0.8592024201004166 and size $|\mathbb{T}|=48 \in \mathbf{X}^3_4$.}
\end{figure}

\begin{figure}[H]
\centering
\includegraphics[width=7.5 cm]{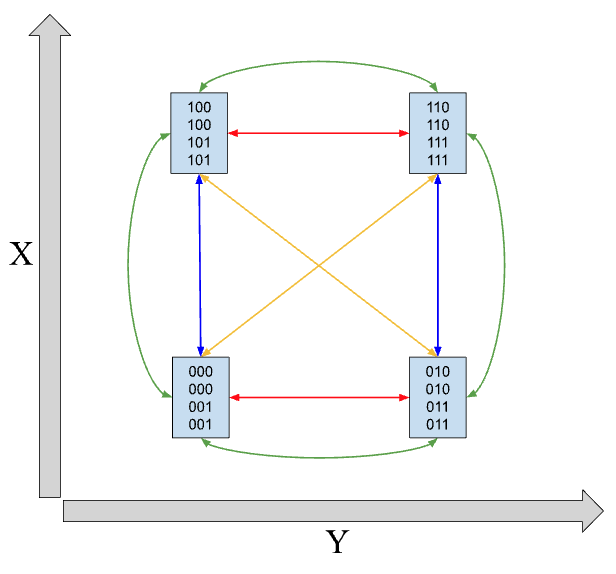}
\caption{This image shows the actions of the group $D_4$ on the elements in the first coplane $\Pi^{XY}_4$ from the subset with radius 0.9273380382140607 and size $|\mathbb{T}|=8 \subset \mathbf{X}^3_4$. The green lines represent rotations, the red and blue lines represent reflections, and the yellow lines represent inversion.}
\end{figure}

The \textbf{rotation operation} is of significant importance in understanding the structure of the subsets. As for every element there exists a coplane where any of the reflection operations is accepted, it implies that the object is an \( n \)-dimensional orthotope (an \( n \)-dimensional rectangle) or possibly a hypercube if all faces are congruent. Let us denote this \( n \)-dimensional orthotope or hypercube by \( \mathcal{O} \). By selecting any point in a subset \( \mathbb{T} \), we can access a total of \( 2^D \) points within the subset through reflection operations, forming an orbit corresponding to \( \mathcal{O} \). Since the total number of points in the subset is \( 2^D \times N \), this suggests that for every subset of size \( 2^D \times N \), there are \( N \) \( n \)-dimensional orthotopes or hypercubes \( \mathcal{O} \). The rotation operation is crucial for exploring the orbits of the distinct \( n \)-dimensional orthotopes \( \mathcal{O} \) within the sphere containing the subset \( \mathbb{T} \). Specifically, it enables transitions between these polytopes by "jumping" across faces that belong to different orthotopes but share a common coplane. Interpreting the subsets \( \mathbb{T} \) as a collection of \( \mathcal{O} \) with \( 2^D \) elements within the set \( \mathbf{X}^D_K \), which contains a total of \( \Upsilon^D_K \) orthotopes, has two important consequences:

\begin{enumerate}

\item \textbf{Sum of Elements in Orthotopes.} As stated in Equation \ref{eq:total_sum}, the sum of all the elements in \( \mathbb{T} \) is \( \Psi \) or a multiple of \( \Psi \). This observation suggests the following conjecture:

\begin{conjecture}
\label{conj:sum_orthotope}
For every orthotope \( \mathcal{O} \subset \mathbb{T} \), the sum of its elements satisfies:
\[S(\mathcal{O}) = \Psi\]
\end{conjecture}

For example, if we take the elements from the subset of 48 elements plotted in the Figure~\ref{fig:subset_48_plot}, $\mathbb{T}$ = $\{$ 7015, 21059, 4130, 22790, 619, 19279, 2350, 16394, 7036, 21080, 4116, 22832, 605, 19321, 2357, 16401, 5882, 24542, 5847, 24563, 1182, 19898, 1203, 19863, 12904, 31564, 12869, 31585, 8204, 26920, 8225, 26885, 16366, 30410, 13446, 32162, 9935, 28651, 11687, 25731, 16373, 30417, 13488, 32148, 9977, 28637, 11708, 25752 $\}$ $\subset \mathbf{X}^3_5$, we can organize them into 6 different orthotopes where the sum of the elements in each orthotope will equal \( \Psi^3_5= 131068 \):
\begin{enumerate}
    \item[--] \(\mathcal{O}_1 = \{605, 4116, 9935, 13446, 19321, 22832, 28651, 32162\}\), where \( S(\mathcal{O}_1) = 131068 \)
    \item[--] \(\mathcal{O}_2 = \{619, 4130, 9977, 13488, 19279, 22790, 28637, 32148\}\), where \( S(\mathcal{O}_2) = 131068 \)
    \item[--] \(\mathcal{O}_3 = \{1182, 5847, 8204, 12869, 19898, 24563, 26920, 31585\}\), where \( S(\mathcal{O}_3) = 131068 \)
    \item[--] \(\mathcal{O}_4 = \{1203, 5882, 8225, 12904, 19863, 24542, 26885, 31564\}\), where \( S(\mathcal{O}_4) = 131068 \)
    \item[--] \(\mathcal{O}_5 = \{2350, 7015, 11708, 16373, 16394, 21059, 25752, 30417\}\), where \( S(\mathcal{O}_5) = 131068 \)
    \item[--] \(\mathcal{O}_6 = \{2357, 7036, 11687, 16366, 16401, 21080, 25731, 30410\}\), where \( S(\mathcal{O}_6) = 131068 \)
\end{enumerate}

\item \textbf{Behavior Under Transformation \( \phi^D_{(S,B)} \).} In \ref{subsubsec:eq_pos}, we observed that applying the transformation \( \phi^D_{(S,B)} \) to a subset \( \mathbb{T} \) in dimension 2 always produces a valid subset in \( \mathbf{X}^D_B \). In dimension 3, however, this is not always the case. Nonetheless, as stated in Conjecture \ref{conj:eq_pos_alw}, we consistently found that the following holds:

For any valid subset \( \mathbb{T} \subset \mathbf{X}^D_S \), is satisfied the equation:
\[
S\big(\phi^D_{(S,B)}(\mathbb{T})\big) = \Psi^D_B \times \frac{|\mathbb{T}|}{2^D}
\]

\begin{conjecture}
\label{conj:orthotope_transformation}
Interpreting the subset \( \mathbb{T} \) as a collection of orthotopes \( \mathcal{O} \), each with \( 2^D \) elements, when we apply the transformation \( \phi^D_{(S,B)} \) to \( \mathbb{T}\subset \mathbf{X}^D_S \), we effectively apply \( \phi^D_{(S,B)} \) separately to each individual orthotope \( \mathcal{O} \). This results in new valid orthotopes \( \mathcal{O}' \) in \( \mathbf{X}^D_B \). While the set of orthotopes \( \{ \mathcal{O} \} \subset \mathbb{T} \) in \( \mathbf{X}^D_S \) belong to the same subset \( \mathbb{T} \), after applying \( \phi^D_{(S,B)} \), they may exist as parts of the same subset or as parts of different subsets in \( \mathbf{X}^D_B \).
\end{conjecture}

\end{enumerate}


\newpage
\subsection{Internal Structure of $\mathbb{T}$} 

In the previous subsections, we observed that the subsets consistently exhibit a wide variety of properties. In this subsection, we will attempt to identify their common factor or internal structure. To identify the internal structure of a subset \( \mathbb{T} \), we proceed as follows:


\begin{enumerate}
    \item \textbf{Sorting and Decomposition:} Sort the elements of $\mathbb{T}$ in ascending order, from the smallest to the largest. Then, represent each number in the subset using its Morton coordinates, separating them into their dimensional components.

    \item \textbf{Identifying Transition Chains:} For each dimensional component and starting in the dimension \( X \), identify the chain of bits that, subjected to the XOR operation, will generate the corresponding dimensional component of the next number. Repeat this step for each consecutive pair of numbers in \( \mathbb{T} \). When the last element is reached, compute the chain of bits that produces the first element.

    \item \textbf{Constructing the Dictionary:} Each time such a chain of bits is identified, record it in a shared \textbf{dictionary}, where each chain of bits will be associated with a variable, starting from \( A \). This dictionary is common to all dimensional components. For example, if the entry \( A = 1011 \) appears in the dimensional component \( X \), other dimensional components can also utilize the variable \( A \). Before including any new chains of bits, the initial dictionary includes the variables \( = \), representing a chain of \( K \) zeros, and \( X \), representing a chain of \( K \) ones. 

\end{enumerate}

\begin{definition}[Subset Generator]
We define the \emph{Subset Generator} (\( SG \)) as the sequence of variables across all dimensional components in a subset, regardless of the specific values assigned to that variables in the dictionary.

\end{definition}

The sequence of variables in a single dimension \( D \) is called \( \text{Path}_D \). We use the term \( SG \) because we can generate every element in the subset using the sequence of variables, the corresponding values stored in the dictionary, and the Morton Coordinates of the first number in the subset. We observe that all the $SG$, regardless of the dimension or subset size, share the following properties:

\begin{enumerate}[label=(\alph*)]
    \item \textbf{Common $SG$ Among Same Size Subsets:} \textit{Each subset has an internal structure that corresponds to one of a limited set of \( SG \) determined by the number of elements in the subset and the dimension. This set of possible \( SG \) structures remains consistent for all subsets with the same number of elements in the same dimension, regardless of the specific values assigned to the variables.} In Appendix~\ref{appendix:subset_generators}, we present the \( SG \) for dimensions 2 and 3 for the subsets of different sizes. We omit the sequences of \( X \) that appear at the beginning or end of all the paths of the \( SG \), connecting the last elements back to the first one, as this is common to all of them.

    In Tables~\ref{tab:dict_size_dim2} and~\ref{tab:dict_size_dim3}, we show how the size of the dictionary and the number of unique \( SG \) vary respect the number of elements in $\mathbb{T}$.

\begin{table}[H]
\centering
\caption{Dictionary size and number of unique \( SG \) in dimension 2 for subsets size 4 and 8.}
\label{tab:dict_size_dim2}
\begin{tabular}{|c|c|c|}
\hline
\( |\mathbb{T}| \) & Dictionary Size & Number of Unique \( SG \) \\
\hline
4 & 2 & 1 \\
8 & 4 & 1 \\
\hline
\end{tabular}
\end{table}

\begin{table}[H]
\centering
\caption{Dictionary size and number of unique \( SG \) in dimension 3 for subsets of different sizes.}
\label{tab:dict_size_dim3}
\begin{tabular}{|c|c|c|}
\hline
\( |\mathbb{T}| \) & Dictionary Size & Number of Unique \( SG \) \\
\hline
8  & 2 & 1 \\
16 & 4 & 1 \\
24 & 4 & 2 \\
32 & 6, 7 & 6 \\
48 & 8 & 2 \\
\hline
\end{tabular}
\end{table}
\end{enumerate}

\begin{enumerate}[label=(\alph*),resume]
    \item \textbf{Unique Paths:} Each path has a distinct sequence of variables, meaning no two paths can share the same variables sequence.
\end{enumerate}

\begin{enumerate}[label=(\alph*),resume]
    \item \textbf{Cyclic Structure:} When the last element of the subset is reached, the sequence of variables that connects it to the first element always is \( X \), common to all paths, creating a cyclic structure. These are referred to as trivial variables, as they are common to all \( SG \) and are sometimes omitted. When not omitted, they are conventionally placed at the beginning of the \( SG \).


\end{enumerate}

\begin{enumerate}[label=(\alph*),resume]
    \item \textbf{Symmetric Structure:} In the middle of every path sequence, the variable \( X \) appears consistently across all paths. The variables that follow this point mirror the variables that preceded it.
\end{enumerate}

\begin{table}[H]
\centering
\begin{minipage}{0.55\textwidth}
\centering
\caption{Example of a Subset Generator (\( SG \)) associated with a subset of 8 elements. The first entries in the \( \text{Path}_X \) and \( \text{Path}_Y \) columns, "$X|X$", represent the dictionary entries that connect the dimensional components of the last element with the first element.}
\label{tab:SG_example}
\begin{tabular}{|c|c|c|c|c|c|}
\hline
\textbf{Decimal} & \textbf{Binary} & \textbf{X} & \textbf{Y} & \( \text{Path}_X \) & \( \text{Path}_Y \) \\
\hline
54  & 00110110 & 0101 & 0110 & \( X \) & \( X \) \\
57  & 00111001 & 0110 & 0101 & \( A \) & \( A \) \\
99  & 01100011 & 0101 & 1001 & \( A \) & \( B \) \\
108 & 01101100 & 0110 & 1010 & \( A \) & \( A \) \\
147 & 10010011 & 1001 & 0101 & \( X \) & \( X \) \\
156 & 10011100 & 1010 & 0110 & \( A \) & \( A \) \\
198 & 11000110 & 1001 & 1010 & \( A \) & \( B \) \\
201 & 11001001 & 1010 & 1001 & \( A \) & \( A \) \\
\hline
\end{tabular}
\end{minipage}%
\hfill
\begin{minipage}{0.4\textwidth}
\centering
\caption{Dictionary associated with the \( SG \) in Table~\ref{tab:SG_example}.}
\label{tab:SG_dictionary}
\begin{tabular}{|c|c|}
\hline
\textbf{Variable} & \textbf{Value} \\
\hline
\( X \) & 1111 \\
\( = \) & 0000 \\
\( A \) & 0011 \\
\( B \) & 1100 \\
\hline
\end{tabular}
\end{minipage}
\end{table}

\begin{enumerate}[label=(\alph*),resume]
    \item \textbf{Path Cumulative XOR:} Given any valid \( SG \) and its corresponding dictionary, we initialize an empty column for each dimensional component with as many rows as variables can be found in the path. To fill the row \( n \) in any of the empty column $D$, we take the value of the variable in the \( SG \) at the dimensional component \( D \) and row \( n \), then perform a bit-wise XOR operation with the value from position \( n-1 \) in the column we want to fill. We repeat this process for each dimension. For the first element in the table (\( n=1 \)), we XOR the corresponding variable with a chain of \( K \) zeros, leaving the first variable unchanged. At the end of each dimensional component, ignoring the trivial \( X \) at the beginning or end of the \( SG \),  always results in chains of \( K \) ones across all dimensions.
    
    In summary, the cumulative XOR of the sequence of variables in an \( SG \), excluding the \( X \) that connects the last element to the first in each dimensional component, results in a chain of \( K \) ones across all the dimensional components.
\end{enumerate}


\begin{table}[H]

\centering

\begin{minipage}{0.6\textwidth}
\centering
\caption{Steps of the cumulative XOR for a given \( SG \).}

\begin{tabular}{|c|c|c|c|c|c|}
\hline
A & A & = & 0100 & 0100 & 0000\\
\hline
A & A & X & 0000 & 0000 & 1111 \\
\hline
A & A & = & 0100 & 0100 & 1111 \\
\hline
A & B & X & 0000 & 1111 & 0000 \\
\hline
A & A & = & 0100 & 1011 & 0000 \\
\hline
A & A & X & 0000 & 1111 & 1111 \\
\hline
A & A & = & 0100 & 1011 & 1111 \\
\hline
X & X & X & 1011 & 0100 & 0000 \\
\hline
A & A & = & 1111 & 0000 & 0000 \\
\hline
A & A & X & 1011& 0100 & 1111 \\
\hline
A & A & = & 1111 & 0000 & 1111 \\
\hline
A & B & X & 1011 & 1011 & 0000 \\
\hline
A & A & = & 1111 & 1111 & 0000 \\
\hline
A & A & X & 1011 & 1011 & 1111 \\
\hline
A & A & = & 1111 & 1111 & 1111 \\
\hline
\end{tabular}
\label{tab:substitute}
\end{minipage}%
\hfill
\begin{minipage}{0.35\textwidth}
\caption{The values of the dictionary associated with the \( SG \) from Table \ref{tab:substitute}.}

\centering
\begin{tabular}{|c|c|}
\hline
A & 0100 \\
\hline
B & 1011 \\
\hline
X & 1111 \\
\hline
= & 0000 \\
\hline
\end{tabular}
\end{minipage}
\end{table}

\subsubsection{Dictionary Isomorphism}

In this subsection, we explore the general structure of the dictionaries associated with the subsets of different sizes. Since the dictionaries are derived by finding the \( SG \) of a given subset \( \mathbb{T} \), and all \( SG \) exhibit structures, it is reasonable to hypothesize that the dictionaries corresponding to these \( SG \) also have an internal structure. To explore the internal structure of the dictionaries and examine whether dictionaries with the same number of elements share any common properties, we will use two valid dictionaries with 7 variables to accompany the explanation of the steps to follow.

\begin{table}[H]
\centering
\caption{Two distinct dictionaries.}
\label{tab:two_dictionaries}
\begin{tabular}{|c|c|c|}
\hline
\textbf{Entry} & \textbf{Dictionary 1} & \textbf{Dictionary 2} \\
\hline
A & 0110100 & 0011011 \\
B & 0001011 & 0101001 \\
C & 0111111 & 0110010 \\
D & 1000000 & 1100100 \\
E & 1110100 & 1010110 \\
X & 1111111 & 1111111 \\
= & 0000000 & 0000000 \\
\hline
\end{tabular}
\end{table}


\begin{enumerate}
    \item \textbf{Constructing the Cayley Table:} We create a Cayley table~\cite{rotman2012introduction} under the bitwise XOR operation (\( \oplus \)) using the variables from the dictionaries. If the result of \( A \oplus B \) is not a variable present in the dictionary, we leave that cell in the table empty. We do this to see how the variables within the dictionaries are related to each other.

\end{enumerate}

\begin{table}[H]
\centering
\begin{tabular}{cc}
\begin{minipage}{0.45\textwidth}
\centering
\caption{Cayley table for Dictionary 1 under XOR operation.}
\label{tab:cayley_dict1}
\begin{tabular}{|c||c|c|c|c|c|c|c|}
\hline
\( \oplus \) & A & B & C & D & E & X & = \\
\hline\hline
A & = & C & B & E & D &    & A \\
\hline
B & C & = & A &   & X & E & B \\
\hline
C & B & A & = &   &   & D & C \\
\hline
D & E &   &   & = & A &   & D \\
\hline
E & D & X &   & A & = & B & E \\
\hline
X &   & E & D &   & B & = & X \\
\hline
= & A & B & C & D & E & X & = \\
\hline
\end{tabular}
\end{minipage}
&
\begin{minipage}{0.45\textwidth}
\centering
\caption{Cayley table for Dictionary 2 under XOR operation.}
\label{tab:cayley_dict2}
\begin{tabular}{|c||c|c|c|c|c|c|c|}
\hline
\( \oplus \) & A & B & C & D & E & X & = \\
\hline\hline
A & = & C & B & X &   & D & A \\
\hline
B & C & = & A &   & X & E & B \\
\hline
C & B & A & = & E & D &   & C \\
\hline
D & X &   & E & = & C & A & D \\
\hline
E &   & X & D & C & = & B & E \\
\hline
X & D & E &   & A & B & = & X \\
\hline
= & A & B & C & D & E & X & = \\
\hline
\end{tabular}
\end{minipage}
\end{tabular}
\end{table}

\begin{enumerate}
\setcounter{enumi}{1}
    \item \textbf{Constructing the Cayley Graph:} Using the Cayley tables, we construct the corresponding Cayley graphs~\cite{rotman2012introduction}. To simplify the graphs, we omit the node corresponding to \( = \). \textbf{In the Cayley table, if we have \( A \oplus B = C \), then the \( A \) in the row is the starting node, \( B \) in the column is the edge label, and the resulting \( C \) is the ending node}. Due to the XOR operation, all edges are bidirectional; thus, an edge from \( A \) to \( C \) labeled \( B \) also represents an edge from \( C \) to \( A \) labeled \( B \).

\end{enumerate}

\begin{figure}[H]
\centering
\begin{minipage}{0.45\textwidth}
    \centering
    \includegraphics[width=\textwidth]{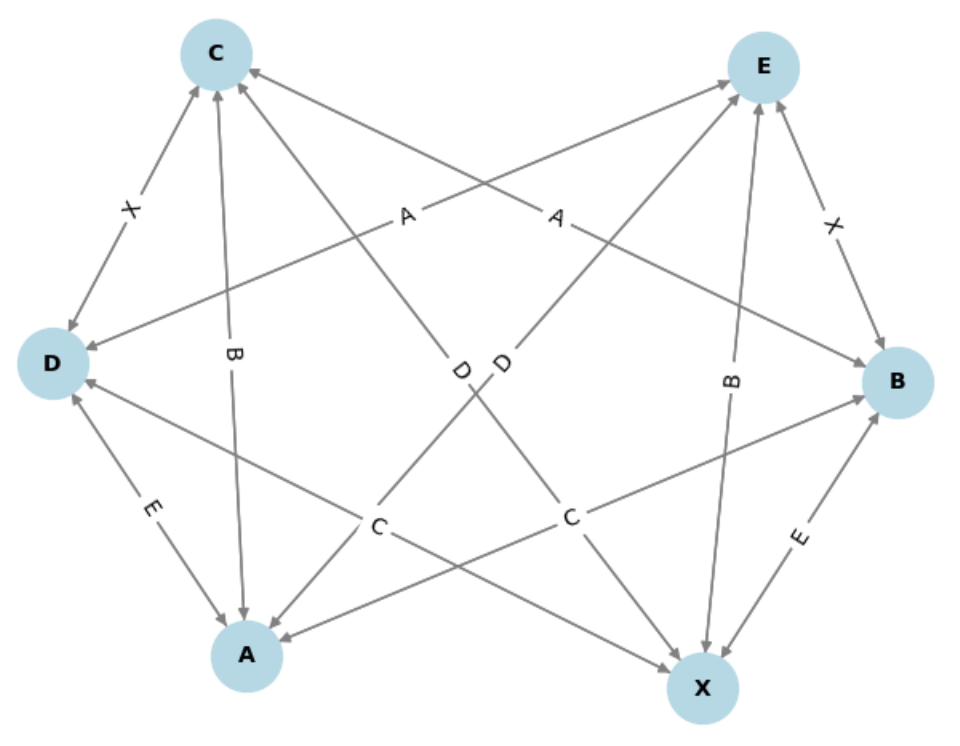}
    \caption{Cayley graph constructed from Dictionary 1.}
    \label{fig:cayley_graph1}
\end{minipage}
\hfill
\begin{minipage}{0.45\textwidth}
    \centering
    \includegraphics[width=\textwidth]{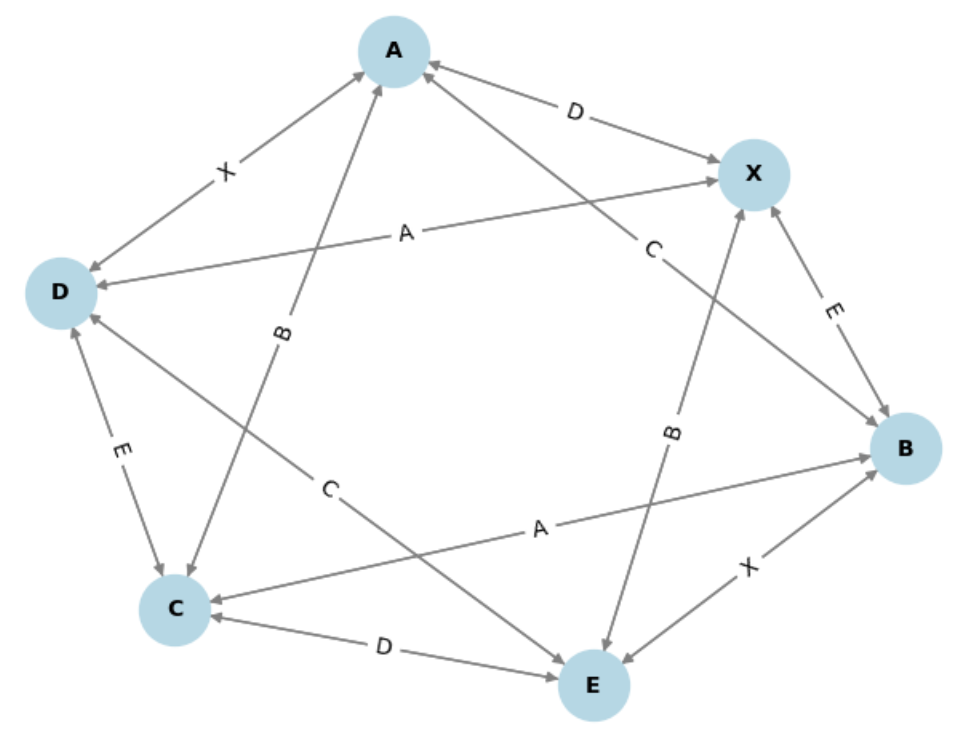}
    \caption{Cayley graph constructed from Dictionary 2.}
    \label{fig:cayley_graph2}
\end{minipage}
\end{figure}

\begin{enumerate}
\setcounter{enumi}{2}
    \item \textbf{Calculating Node Metrics:} We compute a metric for the graph nodes to assign numerical values to each of them. While the choice of metric is not critical, for this example, we use Closeness Centrality~\cite{sabidussi1966centrality}. After computing the values for each node, we organize them into vectors:

    \begin{itemize}
        \item Dictionary 1: \{A: 0.05, B: 0.05, C: 0.05, D: 0.05, E: 0.05, X: 0.05\}
        \item Dictionary 2: \{A: 0.05, B: 0.05, C: 0.05, D: 0.05, E: 0.05, X: 0.05\}
    \end{itemize}

    \item \textbf{Comparing Metric Vectors:} We remove the node identifiers from the metric vectors, leaving only the numerical values, and sort these values. If, after comparing the sorted metric vectors of two graphs from different dictionaries, we find that they are identical, the two graphs are isomorphic~\cite{mckay1981practical}.

    \begin{itemize}
        \item Dictionary 1: \{0.05, 0.05, 0.05, 0.05, 0.05, 0.05\}
        \item Dictionary 2: \{0.05, 0.05, 0.05, 0.05, 0.05, 0.05\}
    \end{itemize}
\end{enumerate}

This method, though not universally applicable, is well-suited to highly symmetric contexts. Due to the XOR operation, the relationships between variables in the graph have the following characteristics:

\begin{enumerate}[label=(\alph*)]
    \item All connections between nodes are bidirectional.
    \item If node \( A \) is connected to node \( B \) through edge \( C \), then node \( C \) is also connected to node \( A \) through edge \( B \) and to node \( B \) through edge \( A \).
\end{enumerate}

After comparing the metric vectors for the dictionaries in dimensions 2 and 3 subsets, up to \( K = 11 \) for dimension 2 and \( K = 7 \) for dimension 3, we observe that \textbf{every unique dictionary with the same number of elements has the same metric vector}. This indicates that their corresponding graphs are isomorphic. Then, we will identify the canonical graphs~\cite{mckay1981practical} of the Cayley graphs associated with dictionaries of different sizes. There are five canonical graphs corresponding to dictionary sizes \(\{2, 4, 6, 7, 8\}\).

To identify the graph's canonical form, we begin by finding two elements in any of the isomorphic dictionaries that, under the XOR operation, produce a third element also contained in the dictionary. Starting from the initial equation:

\[
A \oplus B = C
\]

\noindent We apply \( A \oplus \) and \( B \oplus \) to both sides of the equation to obtain:

\[
A \oplus (A \oplus B) = C \oplus A \implies B = A \oplus C
\]

\[
B \oplus (A \oplus B) = C \oplus B \implies A = B \oplus C
\]

\noindent We encode these relationships as a graph consisting of three nodes, where the edge connecting two nodes always has the value of the third node.

\begin{figure}[H]
    \centering
    \includegraphics[width=0.35\textwidth]{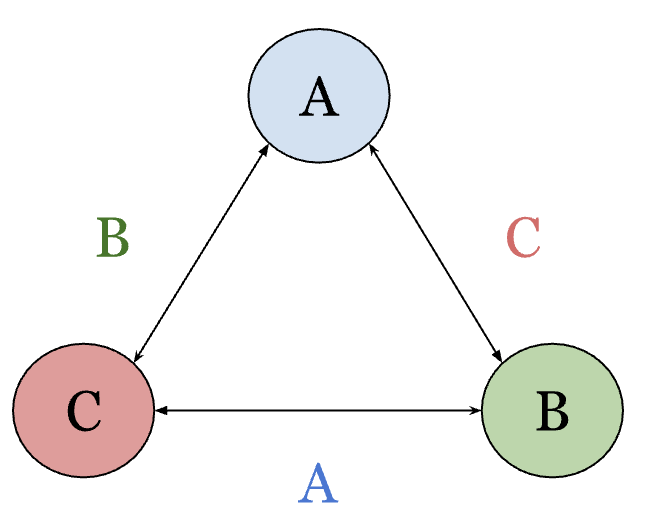}
    \caption{Graph showing the correlation of variables in \( A \oplus B = C \).}
    \label{fig:three_node_graph}
\end{figure}

This approach transforms these relationships into automorphic~\cite{babai1995automorphism} commitments. By combining these commitments from different relations of three variables encoded as graphs, we create a constraint system that allows us to construct the graph's canonical form.

However, the nodes of the canonical graph cannot represent the specific entries of the dictionary because doing so, would prevent the graph from generalizing all possible automorphic configurations of a valid dictionary. Instead, we aim to preserve the graph's overall structure while allowing the dictionary entries to adapt to different valid node configurations.

In dimensions 2 and 3, aside from \( X \) and \( = \) (where \( = \) is omitted because it does not provide new information), we can find up to six entries: \(\{A, B, C, D, E, F\}\), as shown in Table \ref{tab:dict_size_dim3}. We categorize these dictionary entries into three types:

\begin{itemize}
    \item \(\mathcal{A} = \{A, B, C\}\)
    \item \(\mathcal{B} = \{D, E, F\}\)
    \item \( X \) (which is independent)
\end{itemize}

Variables within the same category can permute or switch roles into valid solutions that respect the general structure of the graph after abstraction. We make this abstraction by transforming \(\mathcal{A} = \{A, B, C\}\) into \(\mathcal{A} = \{A_1, A_2, A_3\}\) and \(\mathcal{B} = \{D, E, F\}\) into \(\mathcal{B} = \{B_1, B_2, B_3\}\).

Finally, we substitute the abstract variables into the constraint system. If we assign a specific variable from the dictionary to an abstract variable, we ensure no other variable receives the same abstract variable.

To enhance the readability of the graphs, the connections within subgraphs of three nodes will be color-coded according to the following criteria:

\begin{enumerate}
    \item If the subgraph contains the node \( X \), the connections among the three edges will be colored green.
    \item If the subgraph contains at least one node from the subgroup \(\mathcal{B}\) and does not contain the node \( X \), the edges will be colored red.
    \item If the subgraph contains only nodes from the subgroup \(\mathcal{A}\), the edges will be colored blue.
\end{enumerate}

We now present the canonical graphs for dimension 3, which also include the two graphs from dimension 2. The first graph from dimension 2 consists of the nodes \( X \) and \( = \) connected by the edge \( X \). The second graph from dimension 2, associated with a dictionary of size 4, is also included in the dimension 3 graphs. We omit the node \( = \) because it does not provide new information and makes the graph less comprehensible. Along with the canonical graphs, we present their corresponding constraint systems, a brief description, and examples of valid dictionaries for each graph.

\begin{itemize}

   \item \textbf{Dictionary Size = 4:} All variables, excluding X, can switch roles freely. The entries of the dictionary can permute in a total number of $2$ valid permutations. The way to permute the variables of a valid dictionary into automorphic configurations in the canonical graph is isomorphic to the group action of $S_2$\cite{rotman2012introduction}, having the reflection axis crossing through node X and the middle of edge X.
    
\begin{align*}
\alpha: &\quad A_1 \oplus A_2 = X \\
\end{align*}

    \begin{figure}[H]
    \centering
    \includegraphics[width=0.4\textwidth]{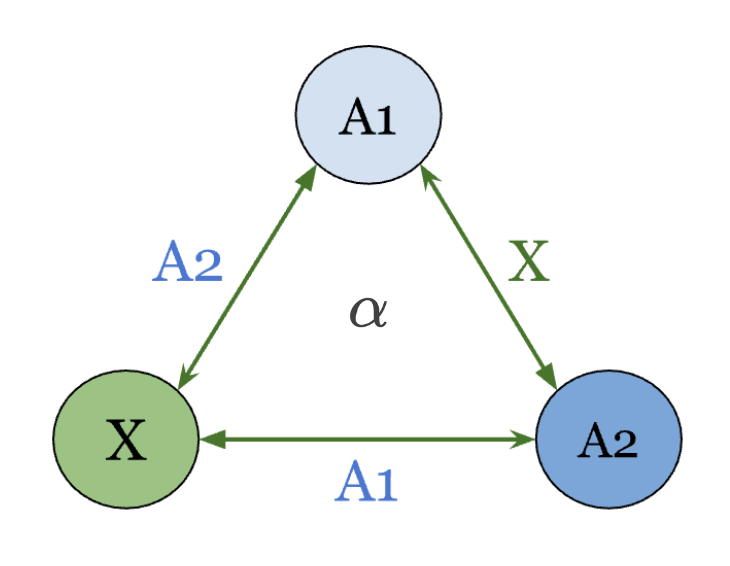}
    \caption{Canonical graph of the dictionaries of 4 elements.}

\end{figure}

\begin{table}[H]
\centering
\caption{Example of valid dictionaries accompanied by valid permutations in the canonical graph. Listed next to each variable are the possible abstract variables that each dictionary entry can take. When viewed in columns, the configurations shown represent valid solutions that satisfy the system. If a variable only has one value, it means that in all permutations, it is consistently placed in the same position.}

\begin{tabular}{|c|l|}
\hline
\textbf{Entry} & \textbf{Value} \\
\hline
A   & 00110($A_1|A_2$) \\
B   & 11001($A_2|A_1$) \\
X   & 11111 \\
=   & 00000 \\
\hline
\end{tabular}
\end{table}

\item \textbf{Dictionary Size = 6:} The variables, X, one variable type $\mathcal{A}$, and the unique variable of type $\mathcal{B}$ can not switch positions.  The way the variables of a valid dictionary permute into automorphic configurations in the canonical graph is isomorphic to the group action of $S_2$, having the reflection axis crossing through node $A_3$ and the two edges $A_3$. 

\begin{align*}
\alpha: &\quad A_1 \oplus A_2 = A_3 \\
\beta: &\quad B_1 \oplus \alpha_1 = X \quad \text{where } \alpha_{1} \in \{ A_1,  A_2 , A_3 \}
\end{align*}

\begin{figure}[H]
    \centering
    \includegraphics[width=0.4\textwidth]{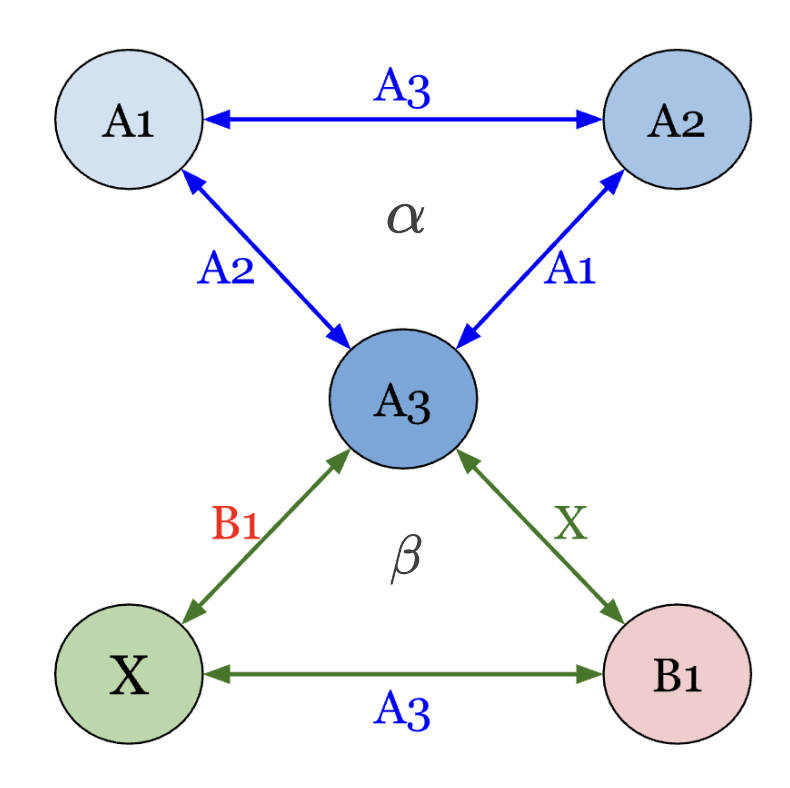}
    \caption{Canonical graph of the dictionaries of 6 elements.}
\end{figure}

\begin{table}[H]
\centering
\caption{Example of valid dictionaries accompanied by valid permutations in the canonical graph. Listed next to each variable are the possible abstract variables that each dictionary entry can take. When viewed in columns, the configurations shown represent valid solutions that satisfy the system. If a variable only has one value, it means that in all permutations, it is consistently placed in the same position.}
\begin{tabular}{|c|l|l|l|l|}
\hline
\textbf{Entry} & \textbf{ Value} & \textbf{ Value} & \textbf{ Value} & \textbf{ Value} \\
\hline
A   & 01001 ($A_1|A_2$)  & 0110010 ($A_1|A_2$) & 000011 ($A_1|A_2$)  & 00011111 ($A_1|A_2$)\\
B  & 01110 ($A_3$)  & 0110001 ($A_3$) & 000010 ($A_3$)& 00011110 ($A_3$)\\
C   & 00111 ($A_2|A_1$)  & 0000011 ($A_2|A_1$) & 000001 ($A_2|A_1$)& 00000001 ($A_2|A_1$)\\
D   & 10001 ($B_1$)  & 1001110 ($B_1$) & 111101 ($B_1$)& 11100001 ($B_1$)\\
X          & 11111 (X)       & 1111111 (X)     & 111111 (X)     & 11111111 (X)\\
=          & 00000 (=)       & 0000000 (=)     & 000000 (=)     & 00000000 (=) \\
\hline
\end{tabular}
\end{table}

\item \textbf{Dictionary Size = 7:} The variables, X and one variable type $\mathcal{A}$, cannot switch positions. There are four remaining variables, two type $\mathcal{A}$ and two type $\mathcal{B}$, such that defining one variable in type $\mathcal{A}$ will define the other variable type $\mathcal{B}$, and vice versa. There is a total of $2$ valid configurations for each valid dictionary.  The way the variables of a valid dictionary permute into automorphic configurations in the canonical graph is isomorphic to the group action of $S_2$, having the reflection axis crossing through node $A_1$ and the two edges $A_1$.

\begin{align*}
\alpha: &\quad A_1 \oplus A_2 = A_3 \\
\beta: &\quad B_1 \oplus B_2 = \alpha_{1} \quad \text{where } \alpha_{1} \in \{ A_1, A_2, A_3 \} \\
\lambda_{1}: &\quad \alpha_{2} \oplus \beta_{1} = X \quad \text{where } \alpha_{2} \in \{ A_1, A_2, A_3 \}, \, \alpha_{2} \neq \alpha_{1}, \text{ ; } \beta_{1} \in \{ B_1, B_2 \} \\
\lambda_{2}: &\quad \alpha_{3} \oplus \beta_{2} = X \quad \text{where } \alpha_{3} \in \{ A_1, A_2, A_3 \}, \, \alpha_{3} \neq \alpha_{1}\neq \alpha_{2}, \text{ ; } \beta_{2} \in \{ B_1, B_2 \}, \, \beta_{2} \neq \beta_{1}
\end{align*}

\begin{figure}[H]
    \centering
    \includegraphics[width=0.65\textwidth]{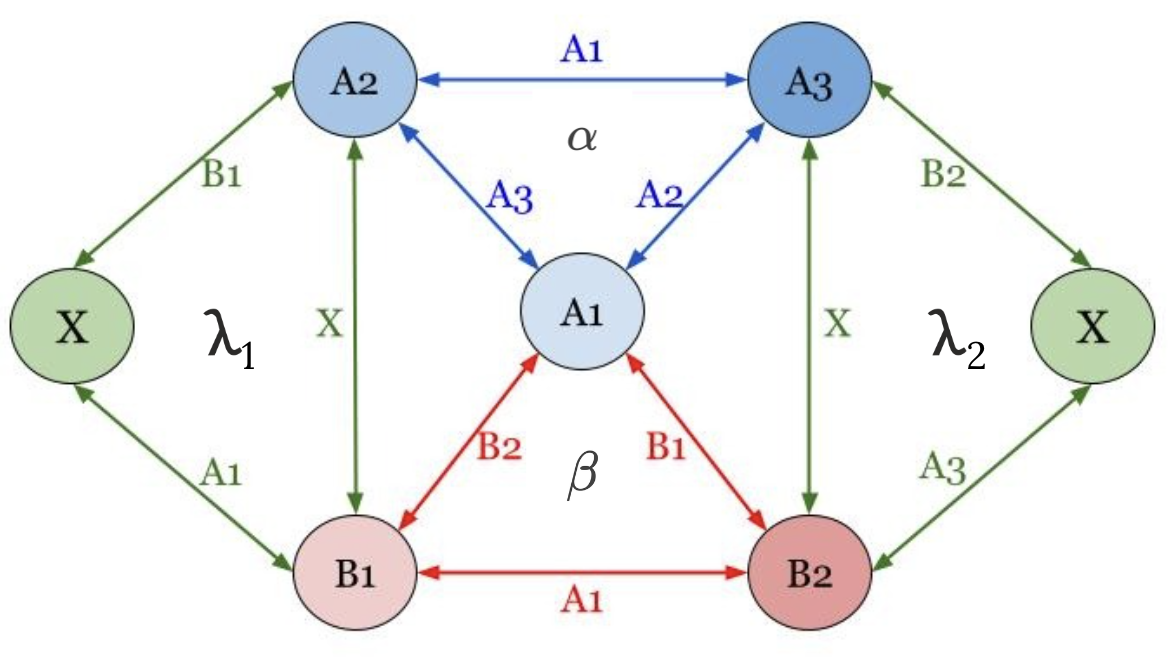}
        \caption{Canonical graph of the dictionaries of 7 elements.}
\end{figure}

\begin{table}[H]
\centering
\caption{Example of valid dictionaries accompanied by valid permutations in the canonical graph. Listed next to each variable are the possible abstract variables that each dictionary entry can take. When viewed in columns, the configurations shown represent valid solutions that satisfy the system. If a variable only has one value, it means that in all permutations, it is consistently placed in the same position.}
\begin{tabular}{|c|l|l|l|l|}
\hline
\textbf{Entry} & \textbf{ Value} & \textbf{ Value} & \textbf{ Value} & \textbf{ Value} \\
\hline
A   & 00000010 ($A_2|A_3$)  &  0010($A_2|A_3$) &  00011($A_2|A_3$)  &  010100($A_1$)\\
B  & 00011110 ($A_1$)      &  0111($A_1$) &  00111($A_3|A_2$)&  001101($A_2|A_3$)\\
C   & 00011100 ($A_3|A_2$)  &  0101($A_3|A_2$) &  00100($A_1$)&  011001($A_3|A_2$)\\
D   & 11111101 ($B_1|B_2$)  &  1101($B_1|B_2$) &  11100($B_1|B_2$)&  100110($B_2|B_1$)\\
E  & 11100011 ($B_2|B_1$)  &  1010($B_2|B_1$) &  11000($B_2|B_1$)&  110010($B_1|B_2$)\\
X          & 11111111 (X)       & 1111 (X)     & 11111 (X)     & 111111 (X)\\
=          & 00000000 (=)       & 0000 (=)     & 00000 (=)     & 000000 (=) \\
\hline
\end{tabular}
\end{table}

    \item \textbf{Dictionary Size = 8}: In this graph, except for \( X \), all possible permutations of \(\mathcal{A}\) or \(\mathcal{B}\) are allowed, resulting in a total of 6 valid configurations for each dictionary. The way the variables of a valid dictionary permute into automorphic configurations in the canonical graph is isomorphic to the group \( D_3 \). This can be visualized by permuting the \(\mathcal{A}\)-type variables in the \(\alpha\) constraint.

    In the system, the constraint named \( \beta \), \( B_1 \oplus B_2 \oplus B_3 = X \), involves four nodes, so we cannot represent it in the graph. However, by operating with it and using the Cayley table of any valid dictionary, we can derive the other constraints, \( \lambda \) and \( \pi \).
     
    \begin{itemize}
        \item The $\lambda$ constraints arise from reducing two variables of type $\mathcal{B}$ within the $\beta$ constraint using XOR. This reduction results in a type $\mathcal{A}$ variable. Each $\lambda$ constraint represents a unique combination of two type $\mathcal{B}$ variables under the XOR operation.

        \item The $\pi$ constraints appear after applying the three variables type $\mathcal{B}$ at bought sides of the constrain $\beta$ under XOR operation. On the left-hand side, one variable cancels out, and on the right-hand side, when combined with $X$, it results in another type $\mathcal{A}$ variable.
    \end{itemize}

\begin{align*}
\alpha: &\quad A_1 \oplus A_2 = A_3 \\
\beta: &\quad B_1 \oplus B_2 \oplus B_3 = X \\
\lambda_1: &\quad \alpha_1 \oplus \beta_1 = X \text{ where } \alpha_1 \in \{A_1,  A_2 , A_3 \} \text{ ; } \beta_1 \in \{B_1,  B_2 , B_3\} \\
\lambda_2: &\quad \alpha_2 \oplus \beta_2 = X \text{ where } \alpha_2 \in \{A_1,  A_2 , A_3 \}, \alpha_{2} \neq \alpha_{1} \text{ ; } \beta_2 \in \{B_1,  B_2 , B_3\}, \beta_{2} \neq \beta_{1} \\
\lambda_3: &\quad \alpha_3 \oplus \beta_3 = X \text{ where } \alpha_3 \in \{A_1,  A_2 , A_3 \}, \alpha_3 \neq \alpha_{2} \neq \alpha_{1} \text{ ; } \beta_3 \in \{B_1,  B_2 , B_3\}, \beta_3 \neq \beta_{2} \neq \beta_{1} \\
\pi_1: &\quad \beta_1 \oplus \beta_2 = \alpha_1 \\
\pi_2: &\quad \beta_2 \oplus \beta_3 = \alpha_2 \\
\pi_3: &\quad \beta_1 \oplus \beta_3 = \alpha_3 \\
\end{align*}


\begin{figure}[H]
    \centering
    \includegraphics[width=0.7\textwidth]{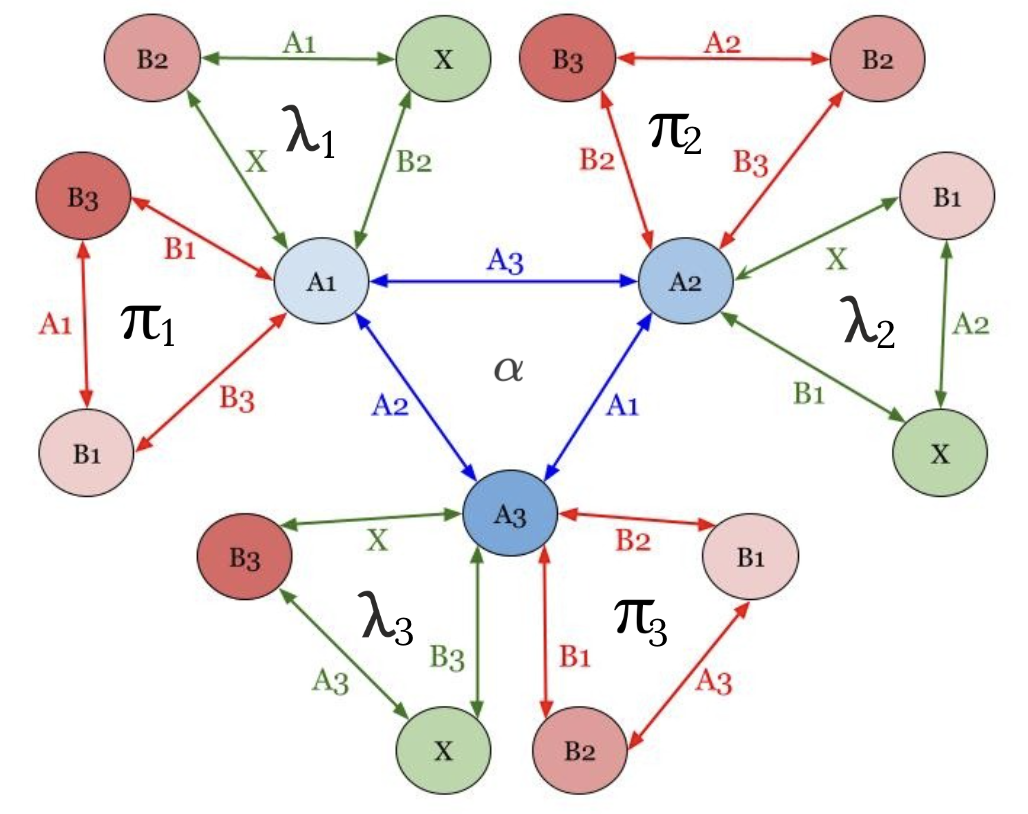}
    \caption{Canonical graph of the dictionaries of 8 elements.}

\end{figure}

\begin{table}[H]
\centering
\caption{Example of valid dictionaries accompanied by valid permutations in the canonical graph. Listed next to each variable are the possible abstract variables that each dictionary entry can take. When viewed in columns, the configurations shown represent valid solutions that satisfy the system. If a variable only has one value, it means that in all permutations, it is consistently placed in the same position.}
\begin{tabular}{|c|l|l|l|l|}
\hline
\textbf{Entry} & \textbf{ Value} & \textbf{ Value} & \textbf{ Value} & \textbf{ Value} \\
\hline
A   &  00111($A_1|A_2|...$) & 00111000($A_1|A_2|...$)  &0010($A_1|A_2|...$)&  00001($A_1|A_2|...$) \\
B  &   01001($A_2|A_3|...$)  &   01011111($A_2|A_3|...$)&  0101($A_2|A_3|...$)& 01111($A_2|A_3|...$) \\
C   & 01110($A_3|A_1|...$)  &   01100111($A_3|A_1|...$)&  0111($A_3|A_1|...$)& 01110($A_3|A_1|...$) \\
D   &  10001($B_3|B_2|...$) &  10100000($B_1|B_3|...$)&  1000($B_3|B_2|...$)& 10000($B_1|B_3|...$)\\
E  & 11000($B_2|B_1|...$)  &   10011000($B_3|B_2|...$)& 1101($B_2|B_1|...$) & 10001($B_3|B_2|...$) \\
F  & 10110($B_1|B_3|...$)  &  11000111($B_2|B_1|...$)&  1010($B_1|B_3|...$)& 11110($B_2|B_1|...$) \\
X          & 11111 (X)       & 11111111 (X)     & 1111 (X)     & 11111 (X)\\
=          & 00000 (=)       & 00000000 (=)     & 0000 (=)     & 00000 (=) \\
\hline
\end{tabular}
\end{table}

\end{itemize}

The dictionaries that satisfy a canonical graph with \textit{n}-number of variables also exhibit cumulative XOR with the subset generator with \textit{n}-number of variables, as shown in the property \textbf{e} of the subset generator.


\begin{conjecture} After substituting the variables from a dictionary that represents a valid solution for any of the isomorphic graphs into an \( SG \) with the same number of variables, the cumulative XOR yields a string of ones across all paths or dimensional components. \textbf{This property holds regardless of the subset size or dimension.} \end{conjecture}

This conjecture can be illustrated by the following case: given the dictionary variables \( A: 0101 \), \( B: 1010 \), \( X: 1111 \), and \( =: 0000 \), we substitute these values and perform the cumulative XOR operation across three subset generators (\( SG \)): two in dimension 3 with subset sizes of 16 and 24, and one in dimension 4 with a subset size of 32.

\begin{table}[H]
\centering
\caption{Three $SG$ of different sizes with four variables, and two of them in the same dimension.}

\begin{tabular}{|c|c|c|}
\hline
 $SG\in \mathbb{|T|}=16 \in \mathbf{X}^3_K$ &  $SG\in \mathbb{|T|}=24 \in \mathbf{X}^3_K$&  $SG\in \mathbb{|T|}=32 \in \mathbf{X}^4_K$\\
\hline

A, A, =  & =, A, A  & A, A, A, A \\
A, A, X  & A, A, =  & A, A, A, B \\
A, A, =  & A, =, B  & A, A, A, A \\
A, B, X  & =, A, A  & A, A, B, B \\
A, A, =  & A, A, =  & A, A, A, A \\
A, A, X  & A, B, X  & A, A, A, B \\
A, A, =  & =, A, A  & A, A, A, A \\
X, X, X  & A, =, A  & A, B, B, B \\
A, A, =  & A, A, X  & A, A, A, A \\
A, A, X  & =, A, A  & A, A, A, B \\
A, A, =  & A, =, A  & A, A, A, A \\
A, B, X  & X, X, X  & A, A, B, B \\
A, A, =  & A, =, A  & A, A, A, A \\
A, A, X  & =, A, A  & A, A, A, B \\
A, A, =  & A, A, X  & A, A, A, A \\
         & A, =, A  & X, X, X, X \\
         & =, A, A  & A, A, A, A \\
         & A, B, X  & A, A, A, B \\
         & A, A, =  & A, A, A, A \\
         & =, A, A  & A, A, B, B \\
         & A, =, B  & A, A, A, A \\
         & A, A, =  & A, A, A, B \\
         & =, A, A  & A, A, A, A \\
         &          & A, B, B, B \\
         &          & A, A, A, A \\
         &          & A, A, A, B \\
         &          & A, A, A, A \\
         &          & A, A, B, B \\
         &          & A, A, A, A \\
         &          & A, A, A, B \\
         &          & A, A, A, A \\
\hline
\end{tabular}
\end{table}

\begin{table}[H]
\centering
\caption{$SG$ with 4 variables after calculating the cumulative XOR with the variables $A: 0101$ and $B: 1010$. At the end of the $SG$, in each dimensional component, we observe a string of 1s.}
\begin{tabular}{|c|c|c|}
\hline
 $SG\in \mathbb{|T|}=16 \in \mathbf{X}^3_K$ &  $SG\in \mathbb{|T|}=24 \in \mathbf{X}^3_K$&  $SG\in \mathbb{|T|}=32 \in \mathbf{X}^4_K$\\
\hline
0101, 0101, 0000  & 0000, 0101, 0101  & 0101, 0101, 0101, 0101 \\
0000, 0000, 1111  & 0101, 0000, 0101  & 0000, 0000, 0000, 1111 \\
0101, 0101, 1111  & 0000, 0000, 1111  & 0101, 0101, 0101, 1010 \\
0000, 1111, 0000  & 0000, 0101, 1010  & 0000, 0000, 1111, 0000 \\
0101, 1010, 0000  & 0101, 0000, 1010  & 0101, 0101, 1010, 0101 \\
0000, 1111, 1111  & 0000, 1010, 0101  & 0000, 0000, 1111, 1111 \\
0101, 1010, 1111  & 0000, 1111, 0000  & 0101, 0101, 1010, 1010 \\
1010, 0101, 0000  & 0101, 1111, 0101  & 0000, 1111, 0000, 0000 \\
1111, 0000, 0000  & 0000, 1010, 1010  & 0101, 1010, 0101, 0101 \\
1010, 0101, 1111  & 0000, 1111, 1111  & 0000, 1111, 0000, 1111 \\
1111, 0000, 1111  & 0101, 1111, 1010  & 0101, 1010, 0101, 1010 \\
1010, 1010, 0000  & 1010, 0000, 0101  & 0000, 1111, 1111, 0000 \\
1111, 1111, 0000  & 1111, 0000, 0000  & 0101, 1010, 1010, 0101 \\
1010, 1010, 1111  & 1111, 0101, 0101  & 0000, 1111, 1111, 1111 \\
1111, 1111, 1111  & 1010, 0000, 1010  & 0101, 1010, 1010, 1010 \\
                  & 1111, 0000, 1111  & 1010, 0101, 0101, 0101 \\
                  & 1111, 0101, 1010  & 1111, 0000, 0000, 0000 \\
                  & 1010, 1111, 0101  & 1010, 0101, 0101, 1010 \\
                  & 1111, 1010, 0101  & 1111, 0000, 0000, 1111 \\
                  & 1111, 1111, 0000  & 1010, 0101, 1010, 0101 \\
                  & 1010, 1111, 1010  & 1111, 0000, 1111, 0000 \\
                  & 1111, 1010, 1010  & 1010, 0101, 1010, 1010 \\
                  & 1111, 1111, 1111  & 1111, 0000, 1111, 1111 \\
                  &                   & 1010, 1010, 0101, 0101 \\
                  &                   & 1111, 1111, 0000, 0000 \\
                  &                   & 1010, 1010, 0101, 1010 \\
                  &                   & 1111, 1111, 0000, 1111 \\
                  &                   & 1010, 1010, 1010, 0101 \\
                  &                   & 1111, 1111, 1111, 0000 \\
                  &                   & 1010, 1010, 1010, 1010 \\
                  &                   & 1111, 1111, 1111, 1111 \\
\hline
\end{tabular}

\end{table}

This simple example shows how the dictionary size seems to be transversal to both the dimension and the size of the $SG$. It suggests that all the $SG$ with the same number of variables, similar to how all valid dictionaries of size $n$ derive from the same canonical graph, share some underlying isomorphism.
\newpage


\section{Acknowledgments}

We would like to \textbf{sincerely thank Professor Andrés Sáez-Schwedt} for his invaluable involvement and feedback in this research. Even though this work falls outside his primary area of expertise, he dedicated significant time to offering us a second opinion and providing thoughtful insights that helped refine our research.\\

We would also like to express our sincere thanks to Professor \textbf{Yen-Chi Roger Lin} for the time he invested in reading and understanding this work, as well as for his valuable suggestions on how to improve its mathematical rigor and potential future directions.

\section{Contributions, Future Work and Conclusion}
\subsection{Contributions}
This research, primarily focused on 2D and 3D, offers contributions that can be grouped into three distinct categories:

\begin{enumerate}
    \item \textbf{Development of the circular projection algorithm:} First, we developed an algorithm to project the elements in \( \mathbf{X}^D_K \) onto an \( n \)-dimensional sphere.
    
    \item \textbf{Study of the subset properties:} We focused on the subsets as our primary object of study, observing properties such as, for any \( D \) and \( K \), the sum of elements in a subset is always \( \Psi^D_K \) or a multiple of it. Furthermore, we observed that subset sizes corresponding to elements with the same radius on an \( n \)-dimensional sphere are consistently \( 2^D \) or a multiple of \( 2^D \). We introduced the \( \phi \) function, which projects a point in \( X^D_S \) to its equivalent position in \( X^D_B \). Additionally, through group actions using the matrix representation of both group actions and Morton coordinates, we demonstrated that each point in a subset could be derived from a single element via symmetries and rotations. Lastly, the rotation operation, together with previous findings, enabled us to reinterpret each subset as a collection of orthotopes, \( \mathcal{O} \), where each \( \mathcal{O} \) contributes to a sum one \( \Psi^D_K \).

    \item \textbf{Identification of structural patterns and isomorphisms:} Finally, to understand the origin of these patterns, we observed that all subsets share a common internal structure, which we refer to as \( SG \). Additionally, the dictionary associated with any \( SG \) exhibits structural patterns as shown in the canonical graphs. We conclude by hypothesizing that all subsets using the same number of variables, regardless of subset size or dimension, are isomorphic.

\end{enumerate}

\subsection{Future Work}

Key questions for future work include why all the \( SG \) with the same number of variables, regardless of dimension and subset size, appear to be isomorphic, why certain subsets, such as the subset with 40 elements, do not exist in dimension 3 while the subset with 48 elements does, or why the sum of all subset elements is \(\Psi^D_K\) or a multiple of it. Another potential direction is to pursue formal mathematical proofs for the observed patterns, which have so far been based on empirical results without rigorous mathematical justification. Lastly, expanding this research to include higher dimensions, beyond the current focus on dimensions 2 and 3, offers another direction for future study.


\subsection{Conclusion}

To conclude, many aspects of the circular projection of the \( n \)-dimensional \( Z \)-curve remain unexplored, offering substantial opportunities for further research. While this work provides a foundational step, it is only the beginning. Ongoing refinements to our methods will be essential for deepening our understanding of this structure, potentially revealing new insights and applications in other fields.


\newpage
\appendix
\section{Appendix A: $SG$ of $\mathbb{T}$ when $D=\{2,3 \}$}
\label{appendix:subset_generators}

In dimension 2  always $|\mathbb{T}|=\{4,8\}$. The $SG$ of them are :\\
\begin{itemize}
    \item $SG(|\mathbb{T}|=4)$ : 
\[
\begin{array}{|c|}
\hline
\text{$SG$ 1} \\
\hline
\text{=, X} \\
\text{X, X} \\
\text{=, X} \\
\hline
\end{array}
\]

    \item $SG(|\mathbb{T}|=8)$ : 

\[
\begin{array}{|c|}
\hline
\text{$SG$ 1} \\
\hline
\text{A, A} \\
\text{A, B} \\
\text{A, A} \\
\text{X, X} \\
\text{A, A} \\
\text{A, B} \\
\text{A, A} \\
\hline
\end{array}
\]

\end{itemize}

In dimension 3 always $|\mathbb{T}|=\{8,16,24,32,48\}$. The $SG$ of them are :\\
\begin{itemize}
    \item $SG(|\mathbb{T}|=8)$ : 
\[
\begin{array}{|c|}
\hline
\text{$SG$ 1} \\
\hline
=, =, X \\
=, X, X \\
=, =, X \\
X, X, X \\
=, =, X \\
=, X, X \\
=, =, X \\
\hline
\end{array}
\]

\newpage
    \item $SG(|\mathbb{T}|=16)$ : 

\[
\begin{array}{|c|}
\hline
\text{$SG$ 1} \\
\hline
A, A, = \\
A, A, X \\
A, A, = \\
A, B, X \\
A, A, = \\
A, A, X \\
A, A, = \\
X, X, X \\
A, A, = \\
A, A, X \\
A, A, = \\
A, B, X \\
A, A, = \\
A, A, X \\
A, A, = \\
\hline
\end{array}
\]

    \newpage
    \item $SG(|\mathbb{T}|=24)$ : 

\[
\begin{array}{|c|c|}
\hline
\text{$SG$ 1} & \text{$SG$ 2} \\
\hline
=,A,A & A,A,= \\
A,A,= & =,A,A \\
A,=,B & A,=,B \\
=,A,A & A,A,= \\
A,A,= & =,A,A \\
A,B,X & A,X,B \\
=,A,A & A,=,A \\
A,=,A & =,A,A \\
A,A,X & A,A,X \\
=,A,A & A,=,A \\
A,=,A & =,A,A \\
X,X,X & X,X,X \\
A,=,A & =,A,A \\
=,A,A & A,=,A \\
A,A,X & A,A,X \\
A,=,A & =,A,A \\
=,A,A & A,=,A \\
A,B,X & A,X,B \\
A,A,= & =,A,A \\
=,A,A & A,A,= \\
A,=,B & A,=,B \\
A,A,= & =,A,A \\
=,A,A & A,A,= \\
\hline

\end{array}
\]

    \item $SG(|\mathbb{T}|=32)$ :
\[
\begin{array}{|c|c|c|c|c|c|}
\hline
\text{$SG$ 1} & \text{$SG$ 2} & \text{$SG$ 3} & \text{$SG$ 4} & \text{$SG$ 5} & \text{$SG$ 6} \\
\hline
A,A,= & A,A,= & A,A,= & =,A,A & A,A,= & A,=,A \\
B,A,C & =,B,B & =,B,B & B,C,A & A,B,C & B,C,A \\
B,B,= & C,C,= & C,C,= & A,=,A & B,B,= & =,A,A \\
A,B,D & B,=,D & B,=,D & C,B,D & B,A,D & C,B,D \\
A,A,= & A,A,= & A,A,= & =,A,A & A,A,= & A,=,A \\
B,A,C & =,B,B & =,B,B & B,C,A & A,B,C & B,C,A \\
B,B,= & C,C,= & C,C,= & A,=,A & B,B,= & =,A,A \\
A,E,D & B,X,D & C,E,X & C,E,X & B,E,X & C,E,D \\
C,=,C & A,C,B & A,C,B & =,A,A & =,C,C & A,=,A \\
B,B,= & C,C,= & A,A,= & B,B,= & A,A,= & B,B,= \\
=,C,C & C,A,B & C,A,B & A,=,A & C,=,C & =,A,A \\
A,A,X & A,A,X & C,C,X & C,C,X & B,B,X & C,C,X \\
C,=,C & A,C,B & A,C,B & =,A,A & =,C,C & A,=,A \\
B,B,= & C,C,= & A,A,= & B,B,= & A,A,= & B,B,= \\
=,C,C & C,A,B & C,A,B & A,=,A & C,=,C & =,A,A \\
X,X,X & X,X,X & X,X,X & X,X,X & X,X,X & X,X,X \\
=,C,C & C,A,B & C,A,B & A,=,A & C,=,C & =,A,A \\
B,B,= & C,C,= & A,A,= & B,B,= & A,A,= & B,B,= \\
C,=,C & A,C,B & A,C,B & =,A,A & =,C,C & A,=,A \\
A,A,X & A,A,X & C,C,X & C,C,X & B,B,X & C,C,X \\
=,C,C & C,A,B & C,A,B & A,=,A & C,=,C & =,A,A \\
B,B,= & C,C,= & A,A,= & B,B,= & A,A,= & B,B,= \\
C,=,C & A,C,B & A,C,B & =,A,A & =,C,C & A,=,A \\
A,E,D & B,X,D & C,E,X & C,E,X & B,E,X & C,E,D \\
B,B,= & C,C,= & C,C,= & A,=,A & B,B,= & =,A,A \\
B,A,C & =,B,B & =,B,B & B,C,A & A,B,C & B,C,A \\
A,A,= & A,A,= & A,A,= & =,A,A & A,A,= & A,=,A \\
A,B,D & B,=,D & B,=,D & C,B,D & B,A,D & C,B,D \\
B,B,= & C,C,= & C,C,= & A,=,A & B,B,= & =,A,A \\
B,A,C & =,B,B & =,B,B & B,C,A & A,B,C & B,C,A \\
A,A,= & A,A,= & A,A,= & =,A,A & A,A,= & A,=,A \\
\hline
\end{array}
\]

    \item $SG(|\mathbb{T}|=48)$ : 

\[
\begin{array}{|c|c|}
\hline
\text{$SG$ 1} & \text{$SG$ 2} \\
\hline

=,A,A & A,A,= \\
B,C,A & A,B,C \\
A,=,A & A,=,A \\
A,C,B & C,B,A \\
A,A,= & =,A,A \\
C,=,D & B,=,D \\
=,A,A & A,A,= \\
B,C,A & A,B,C \\
A,=,A & A,=,A \\
A,C,B & C,B,A \\
A,A,= & =,A,A \\
C,E,F & B,E,F \\
=,A,A & A,=,A \\
B,A,C & A,C,B \\
A,A,= & A,A,= \\
A,B,C & C,A,B \\
A,=,A & =,A,A \\
C,C,X & B,B,X \\
=,A,A & A,=,A \\
B,A,C & A,C,B \\
A,A,= & A,A,= \\
A,B,C & C,A,B \\
A,=,A & =,A,A \\
X,X,X & X,X,X \\
A,=,A & =,A,A \\
A,B,C & C,A,B \\
A,A,= & A,A,= \\
B,A,C & A,C,B \\
=,A,A & A,=,A \\
C,C,X & B,B,X \\
A,=,A & =,A,A \\
A,B,C & C,A,B \\
A,A,= & A,A,= \\
B,A,C & A,C,B \\
=,A,A & A,=,A \\
C,E,F & B,E,F \\
A,A,= & =,A,A \\
A,C,B & C,B,A \\
A,=,A & A,=,A \\
B,C,A & A,B,C \\
=,A,A & A,A,= \\
C,=,D & B,=,D \\
A,A,= & =,A,A \\
A,C,B & C,B,A \\
A,=,A & A,=,A \\
B,C,A & A,B,C \\
=,A,A & A,A,= \\
\hline
\end{array}
\]
\end{itemize}

\newpage

\bibliographystyle{plainnat}   
\bibliography{main}            

\end{document}